\documentclass[a4paper,12pt]{report}
\usepackage[utf8]{inputenc}
\usepackage[brazil]{babel}
\usepackage{epsfig}
\usepackage{subfigure}
\usepackage{float}
\usepackage{color}
\usepackage{latexsym}
\usepackage{graphicx}
\usepackage{amssymb}
\usepackage{amsmath}
\usepackage{amsfonts}
\usepackage{hyperref}
\usepackage{epstopdf}	
\begin{document}
\thispagestyle{empty}
\begin{center}
{\large{Universidade Federal de Itajubá\\
Instituto de Física e Química}}\\
\vspace{2cm}
{\bf \Large{Evolução de campos de matéria em buracos negros com simetria de Lifshitz}}\\
\vspace{2cm}
{\bf João Ider Silva Júnior\\
Orientador: Prof. Dr. Alan Bendasoli Pavan}
\end{center}
\vspace{1cm}
\begin{flushright}
\begin{minipage}[t]{7.5cm}
\hrulefill\\
Dissertação de mestrado apresentada ao Instituto de Física e Química da Universidade Federal de Itajubá para a obtenção do título de Mestre em Ciências

\hrulefill
\end{minipage}
\end{flushright}
\vspace{1cm}
\noindent
{\bf Comissão Examinadora:}\\
Prof. Dr. Alan Bendasoli Pavan (IFQ - UNIFEI)\\
Prof. Dr. Bertha Cuadros-Melgar (EEL - USP/Lorena)\\
Prof. Dr. Alexis Roa Aguirre (IFQ - UNIFEI)\\
Prof. Dr. Eduardo Henrique Silva Bittencourt (IFQ - UNIFEI)
\vspace{1.0cm}
\begin{center}
Itajubá\\
2018
\end{center}
\newpage
\thispagestyle{empty}
\chapter*{Resumo}
Neste trabalho foram analisadas as evoluções de dois campos de matéria em buracos negros planarmente simétricos $D$-dimensionais com simetria de Lifshitz cujo expoente dinâmico é $z$. Os campos investigados foram o campo escalar com acoplamento não-mínimo com o tensor de Einstein e o escalar de Ricci e o campo eletromagnético. Foram escolhidos dois buracos negros, um com 5-dimensões e $z=1$ e outro com 6-dimensões e $z=0$. As equações de movimento para ambos os campos foram desenvolvidas de modo geral e aplicadas a cada buraco negro mencionado. Em alguns casos foram obtidas soluções exatas das equações em termos de funções hipergeométricas e funções de Heun confluente.
Os modos quasinormais (MQN) de evolução dos campos analisados foram calculados numericamente usando dois métodos diferentes: HH e AIM. Em todos os buracos negros estudados os MQN's não indicaram instabilidades nem referente aos acoplamentos nem aos espaços-tempos. De modo geral os modos quasinormais se comportam como os modos de oscilação de um oscilador harmônico amortecido apresentando três regimes: subamortecido $(\omega_R\neq0,\omega_I<0)$, crítico $(\omega_R=0,\omega_I^{crit}<0)$ e supercrítico $(\omega_R=0,\omega_I<0)$. Nos casos analisados foi encontrado um comportamento interessante para $\omega_I$ quando $k\neq0$ e $r_+<10$. Em geral, $\omega_I$ é crescente com $k$, mas aqui encontramos um crescimento quadrático para $k\sim r_+$ com um decréscimo quando $k>>r_+$. Esse comportamento podem ser relevante no contexto da dualidade gravidade/calibre. Como complemento também foram analisados dois buracos negros com simetria de Lifshtiz em $2+1$ dimensões.

\chapter*{Abstract}
In this work the evolution of two fields of matter in planar symmetric black holes $D$-dimensional with symmetry of Lifshitz whose dynamic exponent is $ z $ were analyzed. The fields investigated were a scalar field non-minimal coupled to the Einstein tensor and the Ricci scalar and the electromagnetic field. Two black holes were chosen, one with 5-dimensions and $z=1$ and another with 6-dimensions and $z=0$. The equations of motion for both fields were developed in general and applied to each black hole mentioned. In some cases exact solutions of the equations were obtained in terms of hypergeometric functions and confluent Heun functions.
The quasinormal modes (MQN) of evolution of the analyzed fields were calculated numerically using two different approaches: HH and AIM. In all the black holes studied the NSM's did not indicate instabilities neither regarding the couplings nor the space-times. In general, the quasinormal modes behave like the oscillation modes of a damped harmonic oscillator, presenting three regimes: underdamped $(\omega_R \neq0, \omega_I <0) $, critically damped $ (\omega_R=0,\omega_I^{crit} <0) $ and overdamped $ (\omega_R=0,\omega_I <0) $. In the analyzed cases an interesting behavior was found for $\omega_I $ when $ k \neq0 $ and $ r_+<10 $. In general, $\omega_I $ increase with $k$, but here we find a quadratic growth for $ k\sim r_+$ and then a decrease when $k>>r_+$. This behavior may be relevant in the context of gravity/gauge duality. Additionally, we also analyzed the evolution of the same fields in two black holes with symmetry of Lifshtiz in $ 2 + 1 $ dimensions.

\newpage

\hspace{1cm}\emph{``Os céus declaram a glória de Deus e o firmamento anuncia a obra das suas mãos"  Salmos 19.1; ``Tal ciência é para mim maravilhosíssima; tão alta, que não a posso atingir"  Salmos 139.6}.

\chapter*{Agradecimentos}
Agradeço a Deus, o idealizador de tudo, que tem me dado forças e sabedoria para a cada dia conhecer um pouco mais sobre sua maravilhosa criação.

Aos meus pais João Ider e Maria Angélica pelo apoio e amor incondicionais em todos os momentos em que mais precisei. Aos meus irmãos Pamella, Felipe e Matheus pelas brigas, exemplo e conselhos.

À minha amada esposa Mariana Veloso, pelo carinho e compreensão, por estar comigo todo esse tempo e sua paciência entendendo minhas ausências para conclusão deste trabalho.

Aos meus amigos e irmãos da madrugada Lucas Faria, Robert Oliveira e Thiago Joaquim pelas longas conversas e a P5 de sempre.

Ao meu orientador Alan Pavan, que tem me dado suporte desde a graduação, me ensinando e guiando, e passando um pouco da sua paixão pela física.

À Capes, por todo apoio financeiro e incentivo à pesquisa no Brasil.
\newpage
\tableofcontents
\newpage
\chapter{Introdução}
\label{int}
O ser humano vive sempre em busca de explicações para o mundo ao seu redor. Cada novo cientista contribui um pouco para elas se tornarem cada vez mais precisas. A partir de cada pesquisa são desenvolvidas novas teorias e o conhecimento vai se aprimorando.

Einstein expandiu o conhecimento Newtoniano. As conhecidas leis de Newton funcionavam bem dentro de um certo regime, mas foi necessário ir além, além do que se conhecia ou imaginava. Sua teoria veio para aprimorar o conhecimento newtoniano e ir para muito longe, a distâncias cosmológicas e de maneira muito rápida, à velocidade da luz.

A Teoria da Relatividade revolucionou o nosso entendimento do mundo de diversas maneiras e, uma delas, foi algo que diz respeito a uma certa entidade na qual todos repousam. Com Newton, o local onde todas as coisas estão e por onde se movem era um local rígido e estático. Este espaço no qual todos se locomovem e este tempo o qual passa diante de todos subiram a um novo patamar e se uniram para serem um espaço-tempo, algo tão revolucionário e impressionante até os dias de hoje.

Tal revolução se iniciou com a formulação da Relatividade Restrita \cite{einstein} e cerca de uma década depois com a Teoria da Relatividade Geral, que descreve a Gravitação relativisticamente. Esta teoria transforma agora a entidade em algo que possui dinâmica e movimento, e isso ampliou o entendimento e abriu os nossos olhos a um novo mundo cheio de surpresas e possibilidades.

Algumas da possibilidades são a respeito do próprio Universo que agora pode possuir movimento, ter começo, meio e fim, se expandir ou contrair, ser vazio ou cheio e mais incrível ainda, não ser o único. Além disso, umas das grandes surpresas desta teoria fica por conta de um objeto muito exótico chamado de {\it buraco negro}.

A idéia de um campo gravitacional tão forte que impedisse até mesmo a luz de sair surgiu ainda no contexto da Gravitação newtoniana com os trabalhos independentes de Michell (1783) \cite{michell} e Laplace (1796) \cite{laplace}. Inicialmente ele recebeu o nome de {\it estrela escura} e só em 1968 que Wheeler os batizou como buracos negros.

Buracos Negros são soluções de vácuo da equação de Einstein. Schwarzschild (1916) obteve a primeira solução exata e era esfericamente simétrica. Nela havia uma singularidade física em $r=0$ e uma outra em $r=2M$ ($M$ é a massa do buraco negro) devido ao sistema de coordenadas adotado. Logo, outras soluções apareceram, como a de Reisner (1916) e Nordstrom (1918) para um buraco negro com carga elétrica, por Snyder e Oppenheimer (1939), onde mostram que um buraco negro é obtido como resultado do colapso gravitacional de uma estrela massiva e de Kerr (1963), para um buraco negro em rotação.

Cada dia mais, através de estudos, é visto que buracos negros são objetos cuja física envolvida é riquíssima. Eles não tem tido papel importante apenas astrofisicamente, mas também fundamental do ponto de vista teórico. Eles tem contribuído bastante para o entendimento de aspectos do comportamento quântico da Gravitação e na busca de uma Teoria Quântica da Gravitação.

A relatividade de Einstein, em conjunto com a mecânica quântica, revolucionaram a maneira com que o universo é percebido. Toda a complexidade que se vê no mundo pode surgir do acaso, conforme previsto pela teoria quântica, enquanto nas escalas astronômicas, a própria evolução do universo pode ser descrita a partir de condições iniciais, utilizando-se a relatividade de Einstein.

A teoria quântica de campos se originou do casamento da relatividade especial com a mecânica quântica. A eletrodinâmica quântica (nome que recebe a teoria quântica de campos quando aplicada a fenômenos eletromagnéticos) provou-se a mais bem sucedida das teorias físicas, e, foi o seu sucesso em descrever, de forma unificada, pelo menos em parte, as três interações fundamentais das partículas elementares o que motivou muitos cientistas na busca pela compreensão da gravitação - a quarta força - neste mesmo formalismo.

Na segunda metade do século XX o avanço da cosmologia, através da descrição do universo primordial, acrescentou outro importante fator no estudo da gravitação. O Universo descrito pelas equações de Einstein está em evolução, e se iniciou através de uma explosão universal, o chamado Big Bang.

Neste cenário o universo emerge de um plasma cosmológico de temperatura altíssima, o qual, requer a descrição de um fluido cuja energia média por partícula constituiente (ou seja, a temperatura) é muito alta. Deste modo, a interação se dá no âmago da matéria e requer uma descrição eminentemente quântica. Esta foi uma explosão diferente da convencional, pois, aqui, a explosão é universal, em todos os pontos ao mesmo tempo, e a evolução posterior é dada pela teoria das partículas elementares e dos campos, portanto uma teoria quântica.

Vale ressaltar também que a gravitação de Einstein não pode explicar totalmente a explosão inicial, já que neste caso não haveria como se dar uma causa para tal. Unido a esta questão, quando se quer descrever o comportanto posterior do universo, em seus primeiros instantes (os quais são de fundamental importância para a evolução posterior), é necessário a unificação da teoria para todas as interações, e portanto, sendo as outras interações obrigatoriamente quantizadas, não há como se ter uma gravitação simplesmente clássica. Há outras linhas de argumentação equivalentemente incisivas. O comportamento de buracos negros na presença de campos quantizados também leva a uma obrigatoriedade da quantização dos campos gravitacionais.

Sendo a gravitação uma teoria altamente não-linear ela não pode ser quantizada da mesma maneira que os outros campos, pois, isso geraria quantidades infinitas que não podem ser interpretadas fisicamente. O chamado problema da renormalização de uma teoria de campos, que resolve os infinitos que aparecem devido ao caráter operacional dos campos quantizados, não pode ser resolvido em teorias de campos que contenham a gravitação. Desta forma diz-se que a gravitação é uma {\it teoria não renormalizável}.

A descrição da gravidade e mecânica quântica de forma unificada possui várias dificuldades. Apenas pensar em uma gravitação quântica já requer a descrição da geometria por números discretos, ou seja, tanto o tempo quanto o espaço passariam a ser medidos em termos de unidades fundamentais discretas. Assim, um objetivo antigo, já antevisto por Einstein, de se obter uma teoria unificada dos campos, que foi delineada para as outras interações no decorrer das últimas décadas do século XX, encontra uma alta barreira exatamente na teoria da gravitação.

A gravitação pode ser quantizada de várias maneiras, cada uma com seus problemas e consequências. Serão exibidos aqui alguns exemplos de teorias atualmente estudadas, como visto em \cite{elcio}.

Umas delas é a chamada Teoria de Cordas. Essa teoria pode ser entendida a patir de um conceito relativamente simples: as entidades fundamentais da natureza, partículas constituintes da matéria e das interações, não são objetos pontuais, mas fazem parte de pequenas cordas vibrando no espaço-tempo. Diferentes partículas aparecem como diferentes formas de vibração, mas todas estão incluídas na mesma descrição.

Para o funcionamento da teoria explicando a existência de todas estas partículas e forças, é necessário que hajam mais dimensões do que aquelas que são conhecidas, já que este cenário é muito restritivo. Imagina-se que várias dimensões sejam necessárias para as cordas vibrarem de modo a explicar todas as características das partículas fundamentais.

Outra teoria é a chamada de teoria de Horava-Lifshitz. Esta teoria da gravitação quântica procura solucionar o problema da renormalizabilidade e da unitariedade ao sugerir que o espaço-tempo perca sua estrutura isotrópica de tal modo que as equações de movimento sejam invariantes pelo reescalonamento $x^i\to bx^i$ e $t\to b^zt$.  Essa teoria se torna renormalizável para determinados valores do expoente crítico e do número de dimensões espaciais $D$.

Existem ainda outras teorias alternativas de gravitação que se propõem a ser renormalizáveis, por exemplo, teorias que apresentam, em suas Lagrangeanas correções de ordem superior na curvatura, tais como Gauss-Bonet, $R^2$ e outras. Nestas teorias, como temos mais parâmetros, aumenta a variedade de soluções. Um exemplo desse tipo em $2+1$ dimensões é a New Massive Gravity\cite{bergshoeff}. Em dimensões mais altas temos em \cite{lifshitzhigher}, por exemplo, duas teorias de gravitação com correções quadráticas na curvatura. Essas teorias permitem soluções apresentam uma quebra na simetria de Lorentz, base da Relatividade Geral, e apresentam uma nova simetria, a de Lifshitz. Esta é uma simetria de reescalonamento anisotrópico ($z \neq 1$), com
\begin{eqnarray}
x^i \to bx^i\ \ \ , \ \ \ t \to b^z t ,
\end{eqnarray}
e onde $z$ é expoente crítico. O caso $z=1$ representa a equivalência entre espaço e tempo da Relatividade Geral.
Um reescalonamento consiste na multiplicação das distâncias por um fator constante. Como um reescalonamento é equivalente à escolha de uma unidade de comprimento diferente, as teorias físicas fundamentais devem ser invariantes sob esta transformação. Tal invariância da teoria sob reescalonamento do espaço implica, em geral, que o tempo também seja reescalonado.
As equações que descrevem os sistemas físicos devem levar aos mesmos resultados quando transformadas de um sistema inercial para outro, logo, a Relatividade Geral deve ser covariante.

Nestes cenários pode-se analisar como se comportam buracos negros, que obedecem a alguma teoria quântica de gravitação ainda desconhecida.

Buracos negros podem, à princípio, ser imaginados como objetos muito complexos, todavia, se estiverem na situação de equilíbrio são intrinsicamente simples. Se estiverem isolados, sem uma quantidade de matéria em torno deles, apenas poucos parâmetros são necessários para os descreverem. Estes são: a massa, o momento angular e a carga.

Nos casos onde não há o isolamento, casos reais, isso é bem diferente. Quando buracos negros estão nos centros de galáxias ou possuem massa intermediária a distribuição de massa é bem complexa em torno deles. Como exemplo tem-se núcleos galáticos, discos de acreção, campos magnéticos fortes, outras estrelas, planetas, etc., onde tudo isso está interagindo ativamente com seus arredores.

Eis que surge a pergunta: Pode-se simplesmente considerar que eles estão isolados e estudá-los normalmente? E a resposta é não. Não bastaria apenas retirar todo conteúdo de matéria que está interagindo em torno dele. Mesmo neste caso onde não se tem objetos macroscópicos em seu entorno, um buraco negro interage com o vácuo em torno dele, e ainda criará pares de partículas e irá evaporar devido a radiação Hawking.

Se o objetivo é saber sobre estabilidade desses buracos megros, ondas gravitacionais, evaporação Hawking ou sobre interação de buracos negros com seu meio ambiente astrofísico é necessária a análise de suas perturbações, pois, ele nunca será descrito totalmente por seus parâmetros básicos e, assim sendo, sempre estará em um estado perturbativo.

Sabendo que um buraco negro está em um estado perturbativo, pode se perguntar o que acontece com ele quando é perturbado? A resposta é que ele emite ondas gravitacionais cuja evolução no tempo podem ser divididas em três estágios: (1) Um período relativamente curto de explosão de radiação inicial; (2) Um período geralmente longo de amortecimento das oscilações próprias, dominado pelos então chamados modos quasinormais (`quasi' aqui significa que o sistema é aberto e perde energia através da radiação gravitacional); (3) Um período final onde os modos quasinormais são suprimidos por uma cauda (`late-time tails') que satisfaz uma lei de potência ou uma exponencial.

A maior parte dos artigos que estudam perturbações de buracos negros são sobre o segundo estágio da evolução de perturbação representando os modos quasinormais. E, claro, há um grande número de razões para um interesse tão especial nesses modos.

O trabalho de Regge e Wheeler \cite{reggewheeler} originou o estudo das perturbações de buracos negros onde os autores analisaram a questão da estabilidade de um buraco negro de Schwarzschild que estava sujeito a uma pequena perturbação gravitacional inicial. Vários anos mais tarde Vishveshwara \cite{vishveshwara}, Edelstein e Vishveshwara \cite{edelsteinvishveshwara} e Zerilli \cite{zerilli} deram continuidade a este trabalho. Em particular, Vishveshwara verificou, através de simulações numéricas, que, quando um buraco negro de Schwarzschild é submetido a uma pequena perturbação inicial (um pulso de onda gaussiana, no caso), a resposta da mesma (sua evolução temporal) é dominada por vibrações muito características, exponencialmente amortecidas no tempo, que depois viriam a ser denominadas, por Press \cite{press}, de modos quasinormais (MQNs), sendo independentes da forma inicial da perturbação, dependendo apenas da massa do buraco negro, neste caso, seu único parâmetro característico.

É por meio de um texto de Chandrasekhar que talvez melhor se possa compreender a noção de MQNs de um buraco negro. De acordo com ele \cite{chandrasekhar}, a evolução de uma pequena perturbação inicial de um buraco negro pode, em princípio, ser determinada expressando-a como uma superposição de modos normais. Porém, em seu estágio final, espera-se que qualquer perturbação inicial decaia de maneira característica do próprio buraco negro e independentemente da mesma. Ou seja, pode-se esperar que em estágios finais de uma perturbação, o buraco negro oscile com frequências e taxas de amortecimento características unicamente dele próprio. Tem-se então a seguinte definição: `MQNs são definidos como as soluções próprias das equaçõs de perturbação pertencendo a frequências características complexas que são apropriadas para ondas puramente imergindo no horizonte de eventos e ondas puramente emergindo no infinito'.

Assim, as condições de contorno são dadas por
\begin{eqnarray}
\label{contorno}
\Psi(x)\sim e^{\pm i\omega x} \ \ \ , \ \ \ x\to \pm\infty ,
\end{eqnarray}
sendo que correspondem a frequências complexas
\begin{eqnarray}
\omega = \Re \omega + i\Im \omega ,
\end{eqnarray}
com $\Im\omega<0$, o que caracteriza a estabilidade do buraco negro, quando sujeito a uma pequena perturbação inicial, já que o mesmo oscilará amortecidamente, em função do tempo ($e^{-i\omega t}=e^{-i\Re\omega t}e^{\Im\omega t}\to 0$, para $t\to\infty$, quando $\Im\omega<0$).

Para que melhor se entenda a definição de MQNs, deve-se notar que se deseja estudar a resposta da métrica de um buraco negro, para pequenas perturbações iniciais do sistema. Para isto, não é desejável que haja radiação vindo do infinito, continuando a perturbar o buraco negro. Além disto, assume-se que nada escapa para fora do horizonte de eventos do mesmo.

As frequências $\omega$ que permitem soluções da equação
\begin{eqnarray}
\label{scho}
\frac{d^2\Psi(x)}{dx^2}+(\omega^2-V(x))\Psi(x)=0,
\end{eqnarray}
com as condições de contorno da Eq.(\ref{contorno}), são chamadas de modos (frequências) quasinormais, isto é, as condições de contorno, então, são que haja ondas que são puramente imergentes no horizonte de eventos do buraco negro e ondas puramente emergentes no infinito. É importante notar que tais condições de contorno não se aplicam a MQNs de quaisquer buracos negros, mas sim para aqueles em espaços assintoticamente planos ou assintoticamente de-Sitter (dS), uma vez que é a estrutura assintótica do espaço-tempo que determina tais condições. No caso assintoticamente anti-de-Sitter (AdS), a condição de contorno no infinito é que não haja onda emergente nem imergente \cite{molina} de modo que geralmente se impõe condições de Dirichlet $\Psi(x)\to 0$.

Ainda sobre a perturbações de buracos negros podemos relaciona-las a uma das possíveis características de uma teoria quântica de gravitação: o chamado Princípio Holográfico, segundo o qual, a física num certo volume do espaço-tempo pode ser descrita em termos de uma teoria em sua borda, possuindo um grau de liberdade por área de Planck \cite{hooft}, cuja formulação tem por base a relação entropia-área de buracos negros dada pela fórmula de Bekenstein-Hawking \cite{bekenstein1}, $S=A/4$ (em unidades $c=G=\hbar=k_B=1$), a qual tem natureza holográfica, já que a entropia de um buraco negro seria proporcional à área de seu horizonte de eventos e não ao volume deste e também o chamado limite de Bekenstein \cite{bekenstein2}, um limite superior para a entropia de uma dada região do espaço-tempo que estabelece que a máxima entropia de uma região do espaço.

Visto isso, pode-se utilizar tal princípio na correspondência AdS/CFT, onde uma teoria de gravitação tem uma descrição dual em termos de uma Teoria de Campos Conforme (CFT) de uma dimensão a menos, que pode ser pensada como residente na borda deste espaço. Pode-se dizer então que a informação (física) no interior (bulk) de um espaço AdS está holograficamente codificada em sua borda.

Invocando a correspondencia AdS/CFT, Horowitz e Hubeny (HH) \cite{horowitzhubeny} sugeriram que as frequências quasinormais associadas a perturbações de buracos negros em espaços-tempos AdS possuem uma interpretação direta em termos da correspondente teoria de campos conforme dual que `reside' na borda de tal espaço-tempo. De acordo com tal correspondência, um buraco negro grande num espaço-tempo assintoticamente AdS, isto é, cujo raio de horizonte de eventos é muito maior que o raio cosmológico, corresponderia a um estado (aproximadamente) térmico na teoria de campos conforme dual \cite{witten}.

Assim sendo, perturbar um buraco negro corresponderia a perturbar tal estado térmico, de modo que o decaimento da perturbação no interior do espaço-tempo descreveria o retorno ao equilíbrio da teoria de campo na borda. Isto permite, então, que se obtenha uma predição da escala temporal de termalização da teoria de campo dual, o que seria, em princípio, bastante difícil de ser conseguido diretamente via esta última.

Ao estudarem as perturbações de um buraco negro grande de Schwarzschild-AdS, por meio de um campo escalar não-massivo, minimamente acoplado à curvatura, HH obtiveram que os MQNs resultantes estão em escala com a temperatura do mesmo.

Nos últimos anos tem havido bastante interesse pelo estudo de perturbações e MQNs de buracos negros em espaços-tempos não-assintoticamente planos, a saber, espaços-tempo dS e AdS, motivados pela análise da estabilidade do horizonte de Cauchy de buracos negros \cite{BradyChambersKrivanLaguna} e pela correspondência AdS/CFT \cite{horowitzhubeny}, respectivamente.

No caso de espaços-tempos do tipo dS, o decaimento de perturbações de buracos negros de Schwarzschild e Reissner-Nordstrom, nele imersos, mostrou a influência do horizonte de eventos cosmológico sobre o decaimento das mesmas [\cite{BradyChambersKrivanLaguna},\cite{molina}]. No caso de buracos negros AdS, a análise dos MQNs e do decaimento das perturbações iniciou-se com Chan e Mann \cite{ChanMann}, seguidos de HH \cite{horowitzhubeny}, estes fortemente motivados pela correspondência. O estudo de perturbações nestes espaços-tempos dotados de constante cosmológica é importante não apenas como mera extensão da análise dos casos tratados na literatura, até então em grande maioria para espaços-tempos assintoticamente planos, mas também devido ao grande interesse, nos útlimos anos, por espaçoes do tipo AdS, devido à correspondência.

A busca por soluções exatas em tais situações nem sempre é uma tarefa simples. Visto isso, é necessário a aplicação de métodos numéricos em buscas de tais soluções. Além de HH, outro método numérico utilizado para obtenção de soluções é o chamado Método de Iteração Assintótica AIM). Ele obtém soluções analíticas/numéricas de equações diferenciais ordinárias de segunda ordem com potenciais ligados \cite{hakan} e \cite{cho}.

Portanto, estudar o comportamento de campos de matéria na vizinhança de buracos negros com simetria de Lifshtiz, os quais são soluções de vácuo de teorias de gravitação alternativas com correções quadráticas na curvatura, podem enriquecer nosso entendimento sobre: buracos negros, dualidade gravidade/calibre e gravitação quântica.

No intuito de realizar tal estudo dividimos o presente trabalho com a seguinte maneira: no capítulo 2, intitulado, \emph{Teoria de gravitação alternativa}, será efetuada uma explicação mais detalhada sobre os problemas da Relatividade Geral e a proposta de uma teoria de gravitação com correções quadráticas na curvatura. Uma solução com simetria de Lifshitz puro será exibida e serão abordados detalhes sobre as famílias de soluções do tipo buraco negro com simetria de Lifshitz em $D$-dimensões.
No capítulo 3, intitulado, \emph{Evolução de campos de matéria em espaços-tempos planarmente simétricos e estáticos}, será calculada as equações de evolução do campo escalar com acoplamento com tensor de Einstein e com escalar de Ricci e do campo eletromagnético em um espaço-tempo planarmente simétrico e estático arbitrário.
No capítulo 4, intitulado, \emph{Equações de campo aplicadas aos Buracos Negros}, será aplicada a equação de evolução do campo escalar acoplado a dois buracos negros $2+1$-dimensionais e suas respectivas soluções exatas para os casos com acoplamento cinético e acoplamento potencial. Também serão aplicadas as equações de evolução do campo escalar acoplado ao escalar de Ricci e ao tensor de Einstein e do eletromagnético aos buracos negros com simetria de Lifhitz em $D\geq5$ dimensões.
No capítulo 5, intitulado, \emph{Métodos numéricos}, os métodos númericos HH e AIM serão estudados, assim como, as condições de convergência de cada um.
No capítulo 6, intitulado, \emph{Resultados numéricos da evolução do campo escalar e eletromagnético}, a aplicação dos métodos AIM e HH para o cálculo dos modos quasinormais dos buracos negros AdS $D$-dimensional, BTZ e AGGH e Lifshitz \emph{5Dz1} e \emph{6Dz0} é apresentada. Os resulados obtidos são compilados e discutidos.
No capítulo 7, intitulado, \emph{Conclusões}, serão apresentadas as conclusões obtidas neste trabalho.
Nos Apêndices A, B e C apresentamos os invariantes escalares e as componentes do tensor de Einstein, a transformação de uma equação diferencial de segunda ordem para o formato de Schrödinger necessária para utilização do método AIM e HH, respectivamente.

\chapter{Teoria de gravitação alternativa}
Dez anos após a publicação da Teoria da Relatividade Restrita, Einstein apresentou uma versão generalizada da sua teoria, que ficou conhecida como Teoria da Relatividade Geral(RG), incluindo a gravidade e os efeitos previstos pela teoria restrita, agora aplicados a referenciais acelerados. A interconexão entre espaço e tempo continua desempenhando um papel fundamental na descrição dos fenômenos, já que a gravidade agora é entendida como um efeito da deformação da geometria do espaço-tempo causada pela presença de matéria e energia.

A RG é uma teoria bem estabelecida e se tornou a referência básica para o estudo de fenômenos de larga escala, como astrofísica e cosmologia. Ainda assim é uma teoria que possui limitações. Uma delas é o fato de não se enquadrar entre as forças fundamentais descritas pelo Modelo Padrão, desafiando aqueles que esperam que a natureza seja descrita de maneira unificada e elegante. Outra é a sua incapacidade de descrever cenários extremos em que tanto a curvatura do espaço-tempo como as energias envolvidas são muito grandes (o que significa fenômenos ocorrendo em intervalos de espaço e tempo muito pequenos). Ambas as limitações só serão superadas por uma teoria quântica da gravitação.

Por muitas décadas os físicos têm investido muito esforço na elaboração de uma teoria que descreva a gravitação (essencialmente, que descreva o espaço e o tempo) com base na teoria quântica. Apesar disso, nenhuma teoria completamente satisfatória foi estabelecida. Muito da dificuldade enfrentada vem da ausência de dados experimentais em tais cenários de extremas energia e curvatura. Entretanto, o principal problema é de natureza teórica: a não-renormalizabilidade. Para esclarecer melhor este ponto, considere que a descrição lagrangeana da Relatividade Geral se dá por meio da ação de Einstein-Hilbert,
\begin{equation}
S=-\frac{c^4}{16\pi G_N}\int d^4x\sqrt{-g}\mathcal{R},
\end{equation}
onde $g$ é o determinante da métrica e $\mathcal{R}$ é o escalar de curvatura de Ricci. Uma forma comum de desenvolver a quantização da teoria descrita por esta ação é escrever $g_{\mu\nu}=\eta_{\mu\nu}+h_{\mu\nu}$, onde $\eta_{\mu\nu}$ é a métrica de Minkowski, tratada aqui como um campo de fundo e $h_{\mu\nu}$ é uma perturbação que representa um campo de calibre. A ação prevê a existência do gráviton, partícula sem massa de spin $2$, cujo propagador é dado por $1/k^2$, onde $k^{\mu}$ é o quadrimomento associado.

A análise dimensional possui um papel importante em teoria quântica de campos. Para que uma teoria seja renormalizável é necessário que a dimensão canônica em unidades de massa das constantes de acoplamento envolvidas sejam não negativas, condição que não é atendida pela RG, já que $[G_N]=-2$. A teoria resultante é não-renormalizável, não sendo possível anular as divergências com um número finito de contratermos. Uma maneira de contornar este problema, é a adição de termos de derivada de ordem superior na ação de Einstein-Hilbert.

Neste contexto, serão introduzidas derivadas de ordem mais alta na ação, a saber, derivadas de segunda ordem.

De acordo com \cite{lifshitzhigher} a ação da gravidade mais geral que inclui as correções quadráticas na curvatura em $D$-dimensões é dada por
\begin{eqnarray}
\label{acao}
S=\int d^D x\sqrt{-g}(\mathcal{R}-2\lambda+\beta_1 \mathcal{R}^2+\beta_2 R_{\alpha\beta}R^{\alpha\beta}+\beta_3 R_{\alpha\beta\gamma\nu}R^{\alpha\beta\gamma\nu}),
\end{eqnarray}
onde $\lambda$ é a constante cosmológica, $\beta_1$, $\beta_2$ e $\beta_3$ são constantes de acoplamento, $\mathcal{R}$ é o escalar de Ricci, $R_{\alpha\beta}$ é o tensor de Ricci e $R_{\alpha\beta\gamma\nu}$ é o tensor de Riemann.

Extremizando a ação (\ref{acao}) obtêm-se as seguintes equações de campo
\begin{eqnarray}
\label{Einsteinquadr}
G_{\mu\nu}+\lambda g_{\mu\nu}+A_{\mu\nu}+B_{\mu\nu}=0,
\end{eqnarray}
onde o tensor $A_{\mu\nu}$ representa correções lineares na curvatura nas equações de campo e é definido por
\begin{equation}
A_{\mu\nu}=(\beta_2 +4\beta_3)\Box R_{\mu\nu}+\frac{1}{2}(4\beta_1 +\beta_2)g_{\mu\nu}\Box R-(2\beta_1+\beta_2+2\beta_3)\nabla_{\mu}\nabla_{\nu}R
\end{equation}
e o tensor $B_{\mu\nu}$ representa correções quadráticas na curvatura nas equações de campo e é definido por
\begin{eqnarray}
B_{\mu\nu}&=&2\beta_3R_{\mu\gamma\alpha\beta}R_{\nu}^{\gamma\alpha\beta}+2(\beta_2+2\beta_3)R_{\mu\alpha\nu\beta}R^{\alpha\beta}-4\beta_3 R_{\mu\alpha}R_{\nu}^\alpha+2\beta_1RR_{\mu\nu}-\nonumber\\
&-&\frac{1}{2}(\beta_1R^2+\beta_2R_{\alpha\beta}R^{\alpha\beta}+\beta_3R_{\alpha\beta\gamma\delta}R^{\alpha\beta\gamma\delta})g_{\mu\nu}.
\end{eqnarray}
Por causa do teorema de Gauss-Bonnet em quatro dimensões e do anulamento deste termo em três dimensões é suficiente considerar apenas dois dos três invariantes quadráticos na Lagrangeana mostrada acima. Isto faz necessário dividir a análise de suas possíveis soluções em duas partes, primeiro considerando os casos de dimensão superior, $D\ge 5$,  e depois analisar separadamente os de menor dimensão ($D=3$ e $D=4$).

Analisando, primeiramente, o caso geral da teoria, onde as dimensões são $D\ge 5$, percebe-se que esta ação possui três classes diferentes de soluções de buraco negro assintoticamente Lifshitz que correspondem a diferentes faixas do expoente dinâmico $z$. Como mostrado abaixo, para qualquer valor do expoente dinâmico $z$ existe pelo menos uma família de soluções de buraco negro.

As famílias são as seguintes
\begin{eqnarray}
\label{familia1}
z&>&2-D, \\
\label{familia2}
z&>&1, \\
\label{familia3}
z&<&0.
\end{eqnarray}
Todas as três representam soluções de buracos negros assintoticamente Lifshitz para cada expoente dinâmico em particular.

A seguir é abordado mais especificamente sobre o espaço-tempo de Lifshitz puro e, posteriormente, sobre a família de soluções de buracos negros Lifshitz em altas dimensões escolhida para estudo neste trabalho, a saber, a família \ref{familia1}.

\section{Soluções com simetria de Lifshitz}
Primeiramente serão apresentadas algumas características do espaço-tempo de Lifshitz puro e da família de buracos negros \ref{familia1} assintoticamente Lifshitz os quais são soluções para ação com correções quadráticas na curvatura mencionada anteriormente.
\subsection{Espaço-tempo de Lifshitz puro}
O espaço-tempo Lifshitz puro é representado pelo seguinte elemento de linha
\begin{equation}
\label{lifshitzpuro}
ds^2=-\frac{r^{2z}}{l^{2z}}dt^2+\frac{l^2}{r^2}dr^2+\frac{r^2}{l^2}d\vec{x}^2,
\end{equation}
onde $\vec{x}$ é um vetor ($D-2$)-dimensional\footnote{Uma análise dimensional do elemento de linha nos leva a considerar o caso planar usando coordenadas cartesianas $(x,y,z,...w)$ para as dimensões extras.}, $z$ é o expoente dinâmico e $l$ é o raio de curvatura da geometria. Dentre suas principais propriedades a mais interessante é que o espaço-tempo de Lifshitz admite uma transformação de coordenadas anisotrópica que se segue,
\begin{eqnarray}
\label{isometria}
t\to \lambda^zt, \ \ \ \ \ \ r\to \lambda^{-1}r, \ \ \ \ \ \ \vec{x}\to \lambda \vec{x},
\end{eqnarray}
como parte do seu grupo de isometria. Na verdade a classificação desse espaço-tempo com o nome Lifshitz vem do fato da isometria (\ref{isometria}) ser idêntica à transformação de escala (\ref{isometria}) que mantem invariante a Langrangeana
\begin{eqnarray}
\label{lifshitzlangra}
L=\int d^2x dt [(\partial_t\phi)^2-\kappa(\nabla^2\phi)^2]
\end{eqnarray}
usada por Lifshitz para modelar, via teoria de campos, sistemas de matéria condensada. Esse espaço-tempo tem recebido muita atenção no contexto da dualidade gravidade/calibre. Um trabalho de revisão muito interessante sobre a aplicação desse espaço-tempo e sua relação como modelos de matéria condensada é feito por Hartnoll em \cite{hartnoll} e vale a pena ser estudado.
Essa isometria pode ser confirmada substituindo as transformações de escala (\ref{isometria}) para $t$, $r$ e $\vec{x}$ no elemento de linha (\ref{lifshitzpuro}),
\begin{eqnarray}
ds^2&=&-\frac{(\lambda^{-1}r)^{2z}}{l^{2z}}d(\lambda^zt)^2+\frac{l^2}{(\lambda^{-1}r)^2}d(\lambda^{-1}r)^2+\frac{(\lambda^{-1}r)^2}{l^2}d(\lambda\vec{x})^2,\nonumber\\
ds^2&=&-\frac{r^{2z}}{\lambda^{2z}l^{2z}}\lambda^{2z}dt^2+\frac{\lambda^2l^2}{r^2}\frac{dr^2}{\lambda^2}+\frac{r^2}{\lambda^2l^2}\lambda^2d\vec{x}^2.
\end{eqnarray}
Como pode-se notar, ela recupera exatamente o elemento de linha original, independente do valor do expoente dinâmico $z$. Para informações adicionais sobre o espaço-tempo de Lifshitz ver as discussões apresentadas em \cite{lifshitzhigher},\cite{kachru},\cite{aggh} e \cite{denizozgur}.

A métrica (\ref{lifshitzpuro}) será solução das Eqs.(\ref{Einsteinquadr}) se a constante cosmológica $\lambda$ e a constante de acoplamento $\beta_2$ satisfizerem as seguintes relações,
\begin{eqnarray}
\lambda&=&\frac{-1}{4l^2}\left[2z^2+(D-2)(2z+D-1)-\frac{4z(D-3)(D-4)(z+D-2)\beta_3}{l^2}\right],\ \ \ \ \ \ \ \ \ \\
\nonumber\\
\beta_2&=&\frac{l^2-2\left[2z^2+(D-2)(2z+D-1)\right]\beta_1-4\left[z^2-(D-2)z+1\right]\beta_3}{2(z^2+D-2)}.
\end{eqnarray}
Analisando a métrica apresentada, vê-se que ela é não singular. Todos invariantes locais (Escalar de Ricci, Contração do Tensor de Ricci, Escalar de Kretschmann, Escalar de Einstein e Escalar de Weyl) são finitos em toda geometria. Os detalhes dessas quantidades são apresentados no Apêndice. Essa geometria apresenta muitas semelhanças com o espaço-tempo Anti de-Sitter, mas dependendo do valor de $z$ ela se torna não-relativística.

A partir deste elemento de linha é natural procurar por soluções de buraco negro que apresentem simetria de Lifshitz. Elas podem ser obtidas e são conhecidas como `Buracos Negros de Lifshitz'. Recentemente a busca por tais soluções tem recebido grande atenção.

Um dos primeiros exemplos analíticos foi apresentado em \cite{taylor} sem restrição no valor de $z$. Em \cite{mann} uma solução topológica de buraco negro assintoticamente Lifshitz com $z=2$ foi encontrada. Um exemplo com $z=4$ e topologia esférica foi dado em \cite{bertoldi}. Soluções numéricas para valores mais gerais de $z$ foram exploradas em \cite{mann}-\cite{bertoldi2}. Mais exemplos de buracos negros analíticos de Lifshitz foram estudados em \cite{brynjolfsson}-\cite{pang}. A solução encontrada em \cite{balasubramanian} é particularmente interessante, pois, corresponde a um exemplo analítico notavelmente simples, com $z=2$ e $D=4$ dimensões. A dificuldade de embutir buracos negros de Lifshitz na teoria de cordas também foi investigada em \cite{azeyanagi}-\cite{hartonoll}. A descrição holográfica de espaços-tempos assintoticamente Lifshitz foi estudada em \cite{ross}. Investigações mais recentes relacionadas aos buracos negros de Lifshitz podem ser encontradas em \cite{sin}-\cite{fadafan}.

Em \cite{aggh} uma solução bastante simples com $z=3$ na ausência de campos de matéria foi encontrada para a teoria New Massive Gravity (NMG) \cite{bergshoeff}, que consiste em correções especiais de curvatura quadrada para a gravidade tridimensional. Anteriormente, em \cite{adams}, foi mostrado que correções quadráticas na curvatura para gravidade, genericamente, podem suportar os espaços-tempos de Lifshitz puro, apresentado na Eq.(\ref{lifshitzpuro}). O exemplo da referência \cite{aggh} é o primeiro a mostrar que essas teorias também permitem a existência de buracos negros de Lifshitz. Outro exemplo com $z=3/2$ foi posteriormente encontrado para uma teoria quadridimensional com $R^2$-correções. O próximo passo será procurar soluções do tipo buraco negro nesta teoria de gravidade.

\subsection{Buracos Negros assintoticamente Lifshitz em $D$-dimensões}
Inspirado nos resultados de \cite{aggh} e \cite{cai} é investigado em \cite{lifshitzhigher} como a introdução de correções de ordem superior na curvatura, na ação de Einstein-Hilbert, levam a encontrar várias famílias de soluções analíticas de buracos negros de Lifshitz em $D$-dimensões. Neste trabalho será dada especial atenção a uma dessas famílias de soluções estudando-se a evolução de campos de matéria em sua geometria e averiguando suas características. Curiosamente, esta família a qual foi escolhida, de dimensão superior, também admite como solução crítica o buraco negro de Lifshitz tridimensional com $z=3$, buraco negro AGGH, de \cite{aggh}, no contexto da NMG. As soluções permitidas para este caso mostradas em \cite{aggh} são $z=3$ e $z=1$. Todavia, esta mesma família não leva ao buraco negro tridimensional BTZ \cite{btz}($z=1$). Uma possível explicação para este resultado será dada na próxima seção.

A ação com correções quadráticas mais gerais mostrada por \cite{lifshitzhigher} apresenta três famílias diferentes de soluções para buracos negros assintoticamente Lifshitz com $D\ge 5$. A família a ser estudada neste trabalho é \ref{familia1}, onde o expoente dinâmico satisfaz a condição $z>2-D$. Ela pode ser representada pelo o elemento de linha planarmente simétrico, com funções $A(r)$ e $B(r)$ gerais, escrito de maneira geral como
\begin{eqnarray}
\label{metricagerallifshitz}
ds^2=-A(r)dt^2+B(r)dr^2+\frac{r^2}{l^2}\sum_{i=1}^{D-2} dx_i^2.
\end{eqnarray}
Para o caso estudado, substituindo as funções $A(r)$ e $B(r)$, a métrica fica da seguinte forma
\begin{eqnarray}
\label{metricaDz}
ds^2=-\frac{r^{2z}}{l^{2z}}\left(1-\frac{Ml^{\alpha}}{r^{\alpha}}\right)dt^2+\frac{l^2}{r^2}\left(1-\frac{Ml^{\alpha}}{r^{\alpha}}\right)^{-1}dr^2+\frac{r^2}{l^2}\sum_{i=1}^{D-2} dx_i^2,
\end{eqnarray}
onde define-se o expoente $\alpha$ como
\begin{equation}
\alpha=\frac{z+D-2}{2}.
\end{equation}
Nesta métrica o expoente dinâmico $z$ se relaciona com os parâmetros da ação (\ref{acao}) por meio das seguintes relações
\begin{eqnarray}
\label{lambda}
\lambda&=&\frac{D-2}{4l^2}\{(197D-389)z^4+4(19D^2-200D+325)z^3\nonumber\\
&+&(D-2)[2(5D^2-73D+356)z^2+4(D^3-2D^2+15D-\nonumber\\
&-&62)z+(D+2)(D-1)(D-2)^2]\}/ P_4(z),\\
\label{b1}
\beta_1&=&l^2\{27z^6-18(3D-4)z^5+3(19D^2-168D+356)z^4-12(11D^3-\nonumber\\
&-&84D^2+196D-120)z^3-(D-2)[(19D^3-330D^2+2052D-3640)z^2+\nonumber\\
&+&2(3D^4-30D^3+124D^2-536D+1024)z+(D+2)(D-\nonumber\\
&-&2)^2(D^2-4D+36)]\}/ \left[2(D-3)(D-4)(z+D-2)^2P_4(z)\right],\\
\label{b2}
\beta_2&=&-2l^2[3z^2+(D+2)(D-2)]\{9z^4-6(3D-4)z^3-8(D^2-10)z^2+\nonumber\\
&+&2(D^3-4D^2+32D-80)z-(D-2)[D^3+2D^2-\nonumber\\
&-&12(D-2)]\}/ \left[(D-3)(D-4)(z+D-2)^2P_4(z)\right],\\
\label{b3}
\beta_3&=&l^2[3z^2+(D+2)(D-2)]\{9z^3-3(9D-14)z^2-(D-2)[(5D-62)z+\nonumber\\
&+&D^2-4D+36]\}/ \left[2(D-3)(D-4)(z+D-2)P_4(z)\right].
\end{eqnarray}
A função $P_4$ é um polinômio de grau quatro no expoente dinâmico $z$ dado por
\begin{eqnarray}
P_4(z)&=&27z^4-4(27D-45)z^3-(D-2)[2(5D-116)z^2+\nonumber\\
&+&4(D^2-D+30)z+(D+2)(D-2)^2].
\end{eqnarray}
Inspecionando-se as Eqs.(\ref{b1}, \ref{b2}, \ref{b3}) observa-se que a teoria é válida para $D\ge 5$. Nos casos onde $D=3$ e $D=4$ os termos $\beta_1$, $\beta_2$ e $\beta_3$ divergem, contudo, soluções críticas em dimensões mais baixas são possíveis e serão tratadas adiante. A condição que $z>2-D$ vem também do denominador destas constantes de acoplamento.

Abaixo é apresentada uma tabela com os valores do expoente $\alpha$ em função de $z$ e da dimensão $D$, respeitada a condição que $z>2-D$. Os casos onde aparece o símbolo (*) representam valores que não respeitam a dada condição. O valor $0$ aparece como caso limite, quando $z=2-D$ e também não pode ser utlizado, pois, faria com que o denominador das expressões para $\beta_1$, $\beta_2$ e $\beta_3$ se anulasse. Da tabela é verificado que para cada dimensão $D$ e expoente dinâmico $z$ tem-se um valor igual para $\alpha$ podendo serem vistos nas diagonais da tabela abaixo.
\begin{table}[H]
\centering
\caption{Expoente $\alpha$}
\vspace{0.5cm}
\begin{tabular}{|c|c|c|c|c|c|c|c|}\hline
 & $D$=5 & 6  & 7 & 8 & 9 & 10 & 11 \\ \hline
$z$=-8 & * & * & * & * & * & 0 & $1/2$ \\ \hline
-7  & * & * & * & * & 0 & $1/2$ & 1 \\ \hline
-6   & * & * & * & 0 & $1/2$ & 1& $3/2$  \\ \hline
-5 & * & * & 0 & $1/2$ & 1 & $3/2$ & 2  \\ \hline
-4   & * & 0 & $1/2$ & 1 & $3/2$ & 2& $5/2$ \\ \hline
-3   & 0 & $1/2$  & 1 & $3/2$  & 2 & $5/2$ & 3 \\ \hline
-2   &1/2 & 1 & $3/2$ &  2 & $5/2$ & 3 & $7/2$ \\ \hline
-1   & 1 & $3/2$ &  2 & $5/2$ & 3 & $7/2$ & 4   \\ \hline
0   & $3/2$ &  2 & $5/2$ & 3 & $7/2$ & 4 & $9/2$  \\ \hline
1   &  2 & $5/2$ & 3 & $7/2$ & 4 & $9/2$ & 5 \\ \hline
2   & $5/2$ & 3 & $7/2$ & 4 & $9/2$ & 5 & $11/2$ \\ \hline
3   & 3 & $7/2$ & 4 & $9/2$ & 5 & $11/2$ & 6  \\ \hline
4   & $7/2$ & 4 & $9/2$ & 5 & $11/2$ & 6 & $13/2$  \\ \hline
\end{tabular}
\end{table}
Poderia-se imaginar o porque de não poder utilizar $z<2-D$, e a resposta está no próprio elemento de linha. Nesta região os valores de $\alpha$ são todos negativos e isto faria o buraco negro não ser assintoticamente Lifshitz.

Com o intuito de avaliar a influência do número de dimensões, foram escolhidos dois valores diferentes de $D$. Sendo assim, optou-se por escolher uma dimensão ímpar e uma par, para avaliar uma possível influência diferente entre elas. Visto isto, as dimensões escolhidas foram $D=5$ e $D=6$.

De forma compacta serão colocadas as informações dos buracos negros em uma única tabela. Para $D=5$ são apresentados os casos onde $z=1,2,3$ e $4$ e para $D=6$ os casos onde $z=0,1,2$ e $3$. As funções $A(r)$ e $B(r)$, o horizonte de eventos $r_+$ ($g^{rr}=0$) e a gravidade superficial $\kappa$ foram calculadas para cada um dos buracos negros citados. No caso geral $\kappa$ é calculada por
\begin{eqnarray}
\kappa=\frac{1}{2}\frac{\partial A(r)}{\partial r}\Bigg{|}_{r=r_+}\sqrt{\frac{1}{A(r_+)B(r_+)}}.
\end{eqnarray}
Desta forma as quantidades acima citadas para os buracos negros $5D$ e $6D$  são apresentadas nas tabelas abaixo.
\begin{table}[H]
\centering
\caption{Quantidades dos buracos negros \emph{5D}}
\vspace{0.5cm}
\begin{tabular}{|c|c|c|c|c|c|c|c|}\hline
 & \emph{5Dz1} & \emph{5Dz2} & \emph{5Dz3} & \emph{5Dz4} \\ \hline
$A(r)$ & $\frac{r^2}{l^2}\left(1-\frac{Ml^2}{r^2}\right)$ & $\frac{r^4}{l^4}\left(1-\frac{Ml^{5/2}}{r^{5/2}}\right)$ & $\frac{r^6}{l^6}\left(1-\frac{Ml^3}{r^3}\right)$ & $\frac{r^8}{l^8}\left(1-\frac{Ml^{7/2}}{r^{7/2}}\right)$ \\
$B(r)$ & $\frac{l^2}{r^2}\left(1-\frac{Ml^2}{r^2}\right)^{-1}$ & $\frac{l^2}{r^2}\left(1-\frac{Ml^{5/2}}{r^{5/2}}\right)^{-1}$ & $\frac{l^2}{r^2}\left(1-\frac{Ml^3}{r^3}\right)^{-1}$ & $\frac{l^2}{r^2}\left(1-\frac{Ml^{7/2}}{r^{7/2}}\right)^{-1}$\\
$r_+$ & $l\sqrt{M}$ & $l\sqrt[5]{M^2}$ & $l\sqrt[3]{M}$ & $l\sqrt[7]{M^2}$\\
$\kappa$  & $\sqrt{M}/l$  & $5\sqrt[5]{M^4}/4l$ & $3M/2l$ & $7\sqrt[7]{M^8}/4l$ \\
$\lambda$ & $-\frac{3}{l^2}$ & $-\frac{2197}{551 l^2}$   & $-\frac{6}{l^2}$     & $-\frac{9791}{1109 l^2}$ \\
$\beta_1$ &  $\frac{l^2}{4}$ & $\frac{9839 l^2}{55100}$  & $\frac{41 l^2}{288}$ & $\frac{26213 l^2}{217364}$ \\
$\beta_2$ & $-l^2$           & $-\frac{9471 l^2}{13775}$ & $-\frac{5 l^2}{9}$   & $-\frac{26197 l^2}{54341}$  \\
$\beta_3$ & $\frac{l^2}{4}$  & $\frac{2211 l^2}{11020}$  & $\frac{l^2}{6}$      & $\frac{4531 l^2}{31052}$ \\  \hline
\end{tabular}
\end{table}
\begin{table}[H]
\centering
\caption{Quantidades dos buracos negros \emph{6D}}
\vspace{0.5cm}
\begin{tabular}{|c|c|c|c|c|c|c|c|}\hline
       & \emph{6Dz0} & \emph{6Dz1} & \emph{6Dz2} & \emph{6Dz3} \\ \hline
$A(r)$ & $\left(1-\frac{Ml^2}{r^2}\right)$ & $\frac{r^2}{l^2}\left(1-\frac{Ml^{5/2}}{r^{5/2}}\right)$ & $\frac{r^4}{l^4}\left(1-\frac{Ml^3}{r^3}\right)$ & $\frac{r^6}{l^6}\left(1-\frac{Ml^{7/2}}{r^{7/2}}\right)$ \\
$B(r)$ & $\frac{l^2}{r^2}\left(1-\frac{Ml^2}{r^2}\right)^{-1}$ & $\frac{l^2}{r^2}\left(1-\frac{Ml^{5/2}}{r^{5/2}}\right)^{-1}$ & $\frac{l^2}{r^2}\left(1-\frac{Ml^3}{r^3}\right)^{-1}$ & $-\frac{l^2}{r^2}\left(1-\frac{Ml^{7/2}}{r^{7/2}}\right)^{-1}$ \\
$r_+$ & $l\sqrt{M}$ & $l\sqrt[5]{M^2}$ & $l\sqrt[3]{M}$ & $l\sqrt[7]{M^2}$ \\
$\kappa$ & $1/l$ & $5\sqrt[5]{M^2}/4l$ & $3\sqrt[2]{M^3}/2l$ & $7\sqrt[7]{M^6}/4l$ \\
$\lambda$ & $-\frac{5}{l^2}$  & $-\frac{5}{l^2}$  & $-\frac{101}{17 l^2}$ & $-\frac{61477}{7649 l^2}$ \\
$\beta_1$ & $\frac{l^2}{4}$   & $\frac{l^2}{12}$  & $\frac{45 l^2}{748}$  & $\frac{71707 l^2}{1499204}$ \\
$\beta_2$ & $-\frac{5l^2}{4}$ & $-\frac{l^2}{3}$  & $-\frac{11 l^2}{51}$  & $-\frac{64251 l^2}{374801}$  \\
$\beta_3$ & $\frac{l^2}{4}$   & $\frac{l^2}{12}$  & $\frac{43 l^2}{612}$  & $\frac{12685 l^2}{214172}$ \\  \hline
\end{tabular}
\end{table}

Para todos os casos a constante cosmológica permanece negativa, não alterando assim o caráter assintótico das soluções.

Para cada dimensão escolhida foram utilizados dois valores do expoente dinâmico $z$ que resultam em um expoente $\alpha$ inteiro. Para $D=5$, escolheu-se $z=1$ e $z=3$ (nomeado buraco negro \emph{5Dz1} e \emph{5Dz3}) e para $D=6$, escolheu-se $z=0$ e $z=2$ (nomeado buraco negro \emph{6Dz0} e \emph{6Dz2}).

Nota-se que para os buracos negros com valores iguais de $\alpha$ tem-se horizonte de eventos iguais. Vale frisar aqui que os buracos negros são planarmente simétricos, logo, seu horizonte de eventos é uma superfície planar. Como verificado, os casos \emph{5Dz1} e \emph{6Dz0} tem o mesmo valor de $\alpha$, logo, o mesmo horizonte de eventos. Porém, a semelhança entre eles para por aí, pois, a gravidade superficial é diferente e como veremos a seguir suas quantidades termodinâmicas também.

Agora serão apresentadas algumas propriedades termodinâmicas dos buracos negros escolhidos, tais como, Temperatura $(T)$, Entropia $(S)$ e Energia $(E)$ e Capacidade Térmica $(C)$. Essas propriedades foram discutidas em \cite{denizozgur} e \cite{normansamuel}.

Primeiro define-se as quantidades $p(z)$ e $q(z)$ para facilitar a notação das quantidades termodinâmincas como sendo
\begin{eqnarray}
p(z)&=&-8(D-2)\left\{9z^4-(D+5)z^3-\right.\nonumber\\
&&\left. -(D-2)\left[3(D-5)z^2+(D^2-5D+10)z-(D^2-4)\right]\right\},\nonumber\\
q(z)&=&27z^4-4(27D-45)z^3-\nonumber\\
&&-(D-2)\left[2(5D-116)z^2+4(D^2-D+30)z+(D+2)(D-2)^2\right].\nonumber\\
\end{eqnarray}
Deste modo a temperatura dos buracos negros $D$-dimensionais é definida por
\begin{eqnarray}
T=\frac{z+D-2}{8\pi l}M^{\frac{2z}{z+D-2}}
\end{eqnarray}
que é equivalente a conhecida expressão para temperatura associada ao buraco negro planarmente ou esfericamente simétrico, verificado em \cite{lemos} e \cite{miranda}, chamada temperatura de Hawking, que é calculada por
\begin{eqnarray}
T_H=\frac{\kappa}{2\pi}.
\end{eqnarray}
A entropia $D$-dimensional, seguindo a formulação apresentada por Wald é dada por
\begin{eqnarray}
S_W=-4\pi \Omega_{D-2}\frac{[3z^2+(D^2-4)][D^2-(2-3z)^2]}{q(z)}M^{\frac{2(D-2)}{z+D-2}},
\end{eqnarray}
onde $\Omega_{D-2}$ representa o resultado finito da integração em relação as coordenadas da dimensões extras. A energia é definida por
\begin{eqnarray}
E=\frac{M^2(z-1)p(z)}{l(z+D-2)q(z)}\Omega_{D-2},
\end{eqnarray}
e, por fim, a capacidade térmica, que é apresentada da seguinte maneira
\begin{eqnarray}
C=T\frac{\partial S}{\partial T}.
\end{eqnarray}
Assim foram calculadas as ditas quantidades termodinâmicas para os buracos negros de estudo \emph{5Dz1}, \emph{5Dz3}, \emph{6Dz0} e \emph{6Dz2} e os resultados são apresentados na tabela abaixo.
\begin{table}[H]
\centering
\caption{Quantidades Termodinâmicas dos buracos negros \emph{5D} e \emph{6D}}
\vspace{0.5cm}
\begin{tabular}{|c|c|c|c|c|}\hline
 & \emph{5Dz1} & \emph{5Dz3} & \emph{6Dz0} & \emph{6Dz2}  \\ \hline
$T$  & $\sqrt{M}/2l\pi$  & $3M/4l\pi$  & $1/2l\pi$ & $3\sqrt[3]{M^2}/4l\pi$ \\
$S$  & $4M^{3/2}\pi \Omega[3]$  & $-M\pi\Omega[3]$  & $8M^2\pi\Omega[4]$ & $20M^{4/3}\pi\Omega[4]/17$ \\
$E$  &  0   & $3M^2\Omega[3]/4l$  &  $-2M^2\Omega[4]/l$      & $8M^2\Omega[4]/561 l$   \\
$C$  & $96\pi^4(l^2T^2)^{3/2} \Omega[3]$ & $-4l\pi^2T\Omega[3]/3$  & * & $640l^2T^2\pi^3\Omega[4]/153$  \\ \hline
\end{tabular}
\end{table}
Aqui nota-se algumas características termodinâmicas interessantes destes buracos negros. O buraco negro $6Dz0$ possui uma temperatura constante, independente da massa e não possui uma capacidade térmica definida. $5Dz1$ tem energia nula, o que pode representar algum caso extremo.

\subsection{Buracos Negros de Lifshitz em baixas dimensões}
Nesta subseção serão brevemente discutidos os buracos negros de Lifshitz em dimensões mais baixas. As famílias de soluções \ref{familia1}, \ref{familia2} e \ref{familia3} são proibidas, em princípio, para dimensões inferiores a $5$.

Apesar das famílias serem formalmente definidas para as dimensões superiores, a possibilidade de que novas soluções críticas existam em dimensões inferiores para alguns valores particulares do expoente dinâmico $z$ não é excluída. Uma maneira natural de explorar essa possibilidade é considerar uma continuação dimensional das expressões $D$-dimensionais e estudar se algum cancelamento potencial das divergências das constantes de acoplamento aparece quando se expande em torno de $D=3$ e $D=4$. Usando os resultados desta análise como uma indicação, será confirmada a existência de soluções críticas que, de fato, representam buracos negros de Lifshitz em $D=3$ e em $D=4$.

\subsubsection{Buraco negro assintoticamente Lifshitz $3D$ com $z=3$}
Iniciando com o caso $D=3$ ele é analisado como pertencente a família \ref{familia1}.

A constante cosmológica é regular, e vale
\begin{equation}
\lambda=\frac{10-32z+364z^2-416z^3+202z^4}{4l^2(-5-144z+202z^2+27z^4)},
\end{equation}
porém, as constantes de acoplamento divergem ao infinito em $D=3$, e podem ser expandidas em torno deste ponto. A constante $\beta_1$ pode ser escrita da seguinte forma (com a aproximação em primeira ordem em $D-3$)
\begin{eqnarray}
\beta_1=-\frac{(z-3)(3z^2+5)(9z^2-12z+11)l^2}{2(z+1(27z^4-144z^3+202z^2-144z-5))}\frac{1}{D-3},
\end{eqnarray}
onde $\beta_1=-\frac{1}{4}\beta_2=\beta_3$. Isso indica que a única possibilidade para se ter um comportamento regular do potencial para esta família em $D=3$ é exatamente quando $z=3$.
Isso resulta no seguinte elemento de linha
\begin{equation}
ds^2=-\frac{r^6}{l^6}\left(1-\frac{Ml^2}{r^2}\right)dt^2+\frac{l^2}{r^2}\left(1-\frac{Ml^2}{r^2}\right)^{-1}dr^2+\frac{r^2}{l^2}dx^2,
\end{equation}
que representa exatamente o buraco negro AGGH no contexto da NMG em  \cite{aggh} e \cite{bergshoeff}.

Todavia, no contexto da NMG em $2+1$ dimensões temos o buraco negro BTZ, solução com $z=1$. Como esta teoria é dita uma generalização e recai na NMG esperaríamos encontrar além do AGGH o buraco negro BTZ, todavia, isso não acontece. Esta é uma pergunta ainda tema de pesquisa, pois, não foi encontrada uma respota satisfatória para esta aparente divergência.

Há mais uma curiosidade envolvendo o buraco negro BTZ. Verifica-se que o buraco negro \emph{5Dz1} e \emph{6Dz0} possui os mesmos elementos $B(r)$ do buraco negro BTZ, diferenciando apenas na sua parte espacial e temporal. Isso nos leva a horizonte de eventos e gravidade superficial iguais.

Será verificado evoluindo-se o campo escalar em suas respectivas métricas como se comportam as suas frequências as possíveis relações entre elas.

\subsubsection{Buraco negro assintoticamente Lifshitz  $4D$ com $z=6$}
Em quatro dimensões, de maneira semelhante, a constante cosmológica $\lambda$ possui valor regular igual a
\begin{equation}
\lambda=\frac{-684z^3+399z^4+2(72+120z+288z^2)}{2l^2(-252z^3+27z^4-2(24+168z-192z^2))},
\end{equation}
e as constantes $\beta_1$, $\beta_2$ e $\beta_3$ divergem ao infinito em $D=4$. Expande-se as expressões em torno deste ponto e em primeira aproximação obtêm-se $\beta_1$
\begin{equation}
\beta_1=\frac{3(z-6)(z^2+4)(3z^2-4z+4)l^2}{2(z+2)(9z^4-84z^3+128z^2-112z-16)}\frac{1}{D-4},
\end{equation}
onde $\beta_1=-\frac{1}{4}\beta_2=\beta_3$. Disto vê-se aqui que o único valor possível para o expoente dinâmico $z$ para obter um potencial regular é $z=6$. O elemento de linha da família \ref{familia1} neste caso dá origem a um novo buraco negro de Lifshitz
\begin{eqnarray}
ds^2=-\frac{r^{12}}{l^{12}}\left(1-\frac{Ml^4}{r^4}\right)dt^2+\frac{l^2}{r^2}\left(1-\frac{Ml^4}{r^4}\right)^{-1}dr^2+\frac{r^2}{l^2}(dx^2+dy^2).
\end{eqnarray}

\section{Caso puramente $R^2$}
Fazendo-se $\beta_2$ e $\beta_3$ nulos na ação $D$-dimensional com correções quadráticas é obtido o caso puramente $R^2$. O espaço-tempo de Lifshitz puro é solução destas equações de campo para um valor genérico do expoente dinâmico $z$ em qualquer dimensão, desde que

\begin{eqnarray}
\lambda&=&-\frac{2z^2+(D-2)(2z+D-1)}{4l^2},\\
\beta_1&=&-\frac{1}{8\lambda}.
\end{eqnarray}
Vale notar que as parametrizações acima coincidem com os valores encontrados anteriormente para a NMG em $D=3$ \cite{aggh}, onde $\beta_3=0$ e $\beta_2=-(8/3)\beta_1=-(1/m^2)$. Os autores de \cite{lifshitzhigher} dizem que, visto isso, seu resultado generaliza a NMG, porém, viu-se que para $z=1$, BTZ não é recuperado, e como dito, ainda é tema de estudo tal divergência.

\chapter{Evolução de campos de matéria em espaços-tempos planarmente simétricos e estáticos}
Após ser estudada uma teoria de gravitação com correções quadráticas na curvatura de forma geral, escolhida uma família de soluções para análise e apontado alguns casos particulares de grande interesse, passa-se para a evolução de campos de matéria nestes espaços-tempos planarmente simétricos. Aqui será analisada a evolução do campo escalar com acoplamento não mínimo (acoplamento com escalar de Ricci e tensor de Einstein) e a evolução do campo eletromagnético.

\section{Evolução do Campo Escalar com Acoplamento Não-Mínimo}
Ao evoluir um campo escalar não-massivo na métrica de um buraco negro obtem-se informações muito importantes sobre ele, e alguns conceitos básicos sobre a teoria de perturbações de buracos negros, mesmo que eles não descrevam objetos físicos.

A estabilidade é uma das principais informações obtidas através desta análise. Se um buraco negro se mostrar instável frente a uma perturbação escalar ele, provavelmente, também será instável frente a uma perturbação gravitacional, e como os cálculos neste caso são demasiadamente mais simples pode-se obter valiosas informações sem grandes sofisticações matemáticas, como visto em \cite{sushkov}, \cite{chenjing} e \cite{gao}.

A densidade Lagrangeana para o acoplamento não mínimo, cinético e potencial, será escrita a seguir. No termo que envolve o quadrado do campo escalar $\Psi^2$ temos um escalar, logo, temos que fazer o acoplamento com um escalar, disto vem o acoplamento com o escalar de Ricci $\mathcal{R}$. Este é o chamado acoplamento potencial. Para as derivadas do campo escalar $\nabla_{\mu}\Psi\nabla_{\nu}\Psi$ o acoplamento deve ser feito também com um tensor, disto vem o acoplamento com o tensor de Einstein $G^{\mu\nu}$. Este é o chamado acoplamento cinético. Logo, escreve-se a Lagrangeana para o campo escalar com acoplamentos não-mínimos como
\begin{equation}
\label{lagrangeana}
\mathcal{L}=\frac{1}{2}\sqrt{-g}\left[ g^{\mu\nu}\nabla_{\mu}\Psi\nabla_{\nu}\Psi+\eta G^{\mu\nu}\nabla_{\mu}\Psi\nabla_{\nu}\Psi+\left(m^2+\xi\mathcal{R} \right)\Psi^2\right],
\end{equation}
onde o tensor de Einstein $G^{\mu\nu}$ é dado por
\begin{equation}
G^{\mu\nu}=R^{\mu\nu}-\frac{1}{2}g^{\mu\nu}\mathcal{R}.
\end{equation}
Desta forma a densidade Lagrangeana poderá ser escrita como
\begin{equation}
\mathcal{L}=\frac{1}{2}\sqrt{-g}\left[ g^{\mu\nu}\nabla_{\mu}\Psi\nabla_{\nu}\Psi+\eta\left(R^{\mu\nu}-\frac{1}{2}g^{\mu\nu}\mathcal{R}\right)\nabla_{\mu}\Psi\nabla_{\nu}\Psi+\left(m^2+\xi\mathcal{R} \right)\Psi^2\right].
\end{equation}
Rearranjando os termos da Lagrangeana de maneira conveniente ela pode ser escrita da seguinte maneira
\begin{equation}
\mathcal{L}=\frac{1}{2}\sqrt{-g}\left[g^{\mu\nu}\nabla_{\mu}\Psi\nabla_{\nu}\Psi\left(1-\eta\frac{\mathcal{R}}{2}\right)+
\eta R^{\mu\nu}\nabla_{\mu}\Psi\nabla_{\nu}\Psi+\left(m^2+\xi\mathcal{R} \right)\Psi^2\right],
\end{equation}
e aqui pode-se destacar alguns termos e ver seu significado separadamente. O termo destacado abaixo é chamado de conforme cinético. Vê-se que quando $\eta\mathcal{R}=2$ este termo se anula com o termo original que envolve as derivadas do campo na Lagrangeana,
\begin{equation}
-\eta\frac{\mathcal{R}}{2}g^{\mu\nu}\nabla_{\mu}\Psi\nabla_{\nu}\Psi.
\end{equation}
enquanto o termo aqui destacado é o chamado acoplamento cinético
\begin{equation}
\eta R^{\mu\nu}\nabla_{\mu}\Psi\nabla_{\nu}\Psi.
\end{equation}
Desenvolvendo a equação de Euler-Lagrange para a densidade Lagrangeana (\ref{lagrangeana}) encontra-se a equação de Klein-Gordon para um campo escalar massivo interagindo com escalar de curvatura de Ricci $\mathcal{R}$ e o tensor de Einstein $G^{\mu\nu}$, que é dada por
\begin{eqnarray}
\label{gordon}
[\bar{\Box}-m^2-\xi \mathcal{R}]\Psi(t,r,\theta,\phi)=0,\\
\nonumber\\
\frac{1}{\sqrt{-g}}\partial_{\mu}[\sqrt{-g}(h^{\mu\nu})\partial_\nu \Psi]-m^2\Psi-\xi \mathcal{R}\Psi=0.
\end{eqnarray}
A componente $h^{\mu\nu}=g^{\mu\nu}+\eta G^{\mu\nu}$ pode ser entendida como uma métrica efetiva que interage com o campo.
Utilizando a métrica da Eq.(\ref{metricagerallifshitz}), a qual pertence a família de soluções a ser estudada, a raiz do negativo do determinante da métrica é
\begin{eqnarray}
\label{det}
\sqrt{-g}=\frac{r^{D-2}}{l^{D-2}}\sqrt{AB}.
\end{eqnarray}
Como a métrica depende apenas da coordenada $r$ e tem somente elementos na diagonal o campo escalar pode ser decomposto da seguinte forma $\Psi(t,r,\vec{x})=R(r)e^{-i\omega t}e^{i\vec{x}\cdot\vec{X}}$\footnote{$\vec{x}\cdot\vec{X}=\sum\limits_{j}^{D-2} x_{j} X_{j}$}. Substituindo o campo na Eq.(\ref{gordon}) obtem-se
\begin{eqnarray}
h^{rr}\ddot{R}+\left(h^{rr}\frac{1}{\sqrt{-g}}\partial_r \sqrt{-g}+\dot{h}^{rr}\right)\dot{R}-h^{tt}\omega^2R-V(r)R=0,
\end{eqnarray}
onde o potencial $V(r)$ é dado por
\begin{eqnarray}
V(r)=\sum_{j}^{D-2} h^{jj}X_{j}X_{j}+m^2+\xi\mathcal{R}.
\end{eqnarray}
O símbolo $(\cdot)$ denota a derivada com respeito a coordenada $r$. Agora será definido $\bar{h}^{tt}=-h^{tt}$. Assim a equação se torna
\begin{eqnarray}
h^{rr}\ddot{R}+\left(h^{rr}\frac{1}{\sqrt{-g}}\partial_r \sqrt{-g}+\dot{h}^{rr}\right)\dot{R}+\bar{h}^{tt}\omega^2R-V(r)R=0,
\end{eqnarray}
e o potencial é
\begin{eqnarray}
V(r)=\sum_{j}^{D-2} h^{jj}X_{j}X_{j}+m^2+\xi\mathcal{R}.
\end{eqnarray}
Dividindo toda a equação por $\bar{h}^{tt}$, desta maneira a Eq.(\ref{gordon}) para a métrica geral da Eq.(\ref{metricagerallifshitz}) fica
\begin{eqnarray}
\label{eqgeral}
\frac{h^{rr}}{\bar{h}^{tt}}\ddot{R}+\left(\frac{h^{rr}}{\bar{h}^{tt}}\frac{1}{\sqrt{-g}}\partial_r \sqrt{-g}+\frac{\dot{h}^{rr}}{\bar{h}^{tt}}\right)\dot{R}+\omega^2R-V(r)R=0,
\end{eqnarray}
onde o potencial é
\begin{eqnarray}
\label{potencial}
V(r)=\frac{1}{\bar{h}^{tt}}\left(\sum_{j}^{D-2} h^{jj}X_{j}X_{j}+m^2+\xi\mathcal{R}\right).
\end{eqnarray}
Substituindo o determinante da métrica dado pela Eq.(\ref{det}) obtêm-se a seguinte equação geral
\begin{eqnarray}
\label{eqgerallifshitz}
\frac{h^{rr}}{\bar{h}^{tt}}\ddot{R}+\left[\frac{h^{rr}}{\bar{h}^{tt}}\left(\frac{D-2}{r}+\frac{1}{2}\partial_r\ln A+\frac{1}{2}\partial_r\ln B\right)+\frac{\dot{h}^{rr}}{\bar{h}^{tt}}\right]\dot{R}+\omega^2R-V(r)R=0,\nonumber\\
\end{eqnarray}
onde o potencial fica dado por
\begin{eqnarray}
V(r)=\frac{1}{\bar{h}^{tt}}\left(\sum_{j}^{D-2} h^{jj}X_{j}X_{j}+m^2+\xi\mathcal{R}\right).
\end{eqnarray}
Agora, no caso particularmente estudado, a métrica é dada pela Eq.(\ref{metricaDz}), pertencente a família \ref{familia1}. Neste caso a raiz do negativo do determinante da métrica, para $D\ge 3$, é
\begin{eqnarray}
\sqrt{-g}&=&\frac{r^{z+D-3}}{l^{z+D-3}}.
\end{eqnarray}
Deste modo a Eq.(\ref{eqgerallifshitz}) fica
\begin{eqnarray}
\label{altadimensao}
\frac{h^{rr}}{\bar{h}^{tt}}\ddot{R}+\left(\frac{h^{rr}}{\bar{h}^{tt}}\frac{z+D-3}{r}+\frac{\dot{h}^{rr}}{\bar{h}^{tt}}\right)\dot{R}+\omega^2R-V(r)R=0,
\end{eqnarray}
onde o potencial é
\begin{eqnarray}
\label{potencialaltadimensao}
V(r)=\frac{1}{\bar{h}^{tt}}\left(\sum_{j}^{D-2} h^{jj}X_{j}X_{j}+m^2+\xi\mathcal{R}\right).
\end{eqnarray}
Para escrever essa equação no formato da equação de Schrödinger foram utilizadas as seguintes transformações
\begin{eqnarray}
R&=&\bar{R}(r)b(r),\\
r&=&r(r_*),\\
H&=&\frac{h^{rr}}{\bar{h}^{tt}}.
\end{eqnarray}
A conta feita de maneira detalhada, apresentando-se o valor da função $b(r)$ encontrada, se encontra no Apêndice.

Ao realizar essa transformação o resultado é um novo potencial. O potencial antigo é agora acrescido de um termo oriundo das transformações realizadas. O novo potencial é $\bar{V}=V-C_5$, onde $C_5$ é o termo que subtrai ao potencial e é dado por
\begin{eqnarray}
\label{superc5}
C_5=\frac{H}{b}\left[\ddot{b}+\dot{b}\left(\frac{\dot{h}^{rr}}{h^{rr}}+\frac{z+D-3}{r}\right)\right].
\end{eqnarray}
A equação transformada será dada por
\begin{eqnarray}
\bar{R}^{''}+\omega^2\bar{R}-\bar{V}(r)\bar{R}=0,
\end{eqnarray}
onde o novo potencial tem a seguinte forma
\begin{eqnarray}
\label{superpotencial}
\bar{V}(r)=\frac{1}{\bar{h}^{tt}}\left(\sum_{j}^{D-2} h^{jj}X_{j}X_{j}+m^2+\xi\mathcal{R}\right)-\frac{H}{b}\left[\ddot{b}+\dot{b}\left(\frac{\dot{h}^{rr}}{h^{rr}}+\frac{z+D-3}{r}\right)\right].
\end{eqnarray}

\section{Evolução do Campo Eletromagnético}
Assim como o campo escalar, o campo eletromagnético é de suma importância para o estudo de estabilidade em buracos negros. Ele representa um campo de matéria com algum sentido físico, o que enriquece ainda mais a análise.

Aqui será analisada a evolução de um campo eletromagnético para a métrica geral dada pela Eq.(\ref{metricagerallifshitz}), caso $D$-dimensional planarmente simétrico, com raiz do negativo do determinante da métrica dado pela Eq.(\ref{det}) e decomposição do tensor eletromagnético de acordo com \cite{cardoso}.

O tensor de Maxwell $F_{\mu\nu}$ pode ser escrito em função de $A_\mu$, o potencial vetor, e fica como
\begin{eqnarray}
F_{\mu\nu}=A_{\nu,\mu}-A_{\mu,\nu},
\end{eqnarray}
onde a decomposição de $A_\mu$ respeita a simetria do espaço-tempo e é da seguinte forma
\begin{center}
$A_\mu(t,r,\vec{X})=\left[\begin{matrix}
\mathbf{f}(r) \\
\mathbf{g}(r) \\
\mathbf{X}_{i}(r) \\
\end{matrix}\right]e^{-i\omega t}e^{i\vec{k}\cdot\vec{X}}.$
\end{center}
Aqui $X_i=(x,y,z,...)$ representa as coordenadas espaciais, $k_i=(k_1,k_2,k_3,...)$ os autovalores correspondentes e $i$ um índice de soma que abrange os valores $i=2,...,D-2$. As componentes não-nulas do tensor eletromagnético covariante são dadas por
\begin{eqnarray}
F_{tr}&=&(-i\omega\mathbf{g}-\dot{\mathbf{f}})e^{-i\omega t}e^{i\sum k_i X_i}, \\
F_{tx_{i}}&=&(-i\omega \mathbf{X}_i-ik_i\mathbf{f})e^{-i\omega t}e^{i\sum k_i X_i}, \\
F_{rx_i}&=&(\dot{\mathbf{X}}_i-ik_i\mathbf{g})e^{-i\omega t}e^{i\sum k_i X_i}, \\
F_{x_ix_j}&=&(ik_i\mathbf{X}_j-ik_j\mathbf{X}_i)e^{-i\omega t}e^{i\sum k_i X_i}.
\end{eqnarray}
Para evolução do campo eletromagnético necessita-se do tensor eletromagnético contravariante. Para isso utiliza-se a métrica contravariante para `subir' os índices da seguinte forma
\begin{eqnarray}
F^{\alpha\beta}=F_{\mu\nu}g^{\mu\alpha}g^{\nu\beta}.
\end{eqnarray}
Isto resulta nas seguintes componentes para o tensor eletromagnético contravariante
\begin{eqnarray}
F^{tr}&=&\frac{i\omega\mathbf{g}+\dot{\mathbf{f}}}{AB}e^{-i\omega t}e^{i\sum k_i X_i}, \\
F^{tx_i}&=&\frac{l^2(i\omega \mathbf{X}_i+ik_i\mathbf{f})}{r^2A}e^{-i\omega t}e^{i\sum k_i X_i}, \\
F^{rx_i}&=&\frac{l^2(\dot{\mathbf{X}}_i-ik_i\mathbf{g})}{r^2B}e^{-i\omega t}e^{i\sum k_i X_i}, \\
F^{x_ix_j}&=&\frac{l^4(ik_i\mathbf{X}_j-ik_j\mathbf{X}_i)}{r^4}e^{-i\omega t}e^{i\sum k_i X_i}.
\end{eqnarray}
As equações de Maxwell no espaço curvo para um campo eletromagnético livre de fontes são dadas por
\begin{eqnarray}
\label{maxwell}
F^{\mu\nu}_{\phantom{\mu\nu}{; \nu}}=0 \ \ \ \ \ \to \ \ \ \ \ \frac{1}{\sqrt{-g}}(\sqrt{-g}F^{\mu\nu})_{, \nu}=0.
\end{eqnarray}
Evoluindo os índices da Eq.(\ref{maxwell}), as equações para as componentes $t$, $r$ e $x_i$ são, respectivamente, dadas por
\begin{eqnarray}
\label{t}
(\sqrt{-g}F^{t\nu})_{,\nu}&=&\dot{\Psi}+k_i\frac{r^{D-4}}{l^{D-4}}\sqrt{\frac{B}{A}}(\omega \mathbf{X}_i+k_i\mathbf{f})=0,\\
\nonumber\\
\label{r}
(\sqrt{-g}F^{r\nu})_{,\nu}&=&\omega^2\Psi-\omega k_i\frac{r^{D-4}}{l^{D-4}}\sqrt{\frac{A}{B}}(\dot{\mathbf{X}}_i-ik_i\mathbf{g})=0,\\
\nonumber\\
(\sqrt{-g}F^{x_i\nu})_{,\nu}&=&\omega\sqrt{\frac{B}{A}}r^{D-4}(\omega \mathbf{X}_i+k_i \mathbf{f})+\left[\sqrt{\frac{A}{B}}r^{D-4}(\dot{\mathbf{X}}_i-ik_i \mathbf{g})\right]_{,r}+\nonumber\\
&+&k_j l^2r^{D-6}\sqrt{AB}(k_i\mathbf{X}_j-k_j\mathbf{X}_i)=0.
\end{eqnarray}
Aqui será definido de maneira conveniente a seguinte expressão abaixo
\begin{eqnarray}
\label{psi}
i\omega\mathbf{g}+\dot{\mathbf{f}}=-\Psi\sqrt{AB}\frac{l^{D-2}}{r^{D-2}}.
\end{eqnarray}
Multiplicando a Eq.(\ref{t}) por $\frac{l^{D-4}}{r^{D-4}}\sqrt{\frac{A}{B}}$ tem-se
\begin{eqnarray}
\label{tx}
\frac{l^{D-4}}{r^{D-4}}\sqrt{\frac{A}{B}}\dot{\Psi}+k_i\omega \mathbf{X}_i+k_i^2\mathbf{f}=0.
\end{eqnarray}
Agora derivando a Eq.(\ref{tx}) em relação a $r$ fica
\begin{eqnarray}
\label{t2}
\left[\frac{l^{D-4}}{r^{D-4}}\sqrt{\frac{A}{B}}\dot{\Psi}\right]_{,r}+k_i\omega\dot{\mathbf{X}}_i+k_i^2\dot{\mathbf{f}}=0.
\end{eqnarray}
Agora multiplicando a Eq.(\ref{r}) por $\frac{l^{D-4}}{r^{D-4}}\sqrt{\frac{B}{A}}$ obtém-se
\begin{eqnarray}
\label{r2}
\omega^2\frac{l^{D-4}}{r^{D-4}}\sqrt{\frac{B}{A}}\Psi-k_i\omega\dot{\mathbf{X}}_i+i\omega k_i^2\mathbf{g}=0.
\end{eqnarray}
Somando a Eq.(\ref{t2}) com a Eq.(\ref{r2}) resulta em
\begin{eqnarray}
\label{transformar}
\left[\frac{l^{D-4}}{r^{D-4}}\sqrt{\frac{A}{B}}\dot{\Psi}\right]_{,r}+\omega^2\frac{l^{D-4}}{r^{D-4}}\sqrt{\frac{B}{A}}\Psi+k_i^2(\mathbf{f}+i\omega\mathbf{g})=0.
\end{eqnarray}
Utilizando da definição de $\Psi$ da Eq.(\ref{psi}), a Eq.(\ref{transformar}) se transforma em
\begin{eqnarray}
\label{eletromag}
r^{D-4}\sqrt{\frac{A}{B}}\frac{d}{dr}\left[\frac{1}{r^{D-4}}\sqrt{\frac{A}{B}}\frac{d\Psi}{dr}\right]+\omega^2\Psi-A\frac{k_i^2l^2}{r^2}\Psi=0.
\end{eqnarray}
É objetivo, para aplicação dos métodos numéricos, que a equação resultante da evolução do campo eletromagnético esteja no formato da equação de Schrödinger. Para isto, primeiramente, a Eq.(\ref{eletromag}) deve sofrer uma mudança de coordenadas. A coordenada conveniente para esta transformação é a conhecida como tartaruga e representada por $r_*(r)$. Ela é definida da seguinte forma
\begin{eqnarray}
\label{tartaruga}
\frac{d}{dr}=\sqrt{\frac{B}{A}}\frac{d}{dr_*}.
\end{eqnarray}
Assim, aplicando-se a transformação da Eq.(\ref{tartaruga}), a Eq.(\ref{eletromag}) toma a seguinte forma
\begin{eqnarray}
\label{psitransformado}
\Psi^{''}-r(D-4)\sqrt{\frac{A}{B}}\Psi^{'}+\omega^2\Psi-\frac{k_i^2Al^2}{r^2}\Psi=0.
\end{eqnarray}
Esta equação ainda não esta no formato da equação de Schrödinger. Como segundo passo, o termo de derivada primeira deve ser eliminado. Com esse intuito é que se faz a seguinte transformação para função $\Psi$
\begin{eqnarray}
\label{psib}
\Psi=\bar{\Psi}\mathbf{b},
\end{eqnarray}
de onde, naturalmente, sua segunda derivada é dada por
\begin{eqnarray}
\label{psi2}
\Psi^{''}=\frac{d^2\Psi}{dr_*^2}=\frac{d}{dr_*}\left[\bar{\Psi}^{'}\mathbf{b}+\bar{\Psi}\mathbf{b}^{'}\right]=\bar{\Psi}^{''}\mathbf{b}+2\bar{\Psi}^{'}\mathbf{b}^{'}+\bar{\Psi}\mathbf{b}^{''}.
\end{eqnarray}
Substituindo agora as Eqs.(\ref{psib},\ref{psi2}) na Eq.(\ref{psitransformado}) obtém-se
\begin{eqnarray}
\bar{\Psi}^{''}\mathbf{b}+2\bar{\Psi}^{'}\mathbf{b}^{'}+\bar{\Psi}\mathbf{b}^{''}-r(D-4)\sqrt{\frac{A}{B}}\left[\bar{\Psi}^{'}\mathbf{b}+\bar{\Psi}\mathbf{b}^{'}\right]+\omega^2\bar{\Psi}b-A\frac{k_i^2l^2}{r^2}\bar{\Psi}\mathbf{b}=0.\nonumber\\
\end{eqnarray}
Dividindo toda essa equação por $\mathbf{b}$
\begin{eqnarray}
\label{b}
\bar{\Psi}^{''}+\bar{\Psi}^{'}\left[2\frac{\mathbf{b}^{'}}{\mathbf{b}}-r(D-4)\sqrt{\frac{A}{B}}\right]+\omega^2\bar{\Psi}+\bar{\Psi}\left[\frac{\mathbf{b}^{''}}{\mathbf{b}}-r(D-4)\sqrt{\frac{A}{B}}\frac{\mathbf{b}^{'}}{\mathbf{b}}-\frac{k_i^2Al^2}{r^2}\right]=0.\nonumber\\
\end{eqnarray}
Como o objetivo era eliminar o termo de derivada primeira, ele agora é igualado a zero. Isto traz a seguinte condição para a função $\mathbf{b}$
\begin{eqnarray}
\mathbf{b}=e^{r^2(D-4)/4}.
\end{eqnarray}
Conhecendo $\mathbf{b}$ são necessárias as relações de $\frac{\mathbf{b}^{'}}{\mathbf{b}}$ e $\frac{\mathbf{b}^{''}}{\mathbf{b}^{'}}$ para obter a equação final procurada. Calculando elas separadamente o resultado obtido é apresentado abaixo
\begin{eqnarray}
\frac{\mathbf{b}^{'}}{\mathbf{b}}&=&\frac{1}{\mathbf{b}}\frac{d\mathbf{b}}{dr_*}=\frac{1}{\mathbf{b}}\sqrt{\frac{A}{B}}\frac{d\mathbf{b}}{dr}=\sqrt{\frac{A}{B}}\frac{r(D-4)}{2}, \end{eqnarray}
\begin{eqnarray}
\frac{\mathbf{b}^{''}}{\mathbf{b}}=\sqrt{\frac{A}{B}}\left(\frac{d}{dr}\sqrt{\frac{A}{B}}\right)\frac{r(D-4)}{2}+\frac{A}{B}\frac{r^2(D-4)^2}{4}.
\end{eqnarray}
Agora substituindo as relações obtidas para $\mathbf{b}$ na Eq.(\ref{b}) encontra-se finalmente a equação de evolução eletromagnética no formato da equação de Schrödinger, dada por
\begin{eqnarray}
\label{Maxwellfinal}
\bar{\Psi}^{''}+\omega^2\bar{\Psi}-\bar{V}\bar{\Psi}=0
\end{eqnarray}
onde o potencial efetivo vem dado por
\begin{eqnarray}
\label{Maxpot}
\bar{V}=A\frac{k_i^2l^2}{r^2}+\frac{r(D-4)}{4B}\left[\dot{A}-\frac{A\dot{B}}{B}+rA(D-4)\right]
\end{eqnarray}
Nota-se que o potencial possui um termo que está presente apenas em casos de dimensões mais altas, a saber $D\ge 5$, revelando de maneira explícita a influência das dimensões extras.

\chapter{Equações de campo aplicadas aos Buracos Negros}
Agora com as equações de evolução de dois campos de máteria calculadas, o próximo passo é aplicá-las aos buracos negros de estudo deste trabalho. Será aplicada primeiramente a evolução do campo escalar e eletromagnético para os casos de $2+1$-dimensões, buracos negros BTZ e AGGH. Depois seguiremos para a análise da evolução dos campos escalar e eletromagnético para os buracos negros de Lifshitz em altas dimensões, em dois casos: buracos negros \emph{5Dz1} e \emph{6Dz0}.

\section{Evolução do campo escalar e eletromagnético em $2+1$-dimensões}
A métrica para um espaço-tempo arbitrário estático em $2+1$ dimensões é dada por
\begin{eqnarray}
\label{3dmetric}
ds^2=-A(r)dt2+B(r)dr^2+r^2d\theta^2
\end{eqnarray}
Aplicando a métrica (\ref{3dmetric}) na equacao de Klein-Gordon modificada dada pela Eq.(\ref{gordon}) temos
\begin{eqnarray}
\frac{g^{rr}+\eta G^{rr}}{g^{tt}+\eta G^{tt}}\ddot{R}+\frac{\dot{R}}{g^{tt}+\eta G^{tt}}\left[\left(g^{rr}+\eta G^{rr}\right)\partial_r[\ln(\sqrt{-g})]+\dot{g}^{rr}+\eta\dot{G}^{rr}\right]-\omega^2R-V(r)R=0,\nonumber\\
\end{eqnarray}
onde o potencial é
\begin{eqnarray}
V(r)=\frac{1}{g^{tt}+\eta G^{tt}}\left[\kappa^2 (g^{\theta\theta}+\eta G^{\theta\theta})+m^2+\xi\mathcal{R}\right].
\end{eqnarray}
Substituindo as expressões de $g^{\mu\nu}$, $G^{\mu\nu}$ e $g$ encontra-se a seguinte equação
\begin{eqnarray}
\label{kleingordon}
\frac{2ABr+\eta \dot{A}}{\eta \dot{B}-2B^2r}\ddot{R}&+&\dot{R}\frac{2AB^2r}{\eta \dot{B}-2B^2r}\left\{\frac{2ABr+\eta \dot{A}}{2AB^2r}\partial_r[\ln(r\sqrt{AB})]+\left[\frac{1}{\dot{B}}+\eta\partial_r\left(\frac{\dot{A}}{2AB^2r}\right)\right]\right\}-\nonumber\\
&-&\omega^2R-V(r)R=0,
\end{eqnarray}
e aqui o potencial fica da forma
\begin{eqnarray}
V(r)=\frac{2AB^2r}{\eta \dot{B}-2B^2r}\left\{\kappa^2\frac{4A^2B^2+\eta [2\ddot{A}AB-B(\dot{A})^2-\dot{A}\dot{B}A]}{4A^2B^2r^2}+m^2+\right. \nonumber\\
\left.+\frac{\xi}{2A^2B^2r}\left[-2r\ddot{A}AB+r(\dot{A})^2B+r\dot{A}\dot{B}A-2\dot{A}AB+2\dot{B}A^2\right]\right\}.
\end{eqnarray}
Para o caso $2+1$-dimensional com simetria de Lifshitz e expoente dinâmico $z$ qualquer, a qual é solução de uma teoria da NMG, os termos da métrica $A(r)$ e $B(r)$ são dados por
\begin{eqnarray}
A(r)&=& -\frac{r^{2z}}{l^{2z}}f(r), \\
B(r)&=&\frac{l^2}{r^2f(r)},
\end{eqnarray}
onde têm-se as seguinte definiçõe para a função $f(r)$
 \begin{eqnarray}
f(r)&=& 1-\frac{Ml^2}{r^2}.
\end{eqnarray}
Assim a métrica(\ref{3dmetric}) será escrita como
\begin{eqnarray}
\label{metricaAB}
ds^2=-\frac{r^{2z}}{l^{2z}}\left(1-\frac{Ml^2}{r^2}\right)dt^2+\frac{l^2}{r^2}\left(1-\frac{Ml^2}{r^2}\right)^{-1}dr^2+r^2d\theta^2.
\end{eqnarray}
Há duas possibilidades a serem estudadas a partir desta métrica variando-se o expoente dinâmico $z$. Os dois valores que ele pode assumir são $z=1$, que corresponde ao buraco negro BTZ, e $z=3$, que corresponde ao buraco negro AGGH.
Nota-se aqui que o horizonte de eventos para ambos os buracos negros é o mesmo, $r_+=\sqrt{M}l$.

As expressões para o Escalar de Ricci e as componentes do Tensor de Einstein para este caso são oriundas das Eqs.(\ref{riccigeral}, \ref{Gtgeral}, \ref{Grgeral}, \ref{Gfgeral}), e se tornam
\begin{eqnarray}
\label{ricciAB}
\mathcal{R}&=&\frac{2}{r^2l^2}[r_+^2(z-1)^2-r^2(z^2+z+1)], \\
\nonumber\\
\label{GtAB}
G^{tt}&=&\frac{r^{2(1-z)}l^{2(z-1)}}{r_+^2-r^2}, \\
\nonumber\\
\label{GrAB}
G^{rr}&=&\frac{(r_+^2-r^2)[r_+^2(z-1)-zr^2]}{r^2l^4}, \\
\nonumber\\
\label{GfAB}
G^{\theta\theta}&=&\frac{r_+^2[z(3-z)-2]+z^2r^2}{r^4l^2}.
\end{eqnarray}
Finalmente, a Equação de Klein-Gordon modificada da Eq.(\ref{kleingordon}), para esta métrica, e um $z$ qualquer, dá o acoplamento cinético mais potencial($\eta\neq0$, $m\neq0$ e $\xi\neq0$), a saber
\begin{eqnarray}
\label{eqgeral2+1}
&&\frac{\{l^6+\eta f(r_+^2-r^2)[r_+^2(z-1)-zr^2]\}l^{2z}(r_+^2-r^2)}{r^2l^4f[r^{2z}f(r_+^2-r^2)+\eta r^{2(1-z)}l^{2(2z-1)}]}R^{''}+\nonumber\\
+&&\frac{l^{2z}(r_+^2-r^2)}{r^{2z}f(r_+^2-r^2)+\eta r^{2(1-z)}l^{2(2z-1)}}\left\{\left[\ln\left(\frac{r^z}{l^{z-1}}\right)\right]^{-1}\left(\frac{l^6+\eta f(r_+^2-r^2)[r_+^2(z-1-zr^2)]}{r^2l^4f}\right)+\right.\nonumber\\
&&\left.+\left(\frac{r^2f}{l^2}\right)^{'}+\eta\left(\frac{(r_+^2-r^2)[r_+^2(z-1)-zr^2]}{r^2l^4}\right)^{'}\right\}R^{'}-\omega^2R-V(r)R=0,
\end{eqnarray}
onde o potencial é
\begin{eqnarray}
\label{potencialgeral2+1}
V(r)&=&\frac{l^{2z}(r_+^2-r^2)}{r^{2z}f(r_+^2-r^2)+\eta r^{2(1-z)}l^{2(2z-1)}}\left\{\frac{r^2+\eta\{ r_+^2[z(3-z)-2]+z^2r^2\}}{r^4l^2}\kappa^2+m^2+\right.\nonumber\\
&&\left.+\frac{2}{r^2l^2}[r_+^2(z-1)^2-r^2(z^2+z+1)]\xi\right\}.
\end{eqnarray}
Aqui será analisada separadamente os casos particulares do acoplamento cinético ($\xi=0$) e acoplamento potencial quando ($\eta=0$). No acoplamento potencial existem três diferentes tipos enumerados abaixo
\begin{enumerate}
\item {\bf Acoplamento Minimo} $\to$ $\xi =0$,
\item {\bf Acoplamento Conforme} $\to$ $\xi_c = \frac{1}{4}\left[\frac{D-2}{D-1}\right]$,
\item {\bf Acoplamento Nao-Minimo} $\to$ $\xi\neq 0\neq\xi_c$.
\end{enumerate}
Para as análises que se seguirão será feita a seguinte substituição de variáveis
\begin{eqnarray}
\label{substituicao}
r=\frac{\sqrt{M}l}{y},
\end{eqnarray}

\subsection{Buraco negro BTZ}
\subsubsection{Acoplamento Potencial}
Para o buraco negro BTZ, $z=1$, a Eq.(\ref{eqgeral2+1}) com a substituição proposta pela Eq.(\ref{substituicao}) se torna, para cada caso a seguir:
\begin{itemize}
\item{Acoplamento mínimo:} A equação diferencial é
\begin{eqnarray}
\frac{d^2R}{dy^2}-\frac{y^2+1}{y(1-y^2)}\frac{dR}{dy}+V_{01}(y)\ R=0,
\end{eqnarray}
onde o potencial efetivo é
\begin{eqnarray}
V_{01}(y)=\frac{1}{1-y^2}\left(\frac{\omega^2 l^2}{M(1-y^2)}-\frac{k^2}{M}-\frac{m^2l^2}{y^2}\right).
\end{eqnarray}
A equação possui solução exata e é escrita em função das funções hipergeométricas,
\begin{eqnarray}
R_{01}(y)&=&C_1y^{1+a_1}(y^2-1)^{bM}F\left(\left[b_1+\frac{1+a_1}{2},-b_1-\frac{1-a_1}{2}\right],[1+a_1];y^2\right)+\nonumber\\
&+&C_2y^{1-a_1}(y^2-1)^{bM}F\left(\left[b_1+\frac{1-a_1}{2},-b_1-\frac{1+a_1}{2}\right],[1-a_1];y^2\right), \nonumber \\
\end{eqnarray}
onde define-se
\begin{eqnarray}
a_1&=&\sqrt{1+m^2l^2}\\
b&=&\frac{il\omega}{2M^{3/2}},\\
b_1&=&\frac{ik-il\omega}{2\sqrt{M}}
\end{eqnarray}

\item{Acoplamento Conforme ($\xi=1/8$):} A equação diferencial se torna
\begin{eqnarray}
\frac{d^2R}{dy^2}-\frac{y^2+1}{y(1-y^2)}\frac{dR}{dy}+V_{\xi_c1}(y)\ R=0,
\end{eqnarray}
onde o potencial efetivo é
\begin{eqnarray}
V_{\xi_c1}(y)=\frac{1}{1-y^2}\left(\frac{\omega^2 l^2y}{M(1-y^2)}-\frac{k^2}{M}-\frac{m_2^2l^2}{y^2}\right).
\end{eqnarray}
Este caso possui solução exata e é escrita em funcao das funcoes hipergeométricas,
\begin{eqnarray}
R_{\xi_c1}(y)&=&C_3y^{1+a_2}(y^2-1)^{bM}F\left(\left[b_1+\frac{2+a_2}{4},-b_1+\frac{2-a_2}{4}\right],[1+a_2];y^2\right)+\nonumber\\
&+&C_4y^{1-a_2}(y^2-1)^{bM}F\left(\left[b_1+\frac{2-a_2}{4},-b_1+\frac{2+a_2}{4}\right],[1-a_2];y^2\right), \nonumber\\
\end{eqnarray}
onde são definidas
\begin{eqnarray}
a_2&=&\sqrt{1+m_2^2l^2}\\
b&=&\frac{il\omega}{2M^{3/2}},\\
b_1&=&\frac{ik-il\omega}{2\sqrt{M}}\\
m_2&=&\sqrt{m^2-\frac{3l^2}{4}}
\end{eqnarray}

\item{Acoplamento Não-Mínimo:} A equação diferencial é
\begin{eqnarray}
\frac{d^2R}{dy^2}-\frac{y^2+1}{y(1-y^2)}\frac{dR}{dy}+V_{\xi 1}(y)\ R=0,
\end{eqnarray}
onde o potencial efetivo é
\begin{eqnarray}
V_{\xi 1}(y)=\frac{1}{1-y^2}\left(\frac{\omega^2 l^2}{M(1-y^2)}-\frac{k^2}{M}-\frac{m_3^2l^2}{y^2}\right).
\end{eqnarray}
A solução é exata e escrita em função das funçõs hipergeométricas,
\begin{eqnarray}
R_{\xi 1}(y)&=&C_5y^{1+a_3}(y^2-1)^{bM}F\left(\left[b_1+\frac{1+a_3}{2},-b_1-\frac{1-a_3}{2}\right],[1+a_3];y^2\right)+\nonumber\\
&+&C_6y^{1-a_3}(y^2-1)^{bM}F\left(\left[b_1+\frac{1-a_3}{2},-b_1-\frac{1+a_3}{2}\right],[1-a_3];y^2\right), \nonumber\\
\end{eqnarray}
onde definimos
\begin{eqnarray}
a_3&=&\sqrt{1+m_3^2l^2}\\
b&=&\frac{il\omega}{2M^{3/2}},\\
b_1&=&\frac{ik-il\omega}{2\sqrt{M}}\\
m_3&=&\sqrt{m^2-\frac{6\xi}{l^2}}
\end{eqnarray}
\end{itemize}
Verifica-se que a influência do acoplamento potencial para o buraco negro BTZ é a reparametrização da massa.

\subsubsection{Acoplamento Cinético}
No caso particular do coeficiente dinâmico $z=1$, buraco negro BTZ, a equação de movimento para o campo será
\begin{eqnarray}
\frac{d^2R}{dr^2}+\frac{r^2l^4}{(r^2-Ml^2)(1+\eta r^2)}\left\{\left[\frac{3(r^2-Ml^2)}{l^2}-\eta\frac{Ml^2-r^2}{l^4}\right]\frac{dR}{dr}+\right.\nonumber\\
+\left.\left[-\omega^2\left(\frac{l^2}{r^2-Ml^2}+\frac{\eta}{Ml^2-r^2}\right)-k^2\left(\frac{1}{r^2}+\frac{\eta}{r^2l^2}\right)-m^2\right]R\right\}=0.
\end{eqnarray}

\subsection{Buraco negro AGGH}
\subsubsection{Acoplamento Potencial}
Para o buraco negro AGGH, $z=3$, a Eq.(\ref{eqgeral2+1}) com a substituição proposta pela Eq.(\ref{substituicao}) se torna, para cada caso a seguir.
\begin{itemize}
\item{Acoplamento Mínimo:} A equação diferencial é
\begin{equation}
\frac{d^2R}{dy^2}+\frac{y^2-3}{y(1-y^2)}\frac{dR}{dy}+V_{03}\ R=0,
\end{equation}
onde o potencial efetivo é
\begin{eqnarray}
V_{03}=\frac{1}{1-y^2}\left[\frac{\omega^2 l^2y^4}{M^3(1-y^2)}-\frac{k^2}{M}-\frac{m^2l^2}{y^2}\right].
\end{eqnarray}
Esta equação pode ser resolvida exatamente em termos das funções confluentes de Heun resultando em
\begin{eqnarray}
\label{R03}
R_{03}(y)&=&C_7\ y^{2+a_4}(y^2-1)^{b}HeunC\left(0,a_4,2b,c,d,y^2\right)\nonumber\\
&+&C_8\ y^{2-a_4}(y^2-1)^{b}HeunC\left(0,-a_4,2b,c,d,y^2\right),
\end{eqnarray}
onde define-se
\begin{eqnarray}
a_4&=&\sqrt{4+m^2l^2},\\
b&=&\frac{il\omega}{2M^{3/2}},\\
c&=&\frac{l^2\omega^2}{4M^3},\\
d&=&\frac{m^2l^2M+k^2}{4M}+1
\end{eqnarray}
Esta solução foi apresentada pela primeira vez em \cite{BerthaJef}.

\item{Acoplamento Conforme ($\xi=1/8$):} A equação diferencial é
\begin{eqnarray}
\frac{d^2R}{dy^2}+\frac{y^2-3}{y(1-y^2)}\frac{dR}{dy}+V_{\xi_c3}\ R=0,
\end{eqnarray}
onde o potencial efetivo é
\begin{eqnarray}
V_{\xi_c3}=\frac{1}{1-y^2}\left[\frac{\omega^2 l^2y^4}{M^3(1-y^2)}-\frac{k_5^2}{M}-\frac{m_5^2l^2}{y^2}\right].
\end{eqnarray}
Esta equação possui solução exata que pode ser escrita em termos das funções confluentes de Heun e é dada por
\begin{eqnarray}
R_{\xi_c3}(y)&=& C_9\ y^{2+a_5}(y^2-1)^{b}HeunC\left(0,a_5,2b,c,d_5,y^2\right) \nonumber \\
&+&C_{10}\ y^{2-a_5}(y^2-1)^{b}HeunC\left(0,-a_5,2b,c,d_5,y^2\right),
\end{eqnarray}
onde é definido
\begin{eqnarray}
a_5&=&\sqrt{4+m_5^2l^2},\\
b&=&\frac{il\omega}{2M^{3/2}},\\
c&=&\frac{l^2\omega^2}{4M^3},\\
d_5&=&\frac{m_5^2l^2M+k_5^2}{4M}+1,\\
k_5&=&\sqrt{k^2+M},\\
m_5&=&\sqrt{m^2-\frac{13}{4l^2}}
\end{eqnarray}
O regime de acoplamento conforme ainda não havia sido analisado. Portanto, esta solução é, pela primeira vez, apresentada aqui neste estudo.

\item{Acoplamento Não-Mínimo:} A equação diferencial é
\begin{eqnarray}
\frac{d^2R}{dy^2}+\frac{y^2-3}{y(1-y^2)}\frac{dR}{dy}+ V_{\xi 3}(y)\ R=0,
\end{eqnarray}
onde o potencial efetivo será
\begin{eqnarray}
V_{\xi 3}(y)=\frac{1}{1-y^2}\left[\frac{\omega^2 l^2y^4}{M^3(1-y^2)}-\frac{k_6^2}{M}-\frac{m_6^2l^2}{y^2}\right].
\end{eqnarray}
Novamente, a solução desta equação pode ser escrita em termos das funções confluentes de Heun, sendo
\begin{eqnarray}
R_{\xi 3}(y)&=&C_{11}\ y^{2+a_6}(y^2-1)^{b}HeunC\left(0,a_6,2b,c,d_6,y^2\right) \nonumber  \\
&+&C_{12}\ y^{2-a_6}(y^2-1)^{b}HeunC\left(0,-a_6,2b,c,d_6,y^2\right),
\end{eqnarray}
onde definimos
\begin{eqnarray}
a_6&=&\sqrt{4+m_6^2l^2},\\
b&=&\frac{il\omega}{2M^{3/2}},\\
c&=&\frac{l^2\omega^2}{4M^3},\\
d_6&=&\frac{m_6^2l^2M+k_6^2}{4M}+1,\\
k_6&=&\sqrt{k^2+8\xi M},\\
m_6&=&\sqrt{m^2-{26\xi}{l^2}}.
\end{eqnarray}
\end{itemize}
Aqui é verificado que a influência do acoplamento potencial para o buraco negro AGGH, diferentemente do BTZ, não reparametriza somente a massa, mas também, o número de onda.

\subsubsection{Acoplamento Cinético}
No caso $z=3$, buraco negro AGGH encontra-se
\begin{eqnarray}
\frac{d^2R}{dr^2}&+&\frac{r^2l^4(r^2-Ml^2)^{-1}}{1+\eta(2Ml^2-3r^2)}\left\{\left[\frac{5(r^2-Ml^2)}{l^2}+\eta\left(-\frac{4M^2}{r^3}+\frac{(Ml^2-r^2)(2Ml^2-3r^2)}{r^2l^4}\right)\right]\frac{dR}{dr}+\right.\nonumber\\
&+&\left.\left[-\omega^2\left(\frac{l^6-\eta l^4}{r^4(r^2-Ml^2)}\right)-k^2\left(\frac{1}{r^2}+\eta\frac{9r^2-2Ml^2}{r^4l^2}\right)-m^2\right]R\right\}=0.
\end{eqnarray}

\section{Evolução do campo eletromagnético em $2+1$-dimensões}
As perturbações eletromagnéticas são governadas pelas equações de Maxwell onde define-se $F_{\mu\nu}$ como o tensor de Maxwell e $A_\mu$ como o potencial vetor eletromagnético. Como temos um espaço tridimensional de fundo é aconselhável expandir $A_\mu$ em um vetor harmônico esférico $3$-dimensional
\begin{center}
$A_i(t,r,\theta,\phi)=\left[\begin{matrix}
g^m(t,r) \\
h^m(t,r) \\
k^m(t,r) \\
\end{matrix}\right]\begin{matrix}
e^{i\kappa\theta},
\end{matrix}$ \end{center}
onde $\kappa$ é o número quântico angular. Seguindo os mesmos passos para obter a equação para a perturbação eletromagnética encontramos a mesma equação que no caso da evolução do campo escalar. A razão é que em três dimensões a $2$-forma do campo de Maxwell é dual a $1$-forma do campo escalar.

\section{Evolução do campo escalar e eletromagnético em $D\ge5$ dimensões}

\subsection{Buraco Negro \emph{5Dz1}}

\subsubsection{Evolução do Campo Escalar acoplado ao Escalar de Ricci e ao tensor de Einstein}
Das Eqs.(\ref{altadimensao}) com potencial dado pela Eq.(\ref{potencialaltadimensao}) obtém-se a seguinte equação de perturbação para o buraco negro \emph{5Dz1}
\begin{eqnarray}
\frac{h^{rr}}{\bar{h}^{tt}}\ddot{R}+\left(\frac{3}{r}\frac{h^{rr}}{\bar{h}^{tt}}+\frac{\dot{h}^{rr}}{\bar{h}^{tt}}\right)\dot{R}+\omega^2R-\frac{1}{\bar{h}^{tt}}\left(h^{xx}X^2+m^2+\xi\mathcal{R}\right)R=0.
\end{eqnarray}
Transformando essa equação para o formato de Schrödinger é necessário o termo que subtrai ao potencial, $C_5$, dado pela Eq.(\ref{superc5}) e que fica
\begin{eqnarray}
C_5=\frac{H}{b}\left[\ddot{b}+\dot{b}\left(\frac{\dot{h}^{rr}}{h^{rr}}+\frac{3}{r}\right)\right].
\end{eqnarray}
Isto deixa a equação no formato Schrodinger com potencial dado por
\begin{eqnarray}
\bar{V}=\frac{1}{\bar{h}^{tt}}\left(h^{xx}X^2+m^2+\xi\mathcal{R}\right)-\frac{H}{b}\left[\ddot{b}+\dot{b}\left(\frac{\dot{h}^{rr}}{h^{rr}}+\frac{3}{r}\right)\right].
\end{eqnarray}
onde usamos a seguinte notação $h^{xx}X^2=h^{11}X_1^2+h^{22}X_2^2+h^{33}X_3^2$. Sendo o escalar de Ricci e o tensor de Einstein calculados para este caso dados por
\begin{eqnarray}
\mathcal{R}&=&-\frac{20}{l^2}+\frac{6M}{r^2}, \\
G^{tt}&=&\frac{3Ml^2-6r^2}{r^4-Ml^2r^2},\\
G^{rr}&=&\frac{3M^2}{r^2}+\frac{6r^2}{l^4}-\frac{9M}{l^2},\\
G^{xx}&=&\frac{6r^2-Ml^2}{r^4}.
\end{eqnarray}
pode-se calcular a função $b(r)$. Esta função é dada por
\begin{eqnarray}
b=\frac{r^{-\frac{3}{2}}}{[(1-\alpha(r))(1+\beta(r))]^{1/4}}.
\end{eqnarray}
Para cálculá-la é necessário o conhecimento das funções $\alpha(r)$ e $\beta(r)$ que são obtidas através do tensor de Einstein.

Lembrando que a função $\alpha(r)$ é definida a partir de $\bar{h}^{tt}$, pode-se calculá-la como
\begin{eqnarray}
\bar{h}^{tt}&=&\bar{g}^{tt}-\eta G^{tt}=\bar{g}^{tt}[1-\alpha(r)] \\
&=& \frac{l^2}{r^2-Ml^2}-\eta \frac{3Ml^2-6r^2}{r^2(r^2-Ml^2)} \\
&=& \frac{l^2}{r^2-Ml^2}\left(1-\eta \frac{3Ml^2-6r^2}{r^2l^2}\right)
\end{eqnarray}
de onde vem que
\begin{eqnarray}
\alpha(r)=\eta \frac{3Ml^2-6r^2}{r^2l^2}.
\end{eqnarray}
Agora lembrando da definição da função $\beta(r)$ a partir de $h^{rr}$, e também a calculando, tem-se
\begin{eqnarray}
h^{rr}&=&g^{rr}+\eta G^{rr}=g^{rr}[1+\beta(r)] \\
&=& \frac{r^2-Ml^2}{l^2}+\eta \left(\frac{3M^2}{r^2}+\frac{6r^2}{l^4}-\frac{9M}{l^2}\right) \\
&=& \frac{r^2-Ml^2}{l^2}\left[1+ \frac{\eta l^2}{r^2-Ml^2}\left(\frac{3M^2}{r^2}+\frac{6r^2}{l^4}-\frac{9M}{l^2}\right)\right]
\end{eqnarray}
de onde vem agora que
\begin{eqnarray}
\beta(r)=\frac{\eta l^2}{r^2-Ml^2}\left(\frac{3M^2}{r^2}+\frac{6r^2}{l^4}-\frac{9M}{l^2}\right).
\end{eqnarray}
Usaremos também a função $h^{xx}$, que é dada por
\begin{eqnarray}
h^{xx}=\frac{1}{r^2}\left[l^2+\eta\left(6-\frac{Ml^2}{r^2}\right)\right].
\end{eqnarray}
Como a equação final ficaria muito extensa omitiremos seu formato neste texto. Pode-se estudar aqui o acoplamento potencial ($\xi$) e o acoplamento cinético ($\eta$). O estudo do buraco negro BTZ e AGGH mostrou-nos que quando ocorre somente o acoplamento potencial, com o escalar de Ricci, em geral, apenas faz uma renormalização na massa (BTZ) ou no número de onda e na massa (AGGH). Visto isso, como o caso com o acoplamento cinético parece ser o mais interessante para este estudo, faremos $\xi=0$ e estudaremos apenas o caso quando $\eta$ é não nulo.

\subsubsection{Evolução do Campo Eletromagnético}
A equação diferencial da evolução do campo eletromagnético para este buraco negro é encontrada substituindo as funções $A(r)$ e $B(r)$ nas Eqs.(\ref{Maxwellfinal},\ref{Maxpot}) resultando em
\begin{eqnarray}
\bar{\Psi}^{''}+\omega^2\bar{\Psi}-\bar{V}\bar{\Psi}=0,
\end{eqnarray}
onde o potencial efetivo é dado por
\begin{eqnarray}
\bar{V}=\frac{(Ml^2-r^2)[(4-Ml^2)r^4+r^6+4l^4(X_{1}^2+X_{2}^2+X_{3}^2)]}{4l^4r^2}.
\end{eqnarray}

\subsection{Buraco Negro \emph{6Dz0}}
\subsubsection{Evolução do campo Escalar acoplado ao Escalar de Ricci e ao tensor de Einstein}
Das Eqs.(\ref{altadimensao}) com potencial dado pela Eq.(\ref{potencialaltadimensao}) obtém-se a seguinte equação de perturbação para o buraco negro \emph{6Dz0}
\begin{eqnarray}
\frac{h^{rr}}{\bar{h}^{tt}}\ddot{R}+\left(\frac{3}{r}\frac{h^{rr}}{\bar{h}^{tt}}+\frac{\dot{h}^{rr}}{\bar{h}^{tt}}\right)\dot{R}+\omega^2R-\frac{1}{\bar{h}^{tt}}\left(h^{xx}X^2+m^2+\xi\mathcal{R}\right)R=0.
\end{eqnarray}
Transformando essa equação para o formato de Schrödinger é necessário o termo que subtrai ao potencial, $C_5$, dado pela Eq.(\ref{superc5}) e que fica
\begin{eqnarray}
C_5=\frac{H}{b}\left[\ddot{b}+\dot{b}\left(\frac{\dot{h}^{rr}}{h^{rr}}+\frac{3}{r}\right)\right].
\end{eqnarray}
Isto deixa a equação no formato Schrodinger com potencial dado por
\begin{eqnarray}
\bar{V}=\frac{1}{\bar{h}^{tt}}\left(h^{xx}X^2+m^2+\xi\mathcal{R}\right)-\frac{H}{b}\left[\ddot{b}+\dot{b}\left(\frac{\dot{h}^{rr}}{h^{rr}}+\frac{3}{r}\right)\right].
\end{eqnarray}
Sendo o escalar de Ricci e o tensor de Einstein calculados para este caso dados por
\begin{eqnarray}
\mathcal{R}&=&-\frac{20}{l^2}+\frac{8M}{r^2}, \\
G^{tt}&=&\frac{10r^2-6Ml^2}{Ml^4-l^2r^2},\\
G^{rr}&=&\frac{2M^2}{r^2}+\frac{6r^2}{l^4}-\frac{8M}{l^2},\\
G^{xx}&=&\frac{6r^2-2Ml^2}{r^4}.
\end{eqnarray}
pode-se calcular a função $b(r)$. Esta função é dada por
\begin{eqnarray}
b=\frac{r^{-2}}{[(1-\alpha(r))(1+\beta(r))]^{1/4}}.
\end{eqnarray}
Para cálculá-la é necessário o conhecimento das funções $\alpha(r)$ e $\beta(r)$ que são obtidas através do tensor de Einstein.

Lembrando que a função $\alpha(r)$ é definida da seguinte forma pode-se calculá-la
\begin{eqnarray}
\bar{h}^{tt}&=&\bar{g}^{tt}-\eta G^{tt}=\bar{g}^{tt}[1-\alpha(r)] \\
&=& \frac{r^2}{r^2-Ml^2}-\eta  \frac{10r^2-6Ml^2}{l^2(Ml^2-r^2)} \\
&=& \frac{r^2}{r^2-Ml^2}\left(1- \eta \frac{r^2-Ml^2}{r^2}\frac{10r^2-6Ml^2}{l^2(Ml^2-r^2)}\right) \\
&=& \frac{r^2}{r^2-Ml^2}\left(1- \eta \frac{6Ml^2-10r^2}{l^2r^2}\right)
\end{eqnarray}
de onde vem que
\begin{eqnarray}
\alpha(r)=\eta \frac{6Ml^2-10r^2}{l^2r^2}.
\end{eqnarray}
Agora lembrando da definição da função $\beta(r)$ e também a calculando tem-se
\begin{eqnarray}
h^{rr}&=&g^{rr}+\eta G^{rr}=g^{rr}[1+\beta(r)] \\
&=& \frac{r^2-Ml^2}{l^2}+\eta \left(\frac{2M^2}{r^2}+\frac{6r^2}{l^4}-\frac{8M}{l^2}\right) \\
&=& \frac{r^2-Ml^2}{l^2}\left[1+ \frac{\eta l^2}{r^2-Ml^2}\left(\frac{2M^2}{r^2}+\frac{6r^2}{l^4}-\frac{8M}{l^2}\right)\right]
\end{eqnarray}
de onde vem agora que
\begin{eqnarray}
\beta(r)=\frac{\eta l^2}{r^2-Ml^2}\left(\frac{2M^2}{r^2}+\frac{6r^2}{l^4}-\frac{8M}{l^2}\right).
\end{eqnarray}
Calculando também $h^{xx}$ temos
\begin{eqnarray}
h^{xx}=\frac{1}{r^2}\left[l^2+\eta\left(6-\frac{2Ml^2}{r^2}\right)\right].
\end{eqnarray}
Visto a influência do acoplamento potencial faremos aqui as mesmas considerações que para o buraco negro \emph{5Dz1}, analisando somente a influência do acoplamento cinético.

\subsubsection{Evolução do Campo Eletromagnético}
A equação diferencial da evolução do campo eletromagnético para este caso é dada por
\begin{eqnarray}
\bar{\Psi}^{''}+\omega^2\bar{\Psi}-\bar{V}\bar{\Psi}=0,
\end{eqnarray}
onde o potencial é
\begin{eqnarray}
\bar{V}=\frac{(Ml^2-r^2)[r^4+r^6-Ml^2r^2(r^2-1)+l^4X^2]}{l^2r^4}.
\end{eqnarray}

\chapter{Métodos Numéricos}
Após ser apresentada as equações que governam a evolução de campos escalares e eletromagnéticos na vizinhança dos buracos negros analisados, percebe-se que para alguns casos não é possível a obtenção de soluções exatas. Para isso faz-se necessário o estudo de métodos numéricos, pelos quais as respectivas soluções e os modos quasinormais de oscilação dos campos podem ser obtidos. Os métodos aqui estudados serão o desenvolvido por Horowitz e Hubeny (HH) e o Método de Iteração Assintótica (AIM).

Primeiro irá será exposto o método HH, e em seguida, o método AIM.

\section{Horowitz and Hubeny}
Este método visa obter os modos quasinormais de um buraco negro através da expansão da solução em série de potências sobre o horizonte e imposição da condição de contorno, na qual a solução se anula no infinito. Existem métodos  que só permitem a sua utilização caso as equações para o potencial efetivo tendam a zero ou quando $x$ tende a $-\infty$ e $+\infty$. Tal condição não é satisfeita em espaços-tempos assintoticamente AdS. Visto isso, usa-se o método definido em \cite{horowitzhubeny}, conhecido como Método HH.

A fim de exemplicar o método será utilizado a métrica de Schwarzschild-AdS em termos de coordenadas de Eddington-Finkelstein, $v=t+x$, o que dá
\begin{equation}
ds^2=-f(r)dv^2+2dvdr+r^2d\Omega_2^2.
\end{equation}
Com a separação de variáveis $\Psi(v,r,\theta,\phi)=\frac{R(r)}{r^{(d-2)/2}}e^{-i\omega v}Y_{lm}(\theta,\phi)$ a equação de movimento para $R(r)$ assume a forma
\begin{equation}
\label{hh}
f(r)\frac{d^2R(r)}{dr^2}+[f'(r)-2i\omega]\frac{dR(r)}{dr}-V(r)R(r)=0,
\end{equation}
onde o potencial efetivo $V(r)$, $D$-dimensional, para raio do buraco negro unitário, é dado por
\begin{eqnarray}
V(r)&=&\frac{(D-2)(D-4)}{4r^2}f(r)+\frac{D-2}{2r}f'(r)+\frac{c}{r^2},\\
&=&\frac{D(D-2)}{4}+\frac{(D-2)(D-4)+4c}{4r^2}+\frac{(D-2)^2r_0^{D-3}}{4r^{D-1}},
\end{eqnarray}
onde
\begin{equation}
c=l(l+D-3).
\end{equation}
A fim de mapear toda a região de interesse, $r_+<r<\infty$, em um intervalo de parâmetros finitos, muda-se as variáveis $x=1/r$. Em geral, uma expansão em série de potências terá um raio de convergência pelo menos tão grande como a distância ao polo mais próximo. Reescreve-se, assim, a Eq.(\ref{hh}) como
\begin{equation}
\label{stu}
s(x)\frac{d^2R(x)}{dx^2}+\frac{t(x)}{(x-x_+)}\frac{dR(x)}{dx}+\frac{u(x)}{(x-x_+)^2}R(x)=0,
\end{equation}
com as seguintes definições
\begin{eqnarray}
s(x)&=&\frac{r_0^{D-3}x^{D+1}-x^4-x^2}{x-x_+}, \\
&=&\frac{x_+^2+1}{x_+^{D-1}}x^D+\cdots+\frac{x_+^2+1}{x_+^3}x^4+\frac{x^3}{x_+^2}+\frac{x^2}{x_+},\\
t(x)&=&(D-1)r_0^{D-3}x^d-2x^3-2x^2i\omega, \\
u(x)&=&(x-x_+)V(x).
\end{eqnarray}
O parâmetro $r_0^{D-3}$ deve ser visto como uma função do raio do horizonte, $r_0^{D-3}=(x_+^2+1)/x_+^{D-1}$. Em função de $x_+$ tem-se
\begin{eqnarray}
s(x)&=&\frac{(x_+^2+1)\frac{x^{D+1}}{x_+^{D-1}}-x^4-x^2}{x-x_+}, \\
t(x)&=&(D-1)(x_+^2+1)\frac{x^D}{x_+^{D-1}}-2x^3-2x^2i\omega,\\
u(x)&=&(x-x_+)V(x).
\end{eqnarray}
Os polinômios $s(x)$, $t(x)$ e $u(x)$, expandidos em torno de $x=x_+$, geram uma série finita expressa por
\begin{equation}
s(x)=\sum_{n=0}^{d}s_n(x-x_+)^n,
\end{equation}
e expressões equivalentes para $t_n$ e $u_n$. Isso será útil para notar que
\begin{eqnarray}
s_0&=&2x_+^2\kappa, \\
t_0&=&2x_+^2(\kappa-i\omega), \\
u_0&=&0,
\end{eqnarray}
onde $\kappa$ é a gravidade superficial. Para determinar o comportamento da solução perto do horizonte, primeiro faz-se
\begin{eqnarray}
R(x)=(x-x_+)^\alpha.
\end{eqnarray}
Substituindo na Eq.(\ref{stu}) obtém-se
\begin{eqnarray}
\alpha(\alpha-1)s_0+\alpha t_0&=&2x_+^2\alpha(\alpha\kappa-i\omega)=0,
\end{eqnarray}
com duas soluções, $\alpha=0$ e $\alpha=i\omega /\kappa$. Percebe-se da equação
\begin{eqnarray}
e^{-i\omega(t-r_*)}=e^{-i\omega v}e^{2i\omega r_*}\approx e^{-i\omega v}(r-r_+)^{2i\omega /f^{'}(r_+)},
\end{eqnarray}
que correspondem precisamente aos modos {\it ingoing} e {\it outgoing} perto do horizonte, respectivamente.
Uma vez que querem incluir apenas os modos {\it ingoing}, pegam $\alpha=0$. Isso corresponde a olhar uma solução da forma
\begin{eqnarray}
\label{psian}
R(x)=\sum_{n=0}^{\infty}a_n(x-x_+)^n.
\end{eqnarray}
Substituindo a Eq.(\ref{psian}) em Eq.(\ref{stu}) e igualando coeficientes de $(x-x_+)^n$, para cada $n$, obtém-se a seguinte relação recursiva para $a_n$
\begin{eqnarray}
a_n=-\frac{1}{P_n}\sum_{k=0}^{n-1}[k(k-1)s_{n-k}+kt_{n-k}+u_{n-k}]a_k,
\end{eqnarray}
onde
\begin{eqnarray}
P_n&=&n(n-1)s_0+nt_0, \\
&=&2x_+^2n(n\kappa-i\omega).
\end{eqnarray}
As frequências dos modos quasinormais são calculadas obtendo as raízes da equação $R(x=0)=0$, que é um polinômio de grau infinito em $\omega$. Na prática, truncamos o polinômio em um grau $N$ suficientemente grande e vemos se ao aumentarmos o valor de $N$ a frequência fundamental converge para algum valor. O cálculo das raízes pode ser feito com algum método de busca de soluções, porém, a busca deve levar em conta que as raízes são complexas.

\begin{figure}[H]
\label{convHH}
\centering
{\epsfig{file=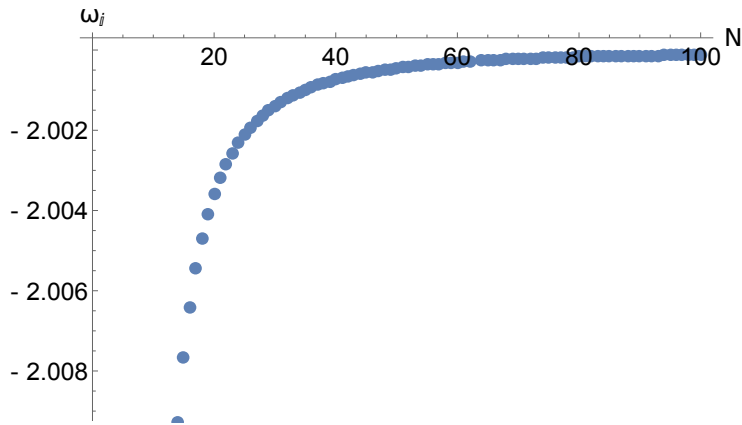, width=0.7\linewidth, height=0.4\linewidth}}
\caption{Gráfico de $\omega_I$ x $N$ para $r_+=1$ com $l=1$ e $\kappa=0$}
\end{figure}

Na Fig.(\ref{convHH}) é exibido um gráfico de $\omega_I\times N$, onde $N$ é o número de iterações. O $N$ escolhido foi 100, mostrando assim explicitamente sua convergência de de $\omega_I$. Esta é uma aplicação do método HH para o buraco negro BTZ no caso onde $\kappa=0$, $r_+=1$ e $l=1$. A parte real da frequência não foi mostrada aqui, pois, é nula.

\section{Método de Iteração Assintótica}
Recentemente um novo método para obter soluções analíticas/numéricas de equações diferenciais ordinárias de segunda ordem com potenciais ligados foi desenvolvido em (AIM)\cite{hakan} e \cite{cho}.

Antes de se usar o método propriamente dito, deve-se realizar algumas mudanças na forma da equação. Como estamos interessados nas frequências quasinormais de oscilação oriundas da evolução de um campo de matéria em uma dada métrica, primeiro evoluímos o campo na métrica desejada. Com este resultado em mãos, a equação de evolução deve ser transformada para ficar no formato da equação de Schrödinger, como mostrado no Apêndice B.

Agora, utiliza-se da coordenada tartaruga $r_*(r)$. Ela deve ser avaliada nos limites $r_*(r\to r_+)$ e $r_*(r\to\infty)$. Estes resultados serão usados na equação de onda escrita como abaixo
\begin{equation}
\Psi=e^{i\omega r_*}e^{-i\omega r_*}\chi.
\end{equation}
Com estes resultados em mãos faz-se uma nova transformação de variáveis de $r$ para $\gamma$
\begin{equation}
\label{transformacao}
\gamma=1-\frac{r_+}{r},
\end{equation}
e assim obteremos uma equação de onda do tipo
\begin{equation}
\Psi=\gamma^{a\omega}(1-\gamma)^{b}\chi,
\end{equation}
onde $a$ e $b$ são constantes que serão determinadas quando forem tomados os limites de $r_*(r\to r_+)$ e $r_*(r\to\infty)$.

A partir deste ponto se realiza o método AIM propriamente dito. Primeiro considera-se a equação diferencial linear de segunda ordem homogênea para a função $\chi(x)$
\begin{eqnarray}
\label{aim}
\chi^{''}=\lambda_0(x)\chi^{'}+s_0(x)\chi,
\end{eqnarray}
onde $\lambda_0$ e $s_0$ são funções em $C_{\infty}(a,b)$. A fim de encontrar uma solução geral para a equação, é analisada a estrutura simétrica no lado direito da Eq.(\ref{aim}) de acordo com \cite{hakan} e \cite{ciftci}. Se a equação for diferenciada com respeito a $x$, será encontrado que
\begin{equation}
\chi^{'''}=\lambda_1(x)\chi^{'}+s_1(x)\chi,
\end{equation}
onde
\begin{eqnarray}
\lambda_1&=&\lambda_0^{'}+s_0+(\lambda_0)^2,\\
s_1&=&s_0^{'}+s_0\lambda_0.
\end{eqnarray}
Tomando a segunda derivada obtém-se
\begin{equation}
\chi^{iv}=\lambda_2(x)\chi^{'}+s_2(x)\chi,
\end{equation}
onde
\begin{eqnarray}
\lambda_2&=&\lambda_1^{'}+s_1+\lambda_0\lambda_1,\\
s_2&=&s_1^{'}+s_0\lambda_1.
\end{eqnarray}
Iterativamente, para $(n+1)$-ésima e $(n+2)$-ésima derivadas, $n=1,2,\cdots$ tem-se
\begin{equation}
\chi^{n+1}=\lambda_{n-1}(x)\chi^{'}+s_{n-1}(x)\chi,
\end{equation}
e, portanto, trazendo para a observação crucial no AIM, de que a diferenciação da equação anterior $n$ vezes em relação a $x$ deixa uma forma simétrica para o lado direito, que é
\begin{equation}
\chi^{n+2}=\lambda_n(x)\chi^{'}+s_n(x)\chi,
\end{equation}
onde
\begin{eqnarray}
\label{lambdan}
\lambda_n(x)&=&\lambda_{n-1}^{'}(x)+s_{n-1}(x)+\lambda_0(x)\lambda_{n-1}(x),\\
\label{sn}
s_n(x)&=&s_{n-1}^{'}(x)+s_0(x)\lambda_{n-1}(x).
\end{eqnarray}
Para $n$ suficientemente grande o aspecto assintótico do método é introduzido, isto é
\begin{equation}
\frac{s_n(x)}{\lambda_n(x)}=\frac{s_{n-1}(x)}{\lambda_{n-1}(x)}\equiv\beta(x),
\end{equation}
onde os MQN são obtidos da condição de quantização
\begin{equation}
\label{quantizacao}
\delta_n=s_n\lambda_{n-1}-s_{n-1}\lambda_n=0,
\end{equation}
que é equivalente a impor um fim ao número de iterações, de acordo com \cite{barakat}. Da razão da $(n+1)$-ésima e $(n+2)$-ésima derivadas, tem-se
\begin{equation}
\frac{d}{dx}\ln(\chi^{n+1})=\frac{\chi^{n+2}}{\chi^{n+1}}=\frac{\lambda_n(\chi^{'}+(s_n/\lambda_n)\chi)}{\lambda_{n-1}(\chi^{'}+(s_{n-1}/\lambda_{n-1}\chi))}.
\end{equation}
Pelo limite assintótico, isso se reduz a
\begin{equation}
\frac{d}{dx}\ln(\chi^{n+1})=\frac{\lambda_n}{\lambda_{n-1}}.
\end{equation}
Uma característica desagradável das relações de recursão das Eqs.(\ref{lambdan},\ref{sn}) é que, a cada iteração, deve-se tomar a derivada dos termos $\lambda$ e $s$ da iteração anterior. Isso pode atrasar a implementação numérica do AIM consideravelmente e também levam a problemas com precisão numérica. Para contornar este problema foi desenvolvida uma versão melhorada do $AIM$, que ignora a necessidade de se tomar derivadas em cada etapa. Isso melhora muito a precisão e velocidade do método. Expande-se $\lambda_n$ e $s_n$ em series de Taylor em torno do ponto $\xi$ em torno do qual o AIM é realizado
\begin{eqnarray}
\lambda_n(\gamma)&=&\sum_{i=0}^{\infty}c_n^i(x-\gamma)^i,\\
s_n(\gamma)&=&\sum_{i=0}^{\infty}d_n^i(x-\gamma)^i,
\end{eqnarray}
onde $c_n^i$ e $d_n^i$ são os iésimos coeficientes de Taylor de $\lambda_n(\gamma)$ e $s_n(\gamma)$, respectivamente. Substituindo essas expressões  nas Eqs.(\ref{lambdan},\ref{sn}) leva a um conjunto de relações de recursão para os coeficientes
\begin{eqnarray}
c_n^i&=&(i+1)c_{n-1}^{i+1}+d_{n-1}^i+\sum_{k=0}^{i}c_0^kc_{n-1}^{i-k},\\
d_n^i&=&(i+1)d_{n-1}^{i+1}+\sum_{k=0}^{i}d_0^kc_{n-1}^{i-k}.
\end{eqnarray}
Em termos desses coeficientes a equação de condição de quantização da Eq.(\ref{quantizacao}) pode ser reexpressa por
\begin{equation}
d_n^0c_{n-1}^0-d_{n-1}^0c_n^0=0.
\end{equation}
Deste modo reduz-se o AIM a um conjunto de relações de recursão que não mais exigem operadores de derivadas.

Um {\it notebook} deste método do programa {\it Mathematica} foi disponibilizado por \cite{cho}, e este foi usado como base para o desenvolvimento de um notebook mais geral, e que abrangesse casos não estudados antes, como AdS.

O método AIM é baseado em 2 parâmetros, como explicitado anteriormente. Existe um valor chamado $h$ que quando aplicado para buracos negros dS representa o ponto mínimo de potencial e $gs$ que é o número de iterações feitas pelo programa. Os buracos negros deste estudo são AdS, o que dificulta a escolha deste valor $h$, então, para isso, foi feita uma varredura de $0\to1$ com passos de $0.01$ para poder cobrir todos os valores e conseguir visualizar algum padrão nos gráficos. Também variou-se o número de iterações $gs$ para verificar em qual região já eram obtidos resultados satisfatórios, visto que, intuitivamente quanto mais iterações mais precisos são os resultados. Todavia, isso dificulta o trabalho computacional, logo, foi procurado um valor mínimo mas que já demonstrasse uma boa precisão.

Afim de confirmar os resultados variou-se o número de onda $\kappa$ ($0 \to 1000$) e também o raio do horizonte de eventos $r_+$ ($0 \to 1000$) para observar o comportamento tanto em buracos negros pequenos quanto grandes (o que geralmente é um problema para outros métodos numéricos).

Foi encontrado que ao variar $h$ é vista sempre uma região de estabilidade em torno do valor exato e logo após esse platô o valor diverge bem rapidamente. E quanto ao $gs$, quanto menor ele for, a região de platô será maior e essa região vai se deslocando para a direita a medida que $gs$ cresce. Com isto é obtido um bom critério para convergência utilizando os gráficos do método AIM.

Nas Figuras da página seguinte serão exibidos gráficos de $\omega_I\times h$  variando $gs$ para o buraco negro BTZ. Foi utilizado aqui o mesmo exemplo que para o método HH a fim de corroborar os resultados ($\kappa=0$, $r_+=1$ e $l=1$). A parte real para este caso é nula e apresentou o mesmo padrão de gráficos mostrado abaixo. Para os buracos negros Scharszchild-AdS, BTZ e AGGH, nos quais foi aplicado este método e serão apresentados no capítulo seguinte, foram observados os mesmo comportamentos.
\newpage
\begin{figure}[H]
\centering
\subfigure[]{\epsfig{file=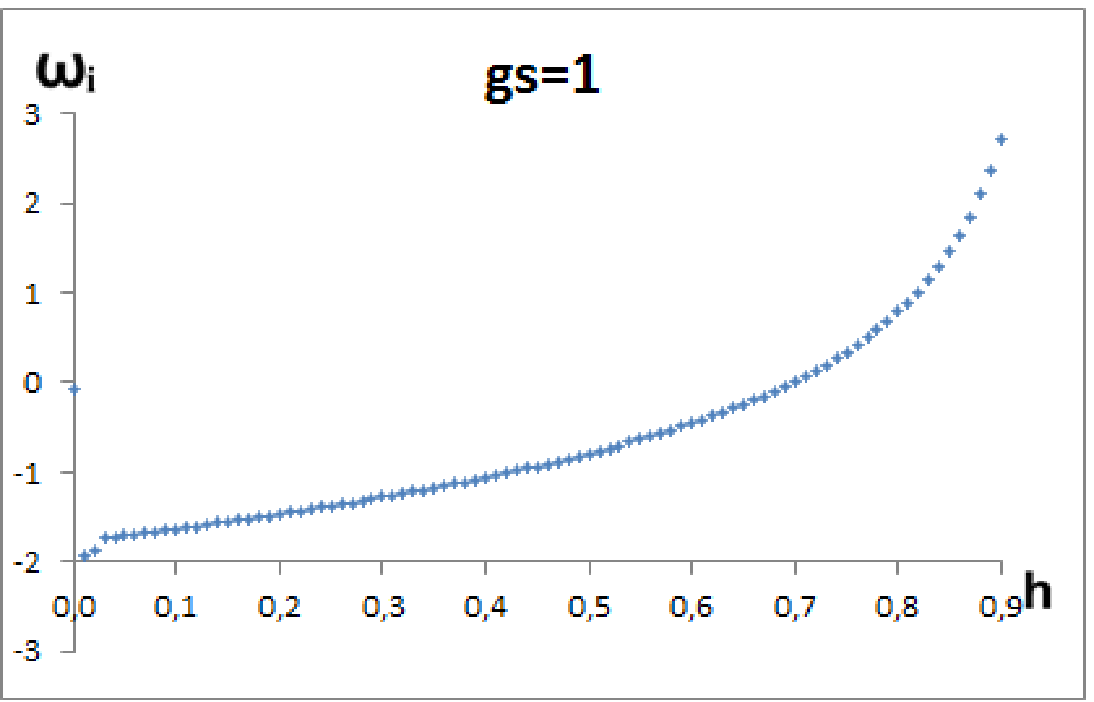, width=0.32\linewidth, height=0.2\linewidth}}
\subfigure[]{\epsfig{file=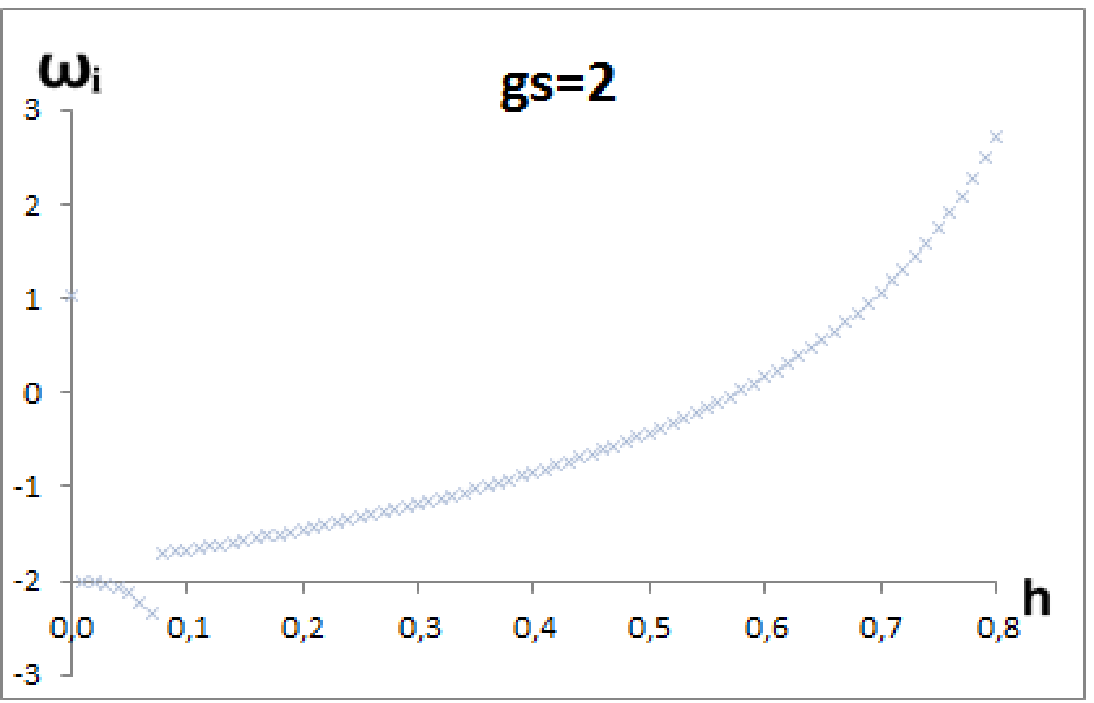, width=0.32\linewidth, height=0.2\linewidth}}
\subfigure[]{\epsfig{file=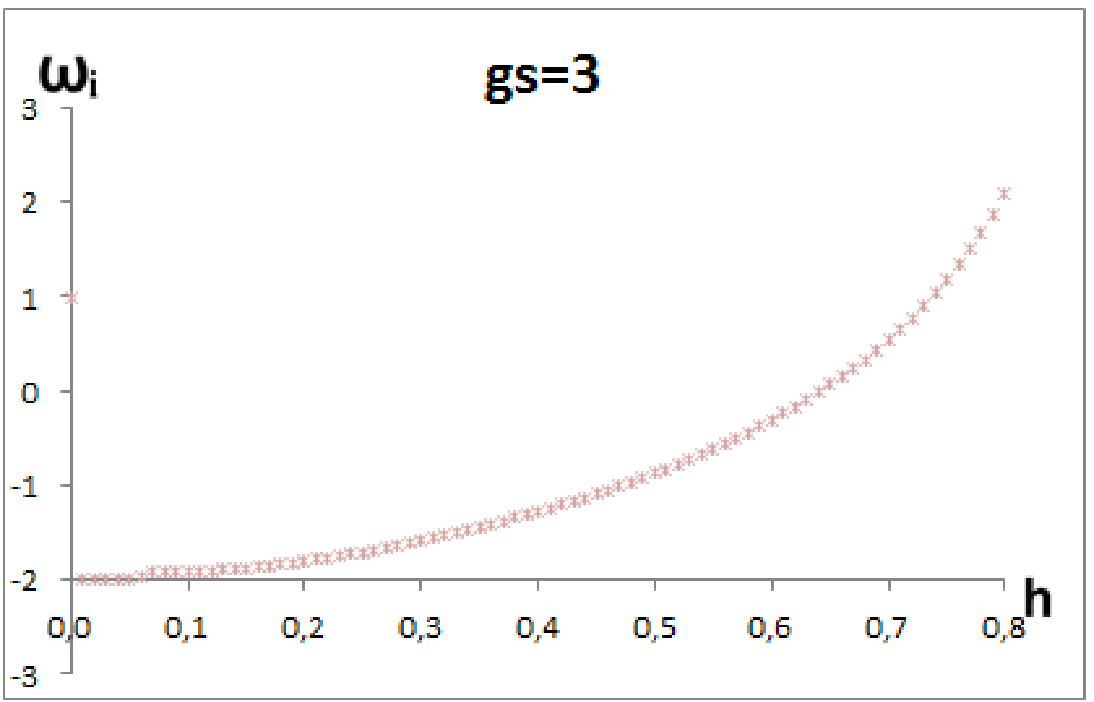, width=0.32\linewidth, height=0.2\linewidth}}
\subfigure[]{\epsfig{file=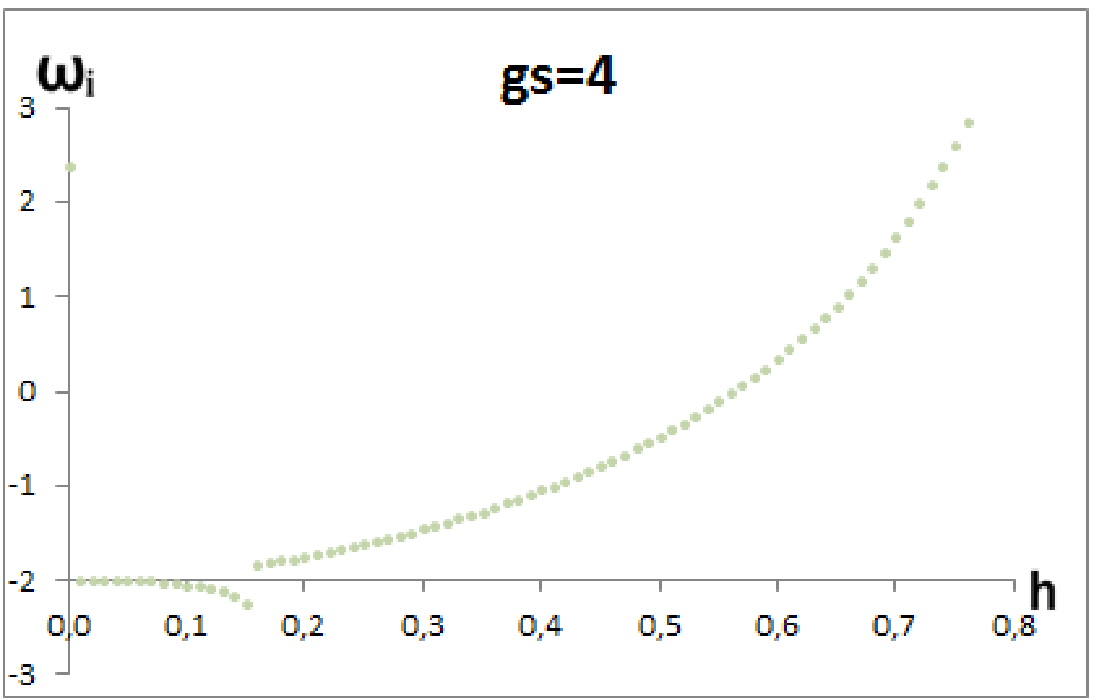, width=0.32\linewidth, height=0.2\linewidth}}
\subfigure[]{\epsfig{file=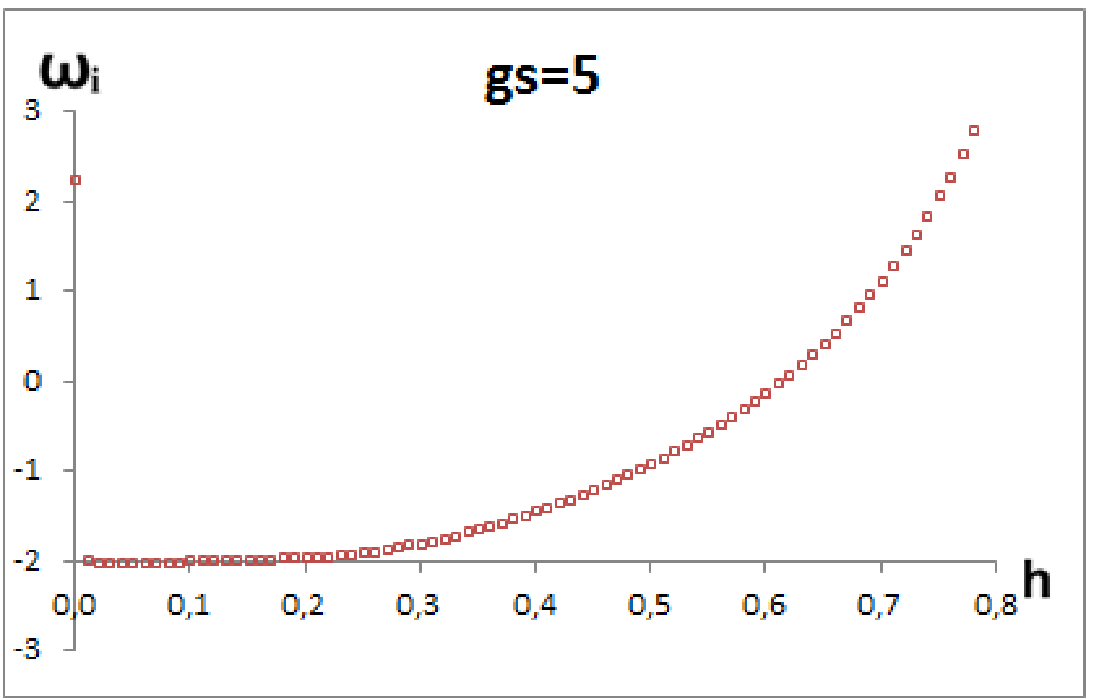, width=0.32\linewidth, height=0.2\linewidth}}
\subfigure[]{\epsfig{file=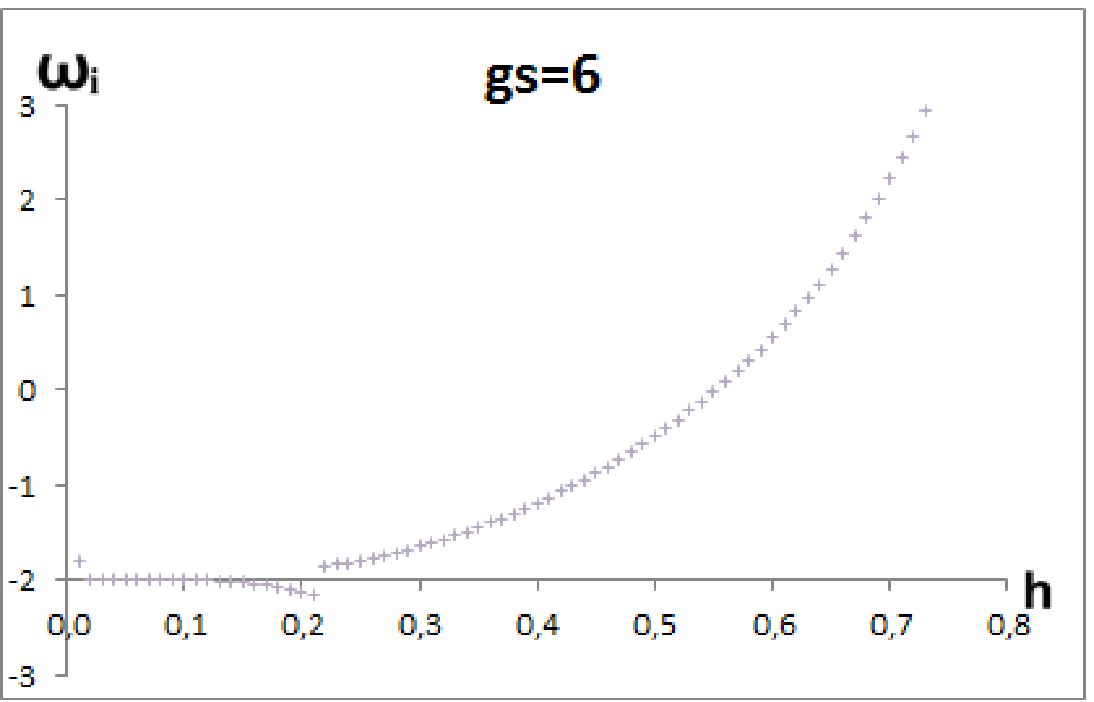, width=0.32\linewidth, height=0.2\linewidth}}
\subfigure[]{\epsfig{file=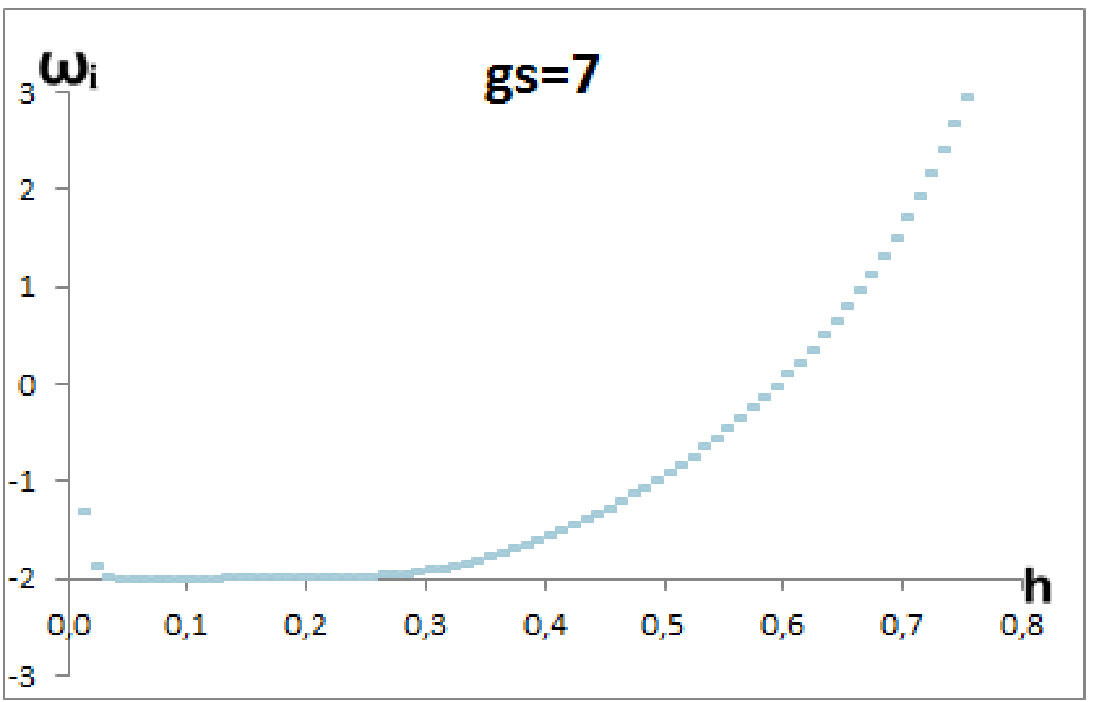, width=0.32\linewidth, height=0.2\linewidth}}
\subfigure[]{\epsfig{file=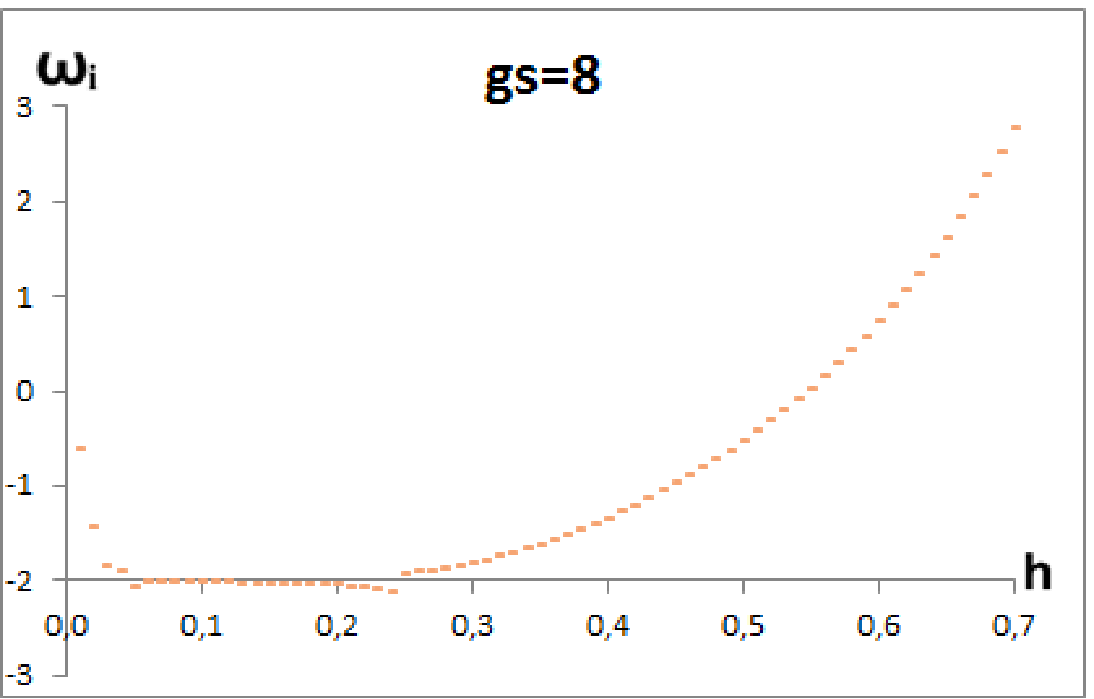, width=0.32\linewidth, height=0.2\linewidth}}
\subfigure[]{\epsfig{file=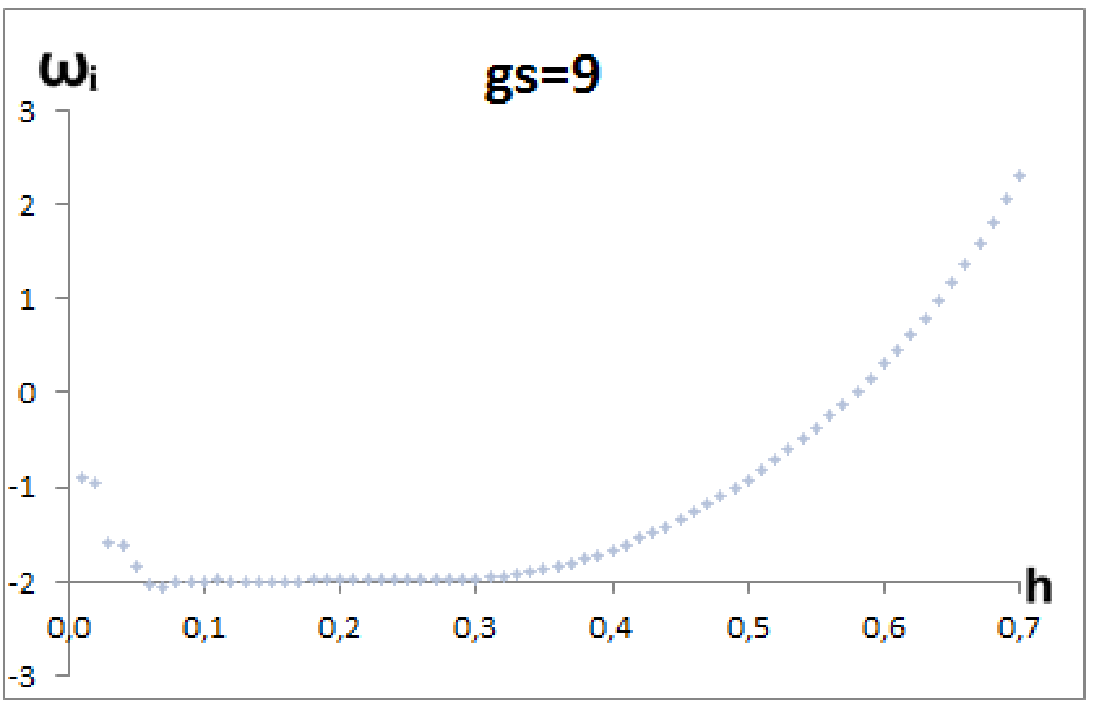, width=0.32\linewidth, height=0.2\linewidth}}
\subfigure[]{\epsfig{file=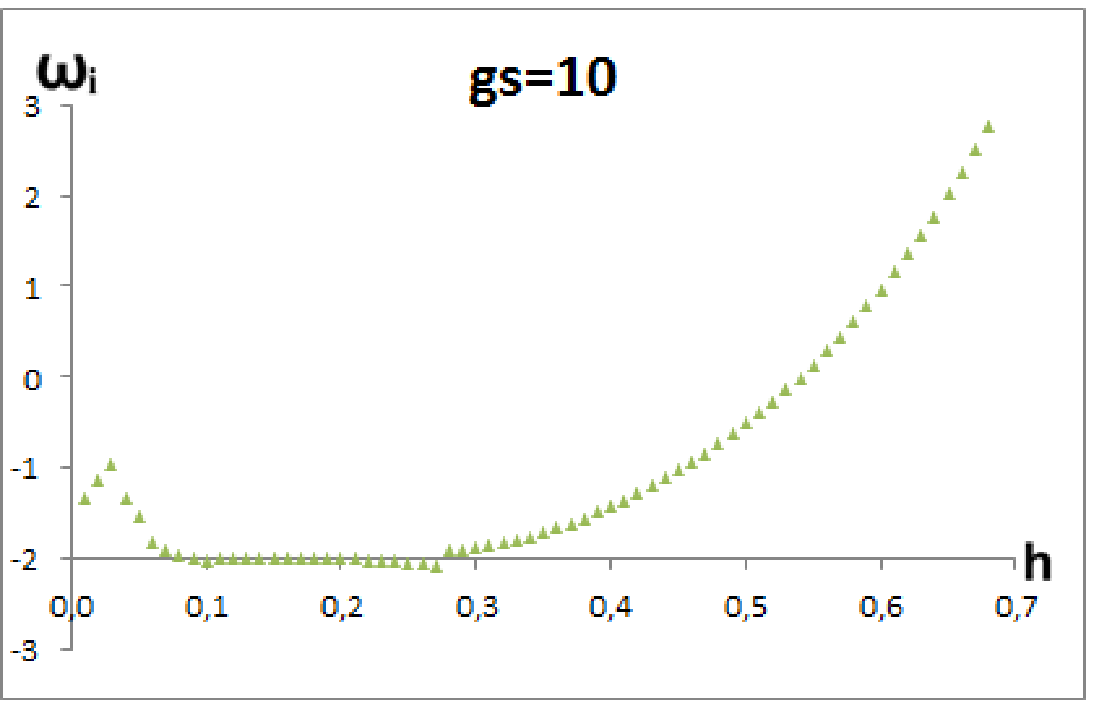, width=0.32\linewidth, height=0.2\linewidth}}
\subfigure[]{\epsfig{file=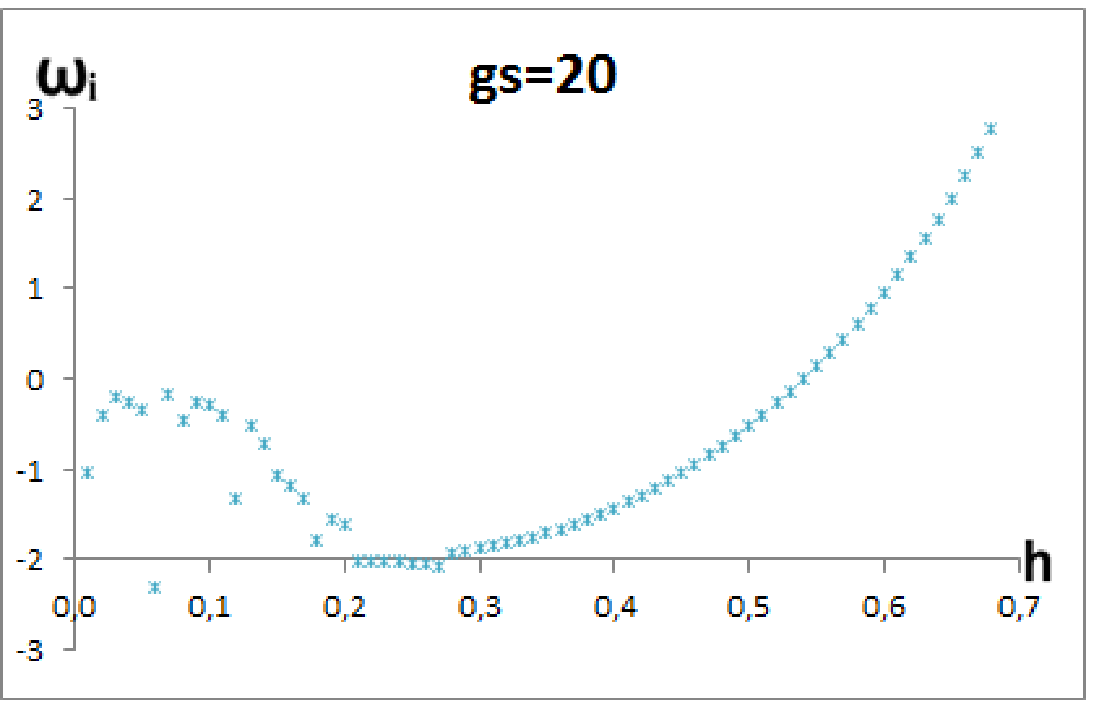, width=0.32\linewidth, height=0.2\linewidth}}
\subfigure[]{\epsfig{file=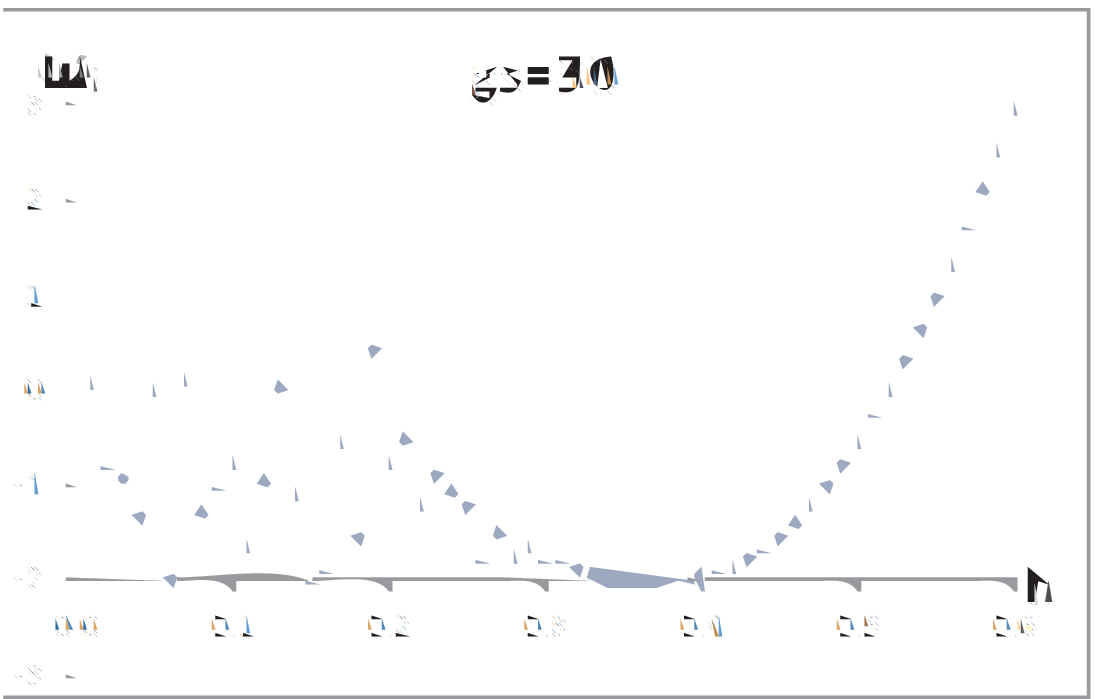, width=0.32\linewidth, height=0.2\linewidth}}
\subfigure[]{\epsfig{file=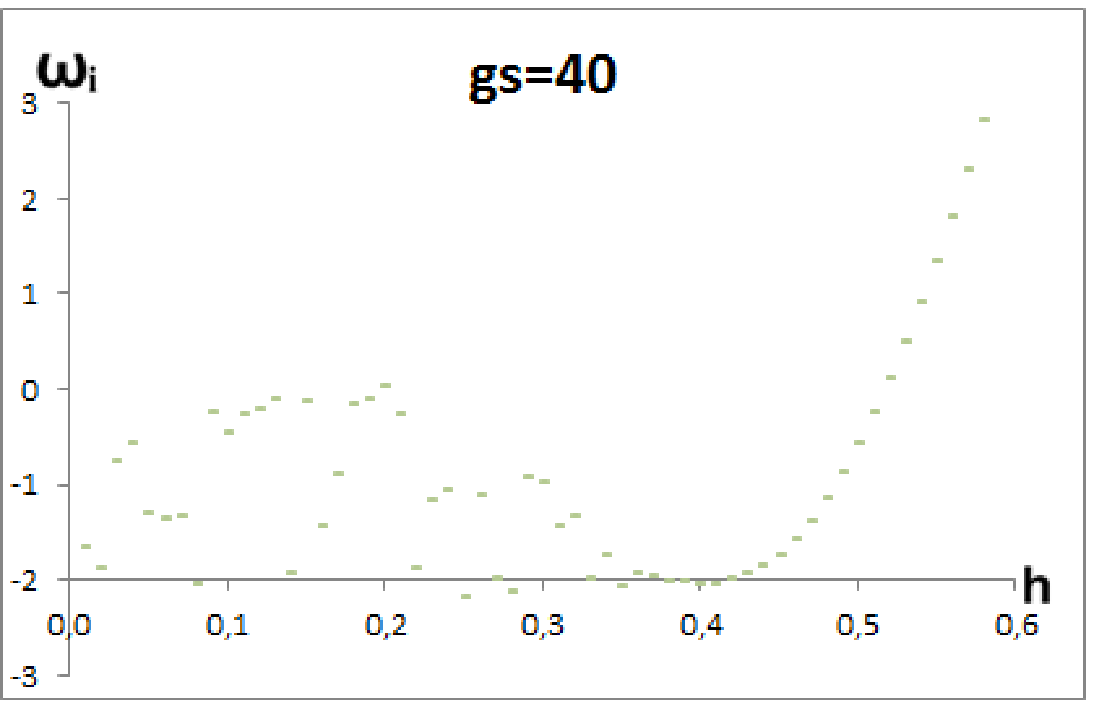, width=0.32\linewidth, height=0.2\linewidth}}
\subfigure[]{\epsfig{file=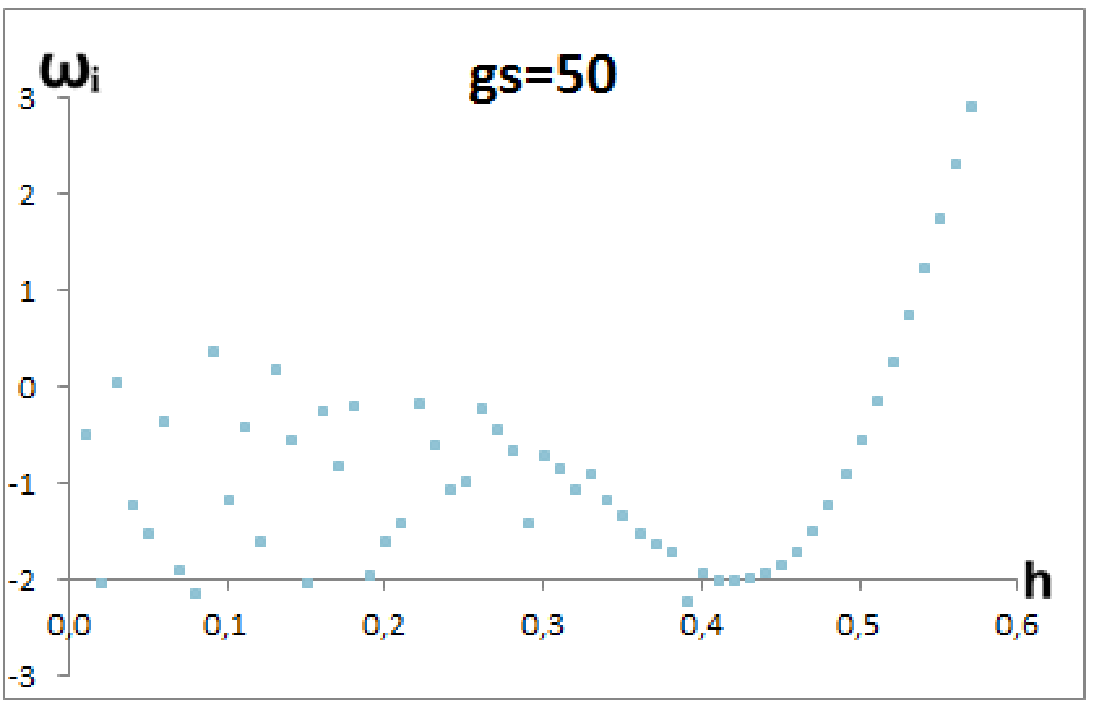, width=0.32\linewidth, height=0.2\linewidth}}
\subfigure[]{\epsfig{file=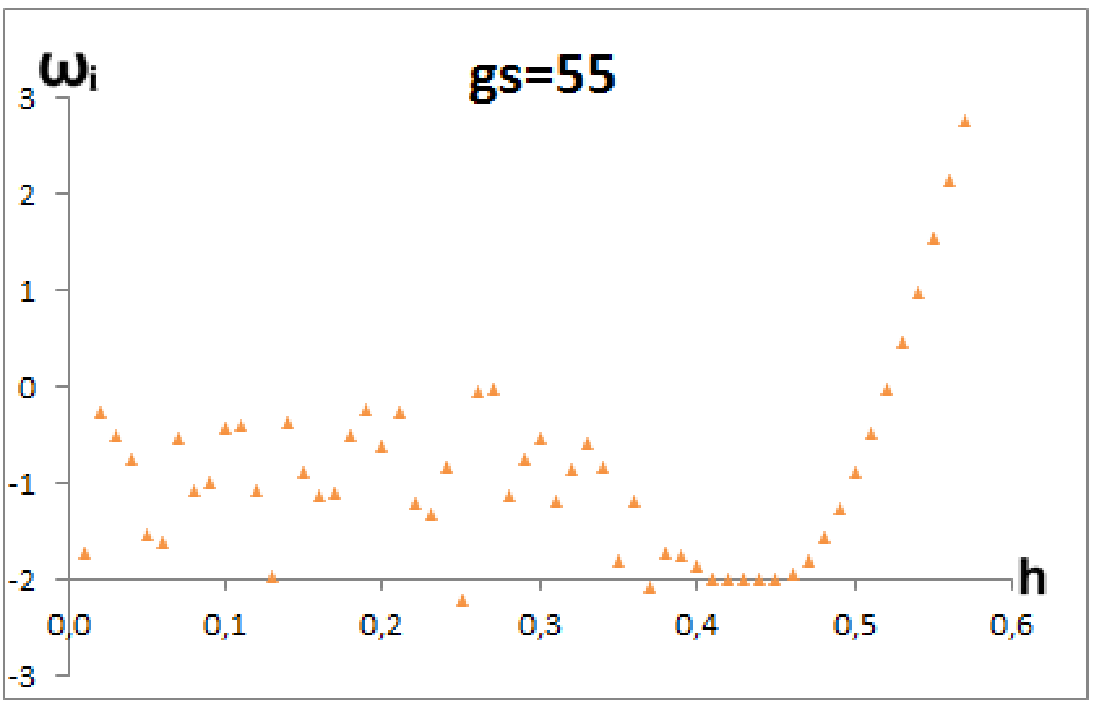, width=0.32\linewidth, height=0.2\linewidth}}
\end{figure}

\chapter{Resultados Numéricos da Evolução do Campo Escalar e Eletromagnético}
Após ser estudada uma teoria de gravitação com correções quadráticas na curvatura de forma geral, escolhida uma família de soluções de buracos negros para análise e apontado alguns casos particulares de grande interesse, evoluído campos de matéria nestes espaços-tempos planarmente simétricos e estudados métodos numéricos para cálculo de frequências quasinormais de oscilações aqui serão exibidos os respectivos resultados numéricos.

Primeiramente será apresentada a aplicação do método AIM para os buracos negros Schwarzschild-AdS, BTZ e AGGH e, por isso, analisada com bastante cuidado a transformação das equações para o formato de Schrodinger e a avaliação nos limites impostos.

Para os buracos negros \emph{5Dz1} e \emph{6Dz0} será aplicado o método HH e feita uma análise da evolução do campo escalar com acoplamento com tensor de Einstein e escalar de Ricci, e a influência de tais acoplamentos, assim como a massa e os números de onda e também a evolução do campo eletromagnético. Em particular para o buraco negro \emph{5Dz1} foi feita também sua análise a luz do método AIM.

\section{Schwarzschild AdS $D$-dimensional}
O buraco negro de Schwarzschild-AdS $D$-dimensional é descrito pela seguinte métrica \cite{horowitzhubeny}
\begin{eqnarray}
ds^2=-\left(1-\frac{r_0^{D-3}}{r^{D-3}}+\frac{r^2}{R^2}\right)dt^2+\left(1-\frac{r_0^{D-3}}{r^{D-3}}+\frac{r^2}{R^2}\right)^{-1}dr^2+r^2d\Omega_{(k,D-2)}^2.\ \ \ \
\end{eqnarray}
onde $R$ é raio AdS e $r_0$ um parâmetro relacionado com a massa $D$-dimensional. Evoluindo-se o campo escalar nesta métrica e reescrevendo no formato de Schrodinger obtém-se
\begin{eqnarray}
\frac{d^2\Psi}{dr_*}+[\omega^2-V(r)]\Psi(r_*)=0,
\end{eqnarray}
onde o potencial é dado por
\begin{eqnarray}
V(r)&=&f(r)\left\{\frac{(D-2)(D-4)+4l(l+D-3)}{4r^2}+\right.\nonumber\\
&+& \left.(1-s^2)\left[\frac{(D-2)^2r_0^{D-3}}{4r^{D-1}}-\frac{D(D-2)-s^2}{12}\Lambda\right]\right\}.
\end{eqnarray}
Aqui tem-se as definições para o raio do horizonte de eventos, constante cosmológica e a função $f(r)$
\begin{eqnarray}
r_0=r_h^{D-3}\sqrt{1+r_h^2} \ \ \ , \ \ \ \Lambda=-\frac{3}{R^2} \ \ \ , \ \ \ f(r)=1-\frac{r_0^{D-3}}{r^{D-3}}+\frac{r^2}{R^2}.
\end{eqnarray}
Serão adotados aqui os seguintes valores a fim de facilitar os cálculos
\begin{eqnarray}
s=0, \ \ \ \ R=1, \ \ \ \ l=0.
\end{eqnarray}
Utilizando aqui das seguintes condições de contorno para a equação de onda
\begin{eqnarray}
\Psi(r_*) &\to&0 \ \ \ \ \ \ \ \ \textrm{quando}\ r_*\to\infty,  \\
\Psi(r_*) &\to& e^{-i\omega r_*} \ \ \ \textrm{quando}\ r_*\to-\infty,
\end{eqnarray}
Transformando agora para a coordenada tartaruga $r_*(r)$, lembra-se que
\begin{eqnarray}
dr_*&=&\frac{dr}{f(r)}, \\
&=&\frac{r^{r-3}dr}{r^{D-3}-r_0^{D-3}+\frac{r^{D-1}}{R^2}},
\end{eqnarray}
onde será escrita a coordenada tartaruga em função de uma função $g(r)$ da maneira como se segue
\begin{eqnarray}
dr_*=\frac{dr}{g(r)(r-r_+)}.
\end{eqnarray}
Tal função $g(r)$ é obtida através de uma expansão em torno de $r_+$ com auxílio do programa Mathematica, que resulta
\begin{eqnarray}
g(r_+)=\frac{r_+}{(D-1)r_+^2+D-3}.
\end{eqnarray}
Logo, a coordenada tartaruga fica aproximadamente
\begin{eqnarray}
r_* \approx \frac{r_+\ln(r-r_+)}{(D-1)r_+^2+D-3} \ \ \ \ \textrm{quando}\ r\to r_+.
\end{eqnarray}
Substituindo-se no campo $\Psi$ tem-se
\begin{eqnarray}
\Psi &\approx & \exp\left(\pm i\omega \frac{r_+\ln(r-r_+)}{(D-1)r_+^2+D-3}\right), \\
&\approx & (r-r_+)^{\pm i\omega \frac{r_+}{(D-1)r_+^2+D-3}}, \\
&=& r\left(1-\frac{r_+}{r}\right)^{\pm i\omega \frac{r_+}{(D-1)r_+^2+D-3}}, \\
&\approx & \left(1-\frac{r_+}{r}\right)^{\pm i\omega \frac{r_+}{(D-1)r_+^2+D-3}}.
\end{eqnarray}
Agora será avaliada a função nos limites da coordenada tartaruga em que $r_*(r\to r_+)$ e $r_*(r\to\infty)$. Primeiramente fazendo o limite quando $r_*(r\to r_+)$ a equação de onda neste limite é
\begin{eqnarray}
\Psi \approx \exp\left(\pm i\omega \frac{r_+}{(D-1)r_+^2+D-3}\right).
\end{eqnarray}
Agora fazendo no limite quando $r_*(r\to\infty)$
\begin{eqnarray}
r_* &=& \int \frac{dr}{[(D-1)r^2+D-3](r-r_+)}\approx \int \frac{rdr}{r^2(D-1)r}, \\
&=& \int \frac{dr}{r^2(D-1)} =-\frac{1}{r(D-1)}.
\end{eqnarray}
Assim a função de onda avaliada no infinito dá
\begin{eqnarray}
\Psi^{''}&+&\omega^2\Psi-V(r\to\infty)\Psi=0, \\
\Psi^{''}&-&\frac{r^2D(D-2)}{4}\Psi=0.
\end{eqnarray}
Supondo uma solução em potências da forma $\Psi= r_*^\alpha$ e substituindo essa solução na função de onda obtém-se a seguinte solução para $\alpha$
\begin{eqnarray}
\alpha=\frac{1\pm\sqrt{1+\frac{D(D-2)}{(D-1)^2}}}{2}.
\end{eqnarray}
Desta maneira a função de onda neste limite fica
\begin{eqnarray}
\Psi=\left(\frac{r_+}{r}\right)^{\frac{1}{2}\pm \frac{\sqrt{1+\frac{D(D-2)}{(D-1)^2}}}{2}}.
\end{eqnarray}
Utilizando a transformação da Eq.(\ref{transformacao}) obtém-se a função de onda completa a fim de aplicar o método AIM
\begin{eqnarray}
\Psi=\gamma^{\pm i\omega \frac{r_+}{(D-1)r_+^2+D-3}}(1-\gamma)^{\frac{1}{2}+ \frac{\sqrt{1+\frac{D(D-2)}{(D-1)^2}}}{2}}\chi.
\end{eqnarray}
Agora será utilizado o método AIM propriamente dito para análise dos MQN e confirmar os resultados obtidos em \cite{horowitzhubeny}. Esses resultados são apresentados na Tabela (\ref{tabqnmschads}).
\begin{table}[H]
\centering
\caption{MQN fundamental $(n=0)$ para Schwarzschild-AdS $D$-dimensional usando o método AIM}
\vspace{0.5cm}
\begin{tabular}{|c|c|c|c|c|c|c|}\hline
       & \multicolumn{2}{c|}{$4D$} & \multicolumn{2}{c|}{$5D$} & \multicolumn{2}{c|}{$7D$} \\ \hline
$r_+$  & $\omega_i$ & $\omega_r$ & $\omega_i$ & $\omega_r$ & $\omega_i$ & $\omega_r$ \\ \hline
$100$  & -266.3856 & 184.9534 & -274.6655 & 311.9624 & -261.2 & 500.8 \\
$50$   &    -      &    -     & -137.3296 & 156.0077 &   -    &  -   \\
$5$    &     -     &   -      & -13.6914  & 15.9454  &  -     & -     \\
$1$    & -2.6712   & 2.7982   & -2.5547   & 4.5788   & -2.16  & 7.27\\
$0.8$  &      -    &  -       &   -       &   -      & -1.52  & 6.62\\
$0.6$  &       -   & -        &    -      &  -       & -1.15  & 6.58\\
$0.4$  & -1.0064   & 2.36     & -0.7462   & 3.7174   & -0.99  & 6.21\\ \hline
\end{tabular}
\label{tabqnmschads}
\end{table}
Para $4D$ foi rodado o programa para um buraco negro pequeno ($r+=0.4)$, um unitário ($r_+=1$) e para um buraco negro grande ($r_+=100$) a fim de confirmar os dados para diferentes regiões de $r+$ e isso foi obtido. O mesmo foi feito para $5D$, porém agora também avaliando com $r+=50$ e $r+=5$ para realmente confirmar que o método se comportava bem e isso foi observado.

Chegando no buraco negro $7D$ foi feita a análise para um buraco grande e o resultado obtido foi correto. O desafio era a análise deste buraco negro para valores de $r+$ menores que 1, o que não era encontrado já que nesta região o HH não funciona bem. Todavia, usando-se do AIM foram obtidos resultados com boa precisão e convergência.

\section{BTZ}
O buraco negro BTZ é descrito pela seguinte métrica \cite{btz}
\begin{eqnarray}
ds^2=-\frac{r^{2}}{l^{2}}\left(1-\frac{Ml^2}{r^2}\right)dt^2+\frac{l^2}{r^2}\left(1-\frac{Ml^2}{r^2}\right)^{-1}dr^2+r^2d\theta^2.
\end{eqnarray}
Evoluindo-se o campo escalar com acoplamento com Tensor de Einstein e Escalar de Ricci e transformando-se para formato Schrodinger obtém-se
\begin{eqnarray}
\bar{R}^{''}+\omega^2\bar{R}-\bar{V}\bar{R}=0,
\end{eqnarray}
onde o potencial é dado por
\begin{eqnarray}
\bar{V}=\left[\frac{m^2}{l^2+\eta}+\frac{3}{4l^4}-\frac{6\xi}{l^2(l^2+\eta)}\right]r^2&+&\left[-\frac{M}{2l^2}+\frac{m^2Ml^2}{l^2+\eta}+\frac{\kappa^2}{l^2+\eta}\left(1+\frac{\eta}{l^2}\right)-\frac{6\xi M}{l^2+\eta}\right]+\nonumber\\
&+&\left[-\frac{M^2}{4}-\frac{\kappa^2Ml^2}{l^2+\eta}\left(1+\frac{\eta}{l^2}\right)\right]r^{-2},
\end{eqnarray}
e a função $f(r)$
\begin{eqnarray}
f(r) \ = \ \frac{r^{2}}{l^{2}}-\frac{r_+^2}{l^2} \ = \ \frac{r^2-r_+^2}{l^2}.
\end{eqnarray}
Transformando para coordenada tartaruga $r_*(r)$ tem-se
\begin{eqnarray}
dr_* &=&\frac{l^2}{r^2-r_+^2}dr  \to  r_*=\int \frac{l^2}{r^2-r_+^2}dr,\\
r_* &\approx & \int \frac{l^2dr}{(r-r_+)(r+r_+)} \approx \int \frac{l^2dr}{2r_+(r-r_+)} =\frac{l^2}{2r_+}\ln(r-r_+).
\end{eqnarray}
Agora será avaliada nos limites onde $r_*(r\to r_+)$ e $r_*(r\to\infty)$. Primeiramente em $r_*(r\to r_+)$ obtém-se
\begin{eqnarray}
r_*\approx \frac{l^2}{2r_+}\ln(r-r_+),
\end{eqnarray}
e a função de onda fica aproximadamente
\begin{eqnarray}
\bar{R} &\approx & \exp\left(\pm i\omega \frac{l^2}{2r_+}\right)\ln(r-r_+) \ \approx \ (r-r_+)^{\pm i\omega l^2/2r_+}, \\
&=& r\left(1-\frac{r_+}{r}\right)^{\pm i\omega l^2/2r_+} \ \approx \ \left(1-\frac{r_+}{r}\right)^{\pm i\omega l^2/2r_+}.
\end{eqnarray}
No limite $r_*(r\to\infty)$ terá
\begin{eqnarray}
r_*=\int \frac{l^2}{r^2-r_+^2}dr \ \approx \  \int \frac{l^2}{r^2}dr \ = \ -\frac{l^2}{r}.
\end{eqnarray}
Resolvendo a equação de movimento no infinito é obtido
\begin{eqnarray}
\bar{R}^{''}&+&\omega^2\bar{R}-V(r\to\infty)\bar{R}=0, \\
\bar{R}^{''}&-&\left[\frac{m^2}{l^2+\eta}+\frac{3}{4l^4}-\frac{6\xi}{l^2(l^2+\eta)}\right]r^2\bar{R}=0.
\end{eqnarray}
Supõe-se aqui a seguinte solução para $\bar{R}$, $\bar{R}=r_*^\alpha$. Substituindo-a na equação de movimento nos dá a seguinte solução para $\alpha$
\begin{eqnarray}
\alpha=\frac{1\pm\sqrt{1+4l^4\left[\frac{m^2}{l^2+\eta}+\frac{3}{4l^4}-\frac{6\xi}{l^2(l^2+\eta)}\right]}}{2}.
\end{eqnarray}
No caso onde $m=0$, $l=1$, $\xi=0$, $\eta=0$ tem-se
\begin{eqnarray}
\alpha_1=3/2, \ \ \ \ \ \ \  \alpha_2=-1/2.
\end{eqnarray}

Agora finalmente fazendo a transformação da Eq.(\ref{transformacao}) a função de onda obtida é
\begin{equation}
\Psi=\gamma^{-i\omega l^2/2r_+}(1-\gamma)^{3/2}\chi.
\end{equation}

Na Tabela(\ref{tabqnmbtz}) são apresentados os resultados do método AIM para o buraco negro BTZ quando $\eta=0$, $\xi=0$, $m=0$ e $l=1$, tentando recuperar os resultados obtidos em \cite{cardoso} onde $\omega=\kappa-i2r_{+}(n+1)$. Os casos onde aparece ``nc", significam que o AIM não convergiu satisfatoriamente.
\begin{table}[H]
\centering
\caption{MQN fundamental $(n=0)$ para BTZ usando AIM}
\vspace{0.5cm}
\begin{tabular}{|c|c|c|c|c|c|c|c|c|}\hline
       & $\kappa=0$ & $\kappa=1$ & $\kappa=5$ & $\kappa=10$ & $\kappa=50$ & $\kappa=100$ & $\kappa=1000$ & \\ \hline
$r_+$  & \multicolumn{7}{c|}{$\omega_r$}    & $\omega_i$ \\ \hline
$0.5$  & 0  & 1 & 5 & 10 & nc & nc  & nc   & -1    \\
$1$    & 0  & 1 & 5 & 10 & 50 & 100 & nc   & -2    \\
$5$    & 0  & 1 & 5 & 10 & 50 & 100 & nc   & -10   \\
$10$   & 0  & 1 & 5 & 10 & 50 & 100 & 1000 & -20      \\
$50$   & 0  & nc& 5 & 10 & 50 & 100 & 1000 & -100    \\
$100$  & 0  & nc& 5 & 10 & 50 & 100 & 1000 & -200    \\
$1000$ & 0  & nc& nc& nc & nc & nc  & nc   & -2000 \\ \hline
\end{tabular}
\label{tabqnmbtz}
\end{table}
Pode-se ver que a medida que se aumentam os valores de $\kappa$ e $r_+$, buracos negros pequenos e grandes não se obtém um resultado satisfatório. Todavia, para buracos negros e $\kappa$ pequenos o programa se comporta muito bem, o que é muito interessantes pois, esta região é exatamente onde o HH não se comporta muito bem.

\section{AGGH}
O buraco negro AGGH é descrito pela seguinte métrica \cite{aggh}
\begin{eqnarray}
ds^2=-\frac{r^{6}}{l^{6}}\left(1-\frac{Ml^2}{r^2}\right)dt^2+\frac{l^2}{r^2}\left(1-\frac{Ml^2}{r^2}\right)^{-1}dr^2+r^2d\theta^2.
\end{eqnarray}
Evoluindo-se o campo escalar com acoplamento com tensor de Einstein e Escalar de Ricci e transformando-se para formato Schrodinger obtém-se
\begin{eqnarray}
\bar{R}^{''}+\omega^2\bar{R}-\bar{V}\bar{R}=0,
\end{eqnarray}
onde o potencial é dado por
\begin{eqnarray}
\bar{V}&=&\left[\frac{7}{4l^8}+\frac{m^2}{l^4(l^2+\eta)}-\frac{26\xi}{l^6(l^2+\eta)}\right]r^6+\nonumber\\
&+&\left[-\frac{5M}{2l^6}-\frac{m^2M}{l^2(l^2+\eta)}+\frac{\kappa^2}{l^4(l^2+\eta)}\left(1+\frac{9\eta}{l^2}\right)+\frac{34\xi M}{l^4(l^2+\eta)}\right]r^4+\nonumber\\
&+&\left[\frac{3M^2}{4l^4}-\frac{\kappa^2M}{l^2(l^2+\eta)}\left(1+\frac{11\eta}{l^2}\right)-\frac{8\xi M^2}{l^2(l^2+\eta)}\right]r^2+\frac{2\eta\kappa^2M^2}{l^2(l^2+\eta)}.\nonumber\\
\end{eqnarray}
Neste caso, diferentemente do buraco negro BTZ, as funções $g^{tt}$ e $g^{rr}$ não guardam entre si a relação onde uma é o inverso da outra. Visto isso, a função $f(r)$ deve ser analisada de forma mais cautelosa. Aqui, avalia-se o caso geral, onde a função é chamada $h(r)$ e definida por
\begin{eqnarray}
h(r) \ = \ \sqrt{\frac{A(r)}{B(r)}} \ = \ \sqrt{\frac{\tilde{g}^{tt}}{g^{rr}}},
\end{eqnarray}
onde as funções $A(r)$ e $B(r)$ são dados por
\begin{eqnarray}
A(r)&=&\frac{r^6}{l^6}\left(1-\frac{r_+^2}{r^2}\right), \\
B(r)&=&\frac{l^2}{r^2}\left(1-\frac{r_+^2}{r^2}\right)^{-1}.
\end{eqnarray}
As dadas funções guardam a seguinte relação entre si
\begin{eqnarray}
A(r)= \frac{r^4}{l^4}\frac{1}{B(r)}.
\end{eqnarray}
Calculando-se a coordenada tartaruga $r_*(r)$ em função de $h(r)$ tem-se
\begin{eqnarray}
r_* &=&\int \frac{dr}{h(r)} = \int \sqrt{\frac{B}{A}}dr =\int \frac{l^2}{r^2} \frac{l^2}{r^2-r_+^2}dr,\\
&=& \int \frac{l^2}{r^2}\frac{l^2}{(r-r_+)(r+r_+)}dr.
\end{eqnarray}
Através do programa {\it Mathematica} obtém-se a solução para a coordenada tartaruga, que confere com a equação de número 9 de \cite{cuadrosoliveirapellicer}, que é
\begin{eqnarray}
r_*=l^4\left(\frac{1}{rr_+^2}-\frac{\arctan(r/r_+)}{r_+^3}\right).
\end{eqnarray}
Agora será avaliada a função nos limites quando $r_*(r\to r_+)$ e $r_*(r\to\infty)$. Primeiramente no limite $r_*(r\to r_+)$ tem-se aproximadamente
\begin{eqnarray}
r_*&=&\int \frac{l^2}{r^2}\frac{l^2}{r^2-r_+^2}dr, \\
&\approx & \int \frac{l^2}{r^2}\frac{l^2}{(r-r_+)(r+r_+)}dr, \\
&\approx &\int\frac{l^4}{r^2}\frac{1}{2r_+(r-r_+)}dr,\\
&\approx & \frac{l^4}{2r_+^3(r-r_+)}dr, \\
&=&\frac{l^4}{2r_+^3}\ln(r-r_+).
\end{eqnarray}
Deste modo a função de onda neste limite fica aproximadamente
\begin{eqnarray}
\bar{R} &\approx &\exp\left(\pm i\omega\frac{l^4}{2r_+^3}\ln(r-r_+)\right),\\
&\approx &(r-r_+)^{\pm i\omega l^4/2r_+^3},\\
&=&r\left(1-\frac{r_+}{r}\right)^{\pm i\omega l^4/2r_+^3},\\
&\approx & \left(1-\frac{r_+}{r}\right)^{\pm i\omega l^4/2r_+^3}.
\end{eqnarray}
Agora no limite $r_*(r\to\infty)$ tem-se
\begin{eqnarray}
r_* &\approx &\int \frac{l^2}{r^2}\frac{l^2}{r^2}dr, \\
&=& -\frac{l^4}{3r^3},
\end{eqnarray}
e aqui função de onda, no infinito, é
\begin{eqnarray}
\bar{R}^{''}&+&\omega^2\bar{R}-V(r\to\infty)\bar{R}=0, \\
\bar{R}^{''}&-&\left[\frac{7}{4l^8}+\frac{m^2}{l^4(l^2+\eta)}-\frac{26\xi}{l^6(l^2+\eta)}\right]r^6\bar{R}=0.
\end{eqnarray}
Supondo uma solução em potência para $\bar{R}$ do tipo $\bar{R}=r_*^\alpha$ e substituindo na equação de movimento, a solução obtida para $\alpha$ é
\begin{eqnarray}
\alpha=\frac{1\pm\sqrt{1+\frac{4l^8}{9}\left[\frac{7}{4l^8}+\frac{m^2}{l^4(l^2+\eta)}-\frac{26\xi}{l^6(l^2+\eta)}\right]}}{2}.
\end{eqnarray}
No caso quando $m=0$, $l=1$ $\xi=0$, $\eta=0$ tem-se
\begin{eqnarray}
\alpha_1=7/6, \ \ \ \ \ \ \alpha_2=-1/6.
\end{eqnarray}
Agora finalmente fazendo a transformação da Eq.(\ref{transformacao})
\begin{equation}
\Psi=\gamma^{-i\omega l^4/2r_+^3}(1-\gamma)^{7/6}\chi.
\end{equation}
Na Tabela(\ref{tabqnmaggh}) são apresentados os valores de $\omega_i$, variando-se $r_+$ e $\kappa$, para $l=1$, $\eta=0$, $\xi=0$ e $m=0$ a fim de corroborar o resultado exato obtido por \cite{BerthaJef}, \cite{Chilenos} e \cite{Coreanos}.
\begin{table}[H]
\centering
\caption{MQN fundamental para AGGH usando AIM}
\vspace{0.5cm}
\begin{tabular}{|c|c|c|c|c|c|c|c|}\hline
                & $\kappa=0$ & 1          & 5          & 10         & 100        & 1000 \\ \hline
$r+=0.1$        & -0.0012    & -0.0099    & -0.0651    & -0.1356    & -1.4082    & -14.1361   \\
$0.2$           & -0.0097    & -0.0328    & -0.2407    & -0.5206    & -5.6091    & -56.5206    \\
$0.5$           & -0.1514    & -0.2182    & -1.2343    & -2.8986    & -34.6168   & -352.8050  \\
$0.9$           & -0.8829    & -1.0062    & -3.4003    & -8.2298    & -110.298   & -1141.1500   \\
$1$             & -1.2111    & -1.3485    & -4.0995    & -9.8745    & -135.6051  & -1408.2319    \\
$10$            & -1211.1026 & -1212.4892 & -1245.6884 & -1348.4692 & -9874.5079 & -135605.0847     \\ \hline
\end{tabular}
\label{tabqnmaggh}
\end{table}
A concordância entre os resultados numéricos e exatos se mostrou satisfatória.

\section{\emph{Buraco Negro 5Dz1}}
O buraco negro \emph{5Dz1} é descrito pela seguinte métrica \cite{lifshitzhigher}
\begin{eqnarray}
ds^2=-\frac{r^{2}}{l^{2}}\left(1-\frac{Ml^2}{r^2}\right)dt^2+\frac{l^2}{r^2}\left(1-\frac{Ml^2}{r^2}\right)^{-1}dr^2+\frac{r^2}{l^2}(dx^2+dy^2+dz^2).\ \ \ \ \
\end{eqnarray}
Evoluindo-se o campo escalar com acoplamento com Tensor de Einstein e Escalar de Ricci e transformando-se para formato Schrodinger obtém-se
\begin{eqnarray}
\bar{R}^{''}+\omega^2\bar{R}-\bar{V}\bar{R}=0,
\end{eqnarray}
onde o potencial é dado por
\begin{eqnarray}
\bar{V}&=&\left\{r^6\left[4l^6m^2+540\eta^2+l^4(15+24m^2\eta-80\xi)-60l^2\eta(-3+8\xi)\right]+\right.\nonumber\\
&+&r^4\left[-4l^8m^2M-1044l^2M\eta^2+4l^8\kappa^2+48l^6\eta\kappa^2-l^6M(15+24m^2\eta-80\xi)-\right.\nonumber\\
&&\left.-3l^6M(1+4m^2\eta-8\xi)+60l^4M\eta(-3+8\xi)+6l^4\eta(-17M+24\eta\kappa^2+64M\xi)\right]+\nonumber\\
&+&r^2\left[639l^4M^2\eta^2-4l^{10}M\kappa^2-64l^8M\eta\kappa^2-96l^6M\eta^2\kappa^2+3l^8M^2(1+4m^2\eta-8\xi)+\right.\nonumber\\
&&\left.+18l^6M^2\eta(1-4\xi)-6l^6M\eta(-17M+24\eta\kappa^2+64M\xi)\right]+\nonumber\\
&+&r^0\left[-126l^6M^3\eta^2+16l^{10}M^2\eta\kappa^2+108l^8M^2\eta^2\kappa^2-18l^8M^3\eta(1-4\xi)\right]+\nonumber\\
&+&\left.r^{-2}\left[-9l^8M^4\eta^2-12l^{10}M^3\eta^2\kappa^2\right]\right\}/\left\{r^4\left[4l^4(l^4+12l^2\eta+36\eta^2)\right]+\right.\nonumber\\
&+&\left.r^2\left[4l^4(-6l^4M\eta-36l^2M\eta^2)\right]+r^0\left[36l^8M^2\eta^2\right]\right\}
\end{eqnarray}
e a função $f(r)$
\begin{eqnarray}
f(r) \ = \ \frac{r^{2}}{l^{2}}-\frac{r_+^2}{l^2} \ = \ \frac{r^2-r_+^2}{l^2}.
\end{eqnarray}
Transformando para coordenada tartaruga $r_*(r)$ tem-se
\begin{eqnarray}
dr_* &=&\frac{l^2}{r^2-r_+^2}dr  \to  r_*=\int \frac{l^2}{r^2-r_+^2}dr,\\
r_* &\approx & \int \frac{l^2dr}{(r-r_+)(r+r_+)} \approx \int \frac{l^2dr}{2r_+(r-r_+)} =\frac{l^2}{2r_+}\ln(r-r_+).
\end{eqnarray}
Agora será avaliada nos limites onde $r_*(r\to r_+)$ e $r_*(r\to\infty)$. Primeiramente em $r_*(r\to r_+)$ obtém-se
\begin{eqnarray}
r_*\approx \frac{l^2}{2r_+}\ln(r-r_+),
\end{eqnarray}
e a função de onda fica aproximadamente
\begin{eqnarray}
\bar{R} &\approx & \exp\left(\pm i\omega \frac{l^2}{2r_+}\right)\ln(r-r_+) \ \approx \ (r-r_+)^{\pm i\omega l^2/2r_+}, \\
&=& r\left(1-\frac{r_+}{r}\right)^{\pm i\omega l^2/2r_+} \ \approx \ \left(1-\frac{r_+}{r}\right)^{\pm i\omega l^2/2r_+}.
\end{eqnarray}
No limite $r_*(r\to\infty)$ terá
\begin{eqnarray}
r_*=\int \frac{l^2}{r^2-r_+^2}dr \ \approx \  \int \frac{l^2}{r^2}dr \ = \ -\frac{l^2}{r}.
\end{eqnarray}
Resolvendo a equação de movimento no infinito é obtido
\begin{eqnarray}
\bar{R}^{''}&+&\omega^2\bar{R}-V(r\to\infty)\bar{R}=0, \\
\bar{R}^{''}&-&\frac{4l^2m^2+540\eta^2l^{-4}-60\eta l^{-2}(8\xi-3)+15+24m^2\eta-80\xi}{4(l^4+12l^2\eta+36\eta^2)} r^2\bar{R}=0.\nonumber\\
\end{eqnarray}
Supõe-se aqui a seguinte solução para $\bar{R}$, $\bar{R}=r_*^\alpha$. Substituindo-a na equação de movimento nos dá a seguinte solução para $\alpha$
\begin{eqnarray}
\alpha=\frac{1\pm \sqrt{1+\frac{4l^6m^2+540\eta^2-60\eta l^{2}(8\xi-3)+l^4(15+24m^2\eta-80\xi)}{(l^4+12l^2\eta+36\eta^2)}}}{2}.
\end{eqnarray}
Agora finalmente fazendo a transformação da Eq.(\ref{transformacao}) a função de onda obtida é
\begin{equation}
\Psi=\gamma^{-i\omega l^2/2r_+}(1-\gamma)^{\alpha}\chi.
\end{equation}

\subsection{Evolução Escalar}
Aqui serão apresentados os resultados da evolução do campo escalar se utilizando do método HH. Nota-se que, assim como no AGGH, o acoplamento com $\xi$ causam apenas uma redefinição na massa e no número de onda. Visto isso, todos os cálculos abaixo serão feitos para $\xi=0$. Primeiramente avaliou-se o caso sem o acoplamento com tensor de Einstein, fazendo $\eta=0$ e variando o número de onda $\kappa$ e, em seguida, avaliou-se o caso com acoplamento não nulo. Por simplicidade também faremos sempre a massa do campo escalar $m=0$, já que estamos interessados apenas com $\eta$.

\subsubsection{Sem acoplamento $\eta=0$}
O buraco negro \emph{5Dz1} apresentou vários desafios computacionais. Para vários valores de $r_+$ suas frequências não convergiam, fazendo-se necessário algumas combinações específicas de $r_+$ e $l$. Diferente do que se espera do HH, para buracos negros pequenos foram obtidos bons resultados. Abaixo são apresentados 3 gráficos de -$\omega_I$ x $\kappa$. O primeiro para $r_+=0.01$ e $l=0.01$, o segundo para $r_+=0.1$ e $l=0.01$ e o terceiro para $r_+=1$ e $l=0.1$.
\begin{figure}[H]
\centering
{\epsfig{file=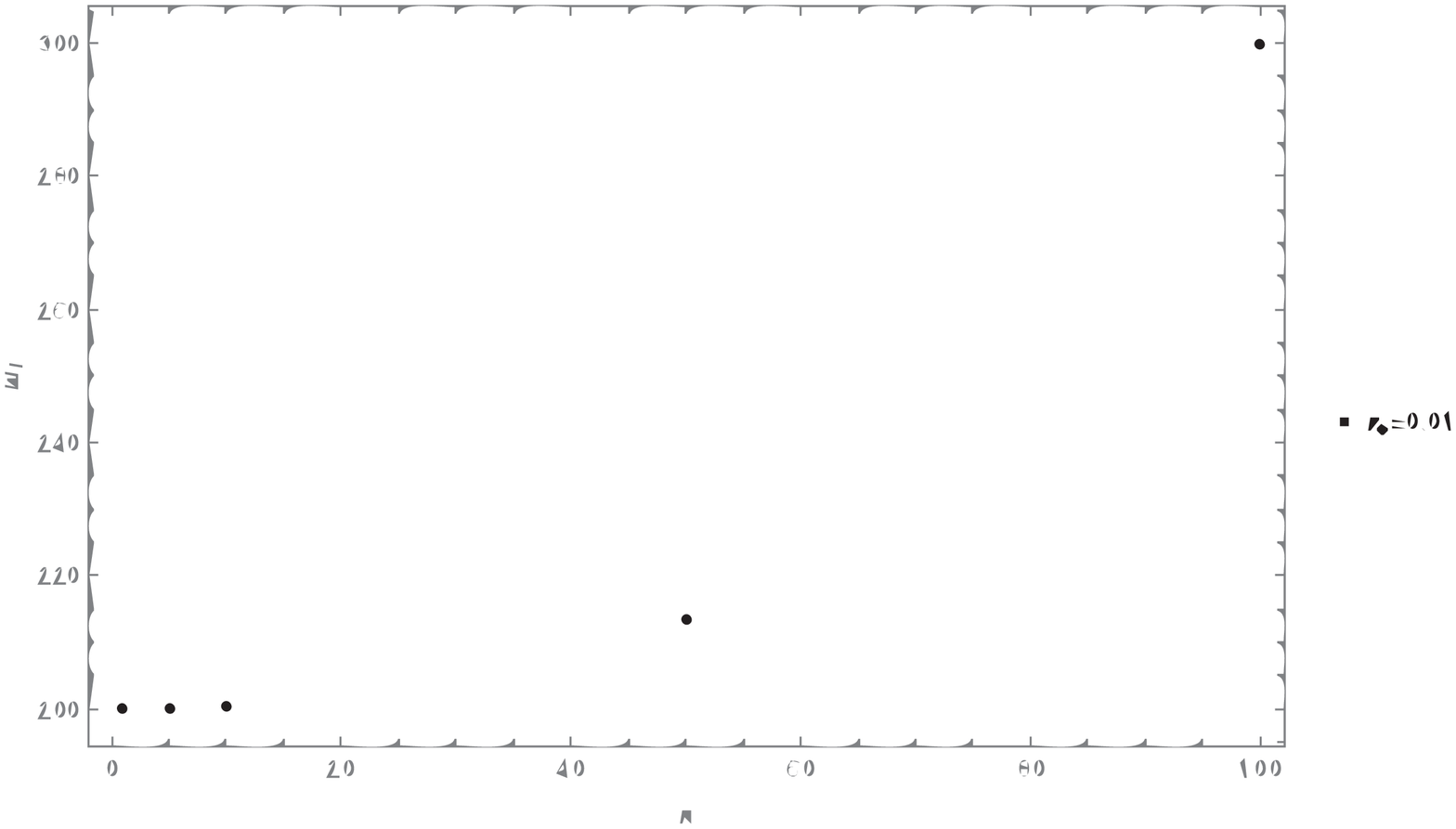, width=1.0\linewidth, height=0.6\linewidth}}
\caption{Gráfico de -$\omega_I$ x $\kappa$ para $r_+=0.01$ com $l=0.01$}
\end{figure}
\begin{figure}[H]
\centering
{\epsfig{file=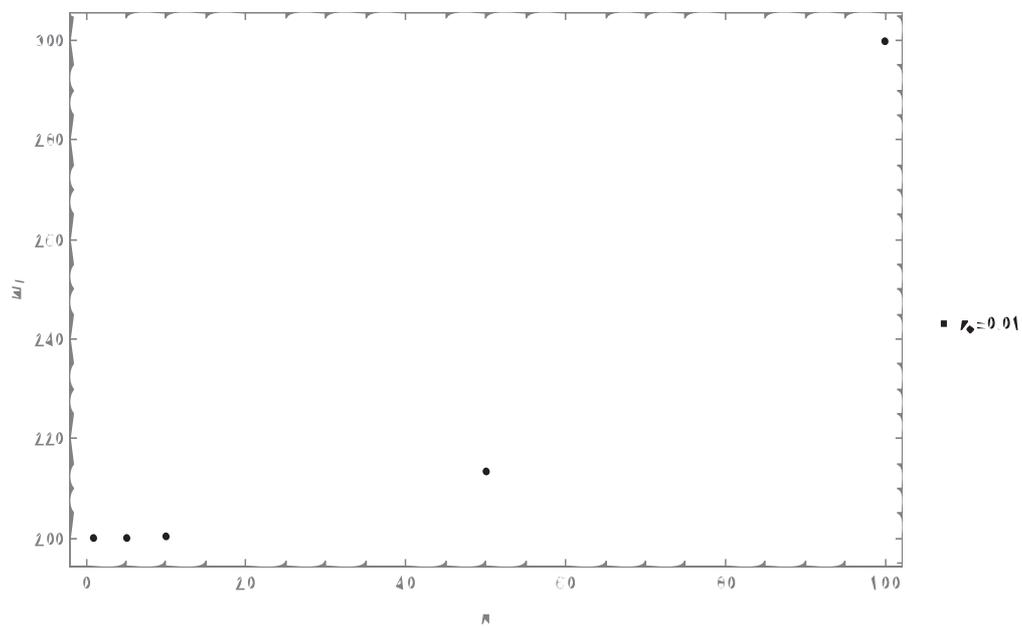, width=1.0\linewidth, height=0.6\linewidth}}
\caption{Gráfico de -$\omega_I$ x $\kappa$ para $r_+=0.1$ com $l=0.01$}
\end{figure}
\begin{figure}[H]
\centering
{\epsfig{file=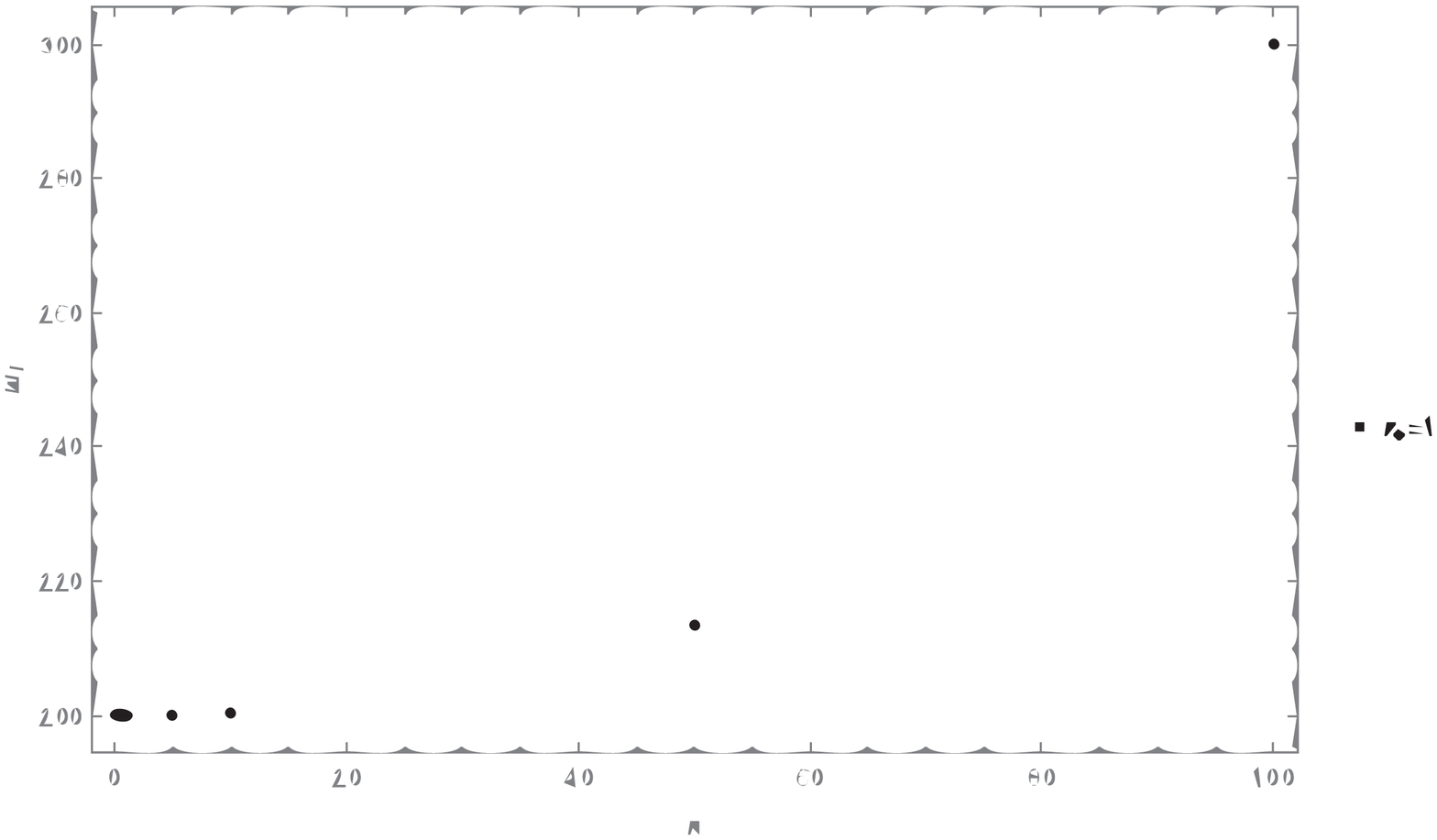, width=1.0\linewidth, height=0.6\linewidth}}
\caption{Gráfico de -$\omega_I$ x $\kappa$ para $r_+=1$ com $l=0.1$}
\end{figure}
Todos os gráficos apresentam um crescimento de $\omega_I$ à medida que $\kappa$ cresce, porém, esse aumento é muito pequeno. O número de onda tem que ser aumentado consideravelmente e isto não é sentido fortemente por $\omega_I$.

O que se percebeu em geral é que há uma transição nas frequências de puramente imaginário para um número complexo, com parte real e imaginária. Porém, à medida que se aumenta $\kappa$, a convergência do método se torna mais complicada. Para estes três casos apresentados abaixo, até onde conseguiu-se obter resultados convergentes, a parte real ainda era toda nula, não sendo manifesta esta transição.

A seguir mostramos o gráfico de $\omega_I$ por $r_+$.
\begin{figure}[H]
\centering
{\epsfig{file=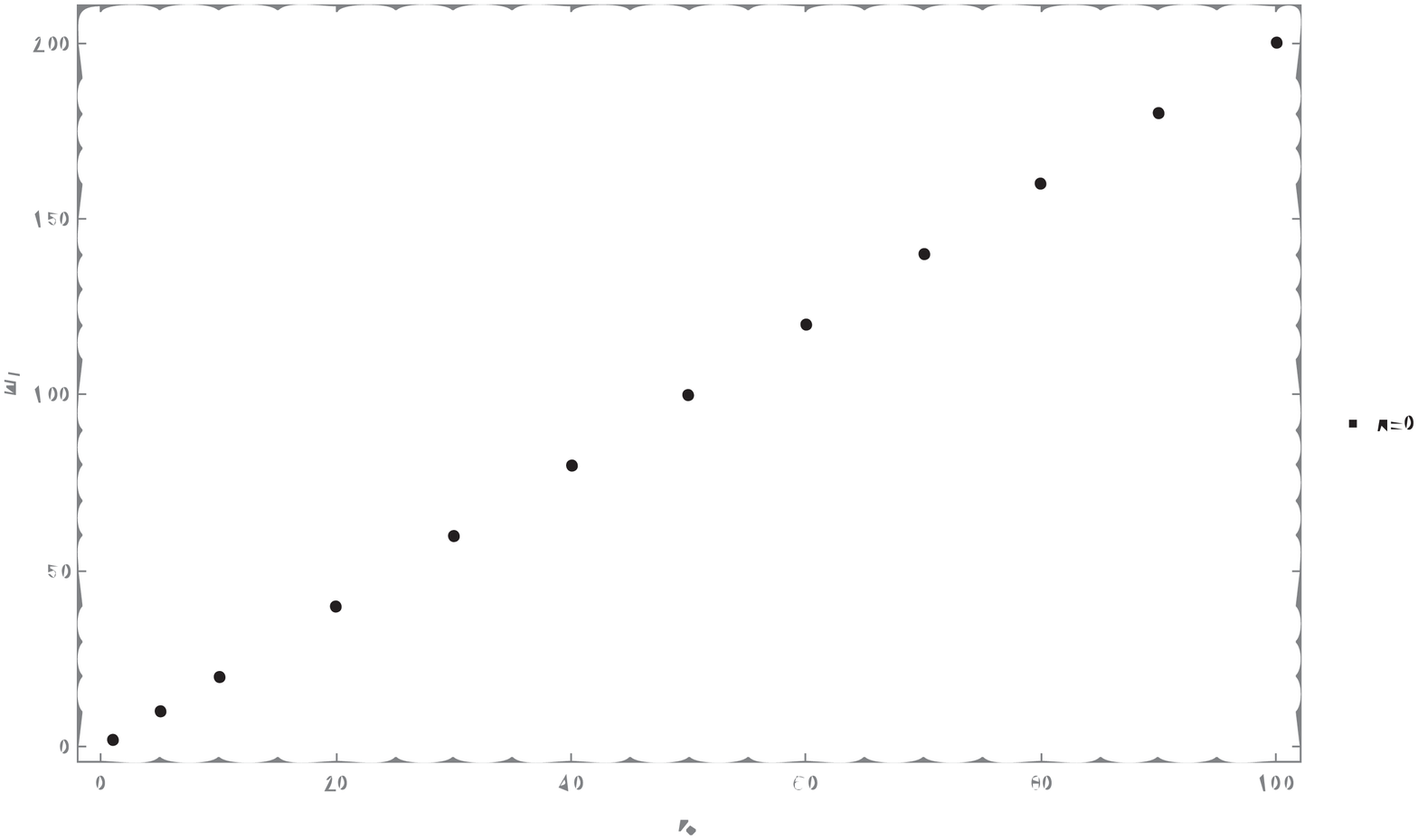, width=0.8\linewidth, height=0.6\linewidth}}
\caption{Gráfico de -$\omega_I$ x $r_+$ para $\kappa=0$ com $l=1$}
\end{figure}
Vê-se daqui que a dependência da frequência imaginária com o raio do horizonte de eventos é linear, e, para este buraco negro, como visto no capítulo 2, sua temperatura também é. Este comportamento é semelhante ao descrito por \cite{horowitzhubeny}. De acordo com a dualidade gravidade/calibre, o inverso desta frequência, $\tau=1/\omega_I$, é a escala de tempo para a abordagem do equilíbrio térmico. No citado artigo isso vale apenas para buracos negros grandes. Para o \emph{5Dz1}, e também para BTZ, este comportamento se estende desde buracos negros pequenos até grandes.

Isso, curiosamente, não acontece também para o \emph{6Dz0}.

\subsubsection{Com acoplamento $\eta\neq 0$}
Aqui são apresentados os gráficos, fazendo-se $\kappa=0$ e analisando agora a influência do acoplamento. O que se percebeu aqui é de fato muito interessante. Foi feita a análise para buracos negros com $r_+=5$, $10$, $50$ e $100$, todos com $l=10$ e com $M=0.25$, $1$, $25$ e $100$, respectivamente. Vamos colocar, primeiramente, uma tabela com alguns dos dados obtidos, para uma melhor visualização dos dados, e, logo após, os gráficos.
\begin{table}[H]
\centering
\caption{MQN escalar \emph{5Dz1}}
\vspace{0.5cm}
\begin{tabular}{|c|c|c|c|c|c|c|}\hline
 & \multicolumn{2}{|c|}{$r_+=5$} & \multicolumn{2}{|c|}{$r_+=10$} & \multicolumn{2}{|c|}{$r_+=100$}\\ \hline
$\eta$ & $\omega_R$ & $\omega_I$ & $\omega_R$ & $\omega_I$  & $\omega_R$& $\omega_I$  \\ \hline
0      & 0          & -0.1000000 & 0          & -0.20000000 & 0          & -2.000000 \\ \hline
1      & 0          & -0.1015125 & 0          & -0.20302520 & 0          & -2.030252 \\ \hline
5      & 0          & -0.1079890 & 0          & -0.21597934 & 0          & -2.159793 \\ \hline
10     & 0          & -0.1188533 & 0          & -0.23770684 & 0          & -2.377068 \\ \hline
13     & 0          & -0.1332953 & -0.016276  & -0.26659060 & -0.162766  & -2.665906 \\ \hline
15     & -0.01618700& -0.1318831 & -0.03237   & -0.26376600 & -0.323741  & -2.637660 \\ \hline
\end{tabular}
\end{table}
\begin{figure}[H]
\centering
{\epsfig{file=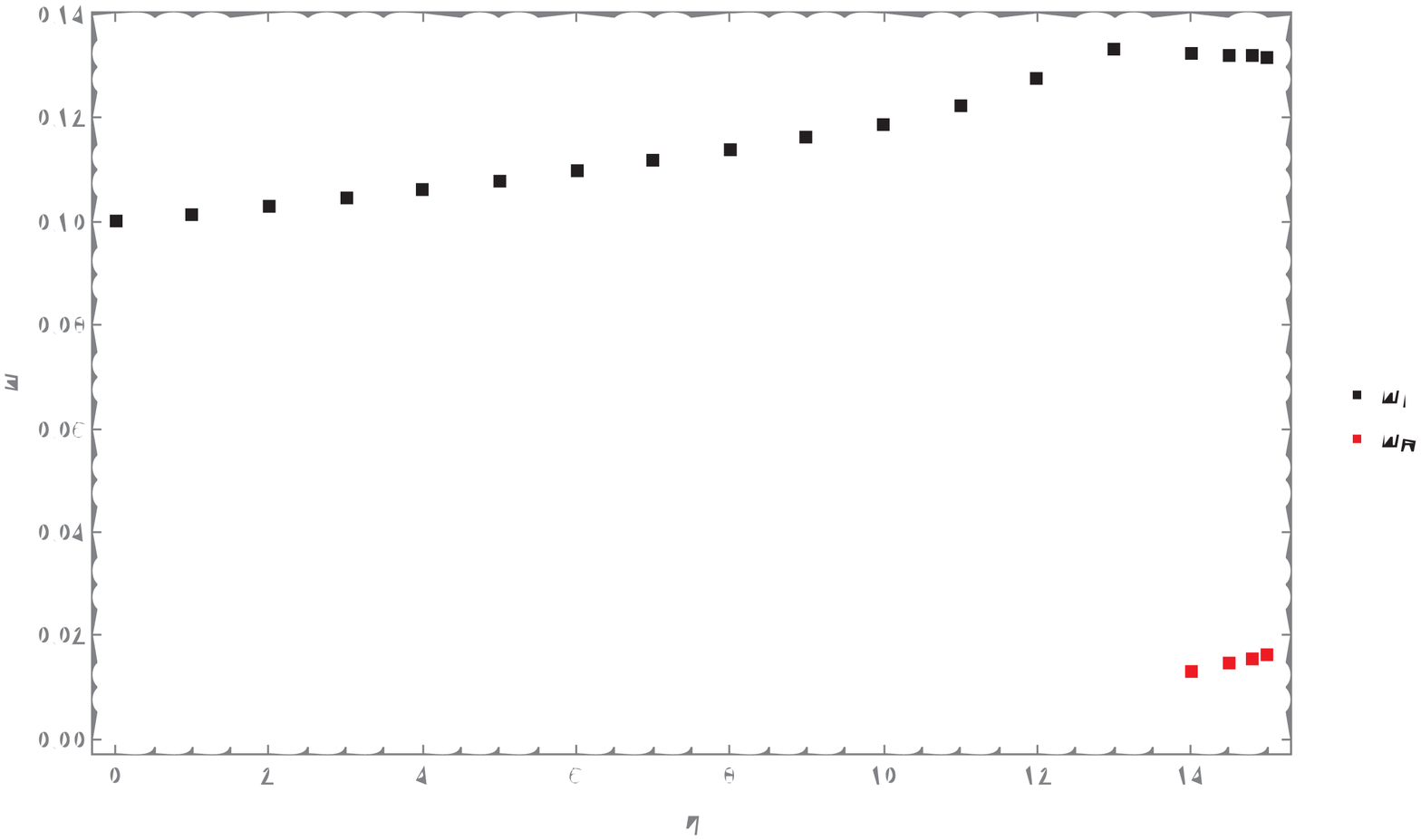, width=0.8\linewidth, height=0.6\linewidth}}
\caption{Gráfico de -$\omega_I$,-$\omega_R$ x $\eta$ para $r_+=5$, $l=10$ e $M=0.25$ }
\end{figure}
\begin{figure}[H]
\centering
{\epsfig{file=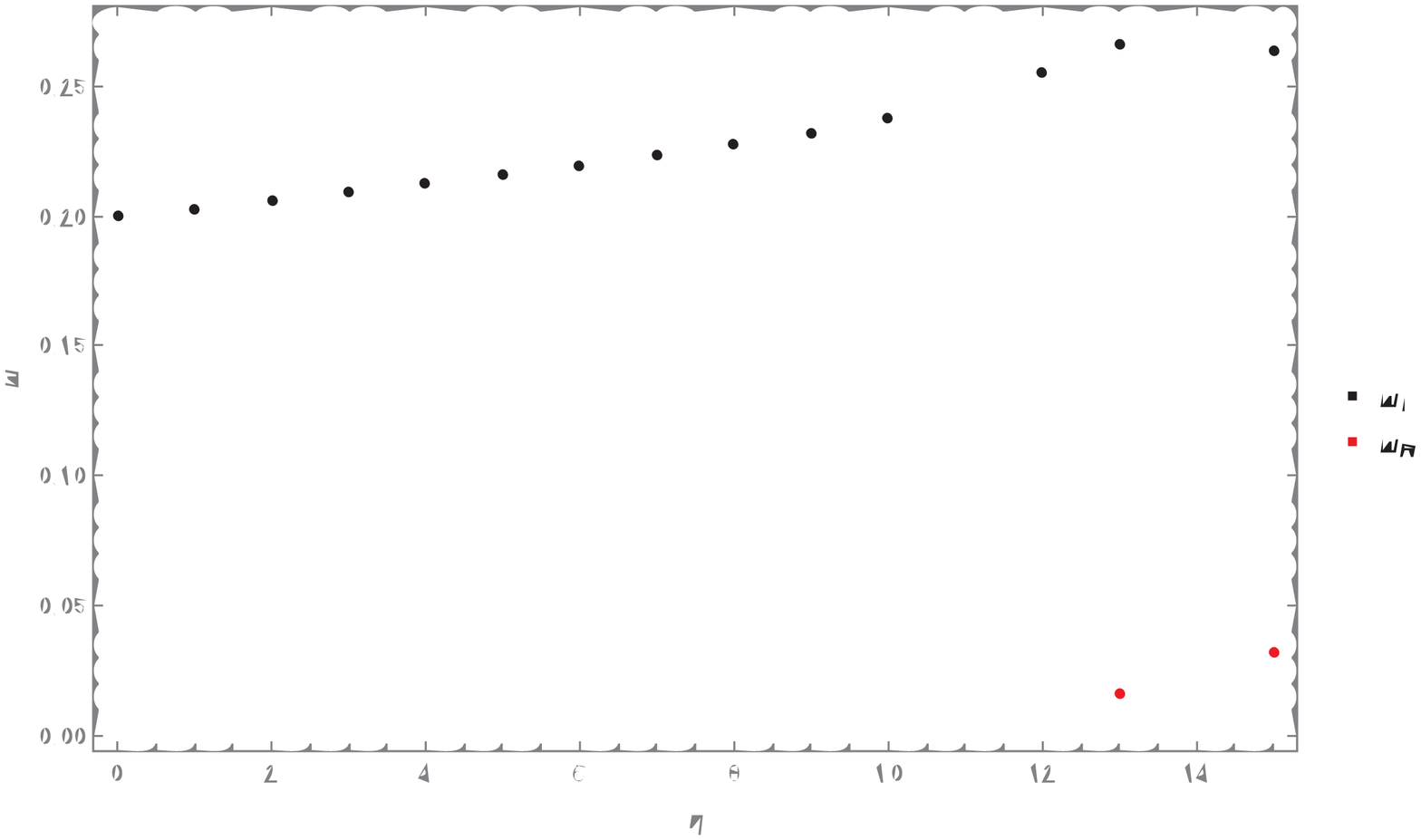, width=0.8\linewidth, height=0.6\linewidth}}
\caption{Gráfico de -$\omega_I$,-$\omega_R$ x $\eta$ para $r_+=10$, $l=10$ e $M=1$ }
\end{figure}
\begin{figure}[H]
\centering
{\epsfig{file=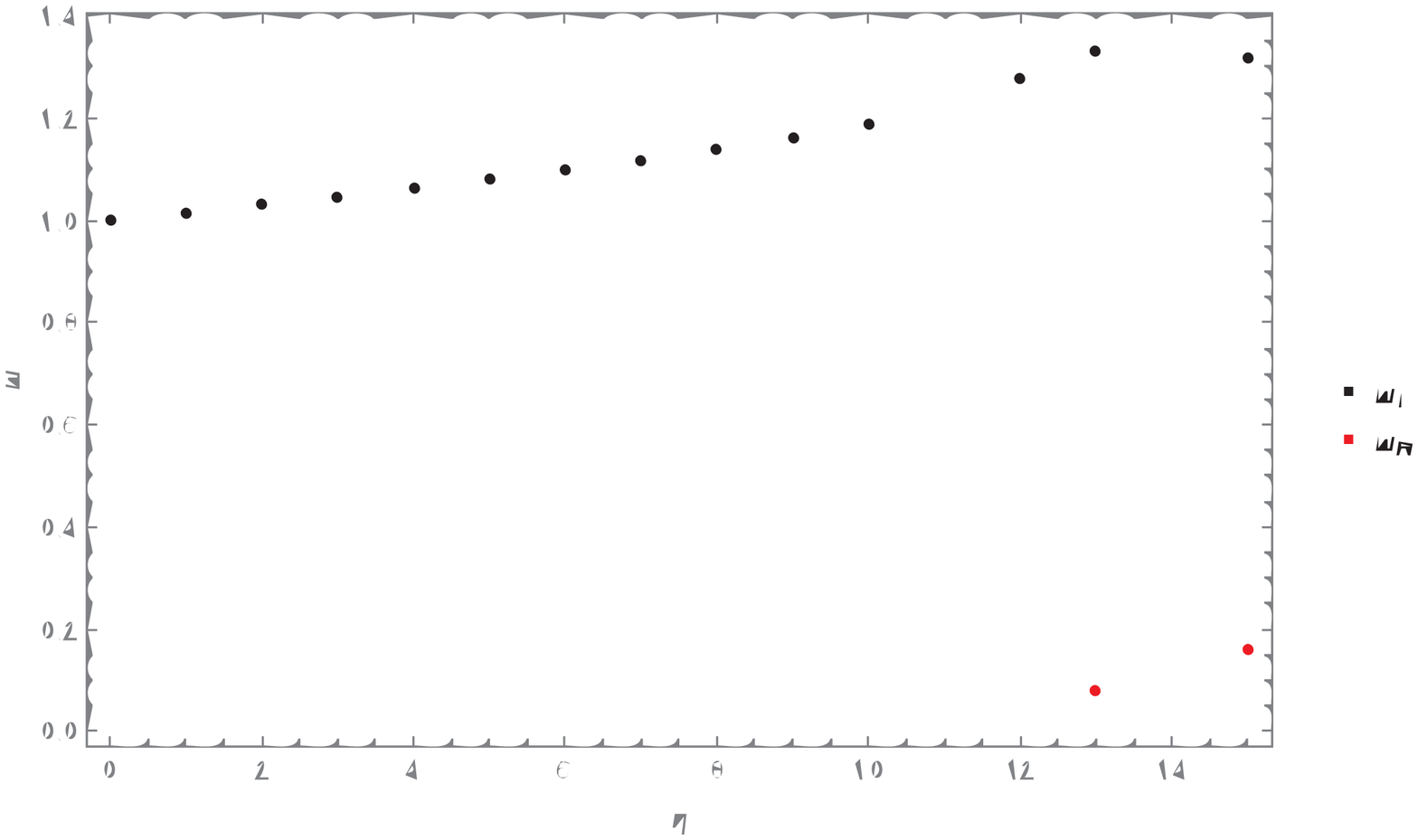, width=0.8\linewidth, height=0.6\linewidth}}
\caption{Gráfico de -$\omega_I$,-$\omega_R$ x $\eta$ para $r_+=50$, $l=10$ e $M=25$ }
\end{figure}
\begin{figure}[H]
\centering
{\epsfig{file=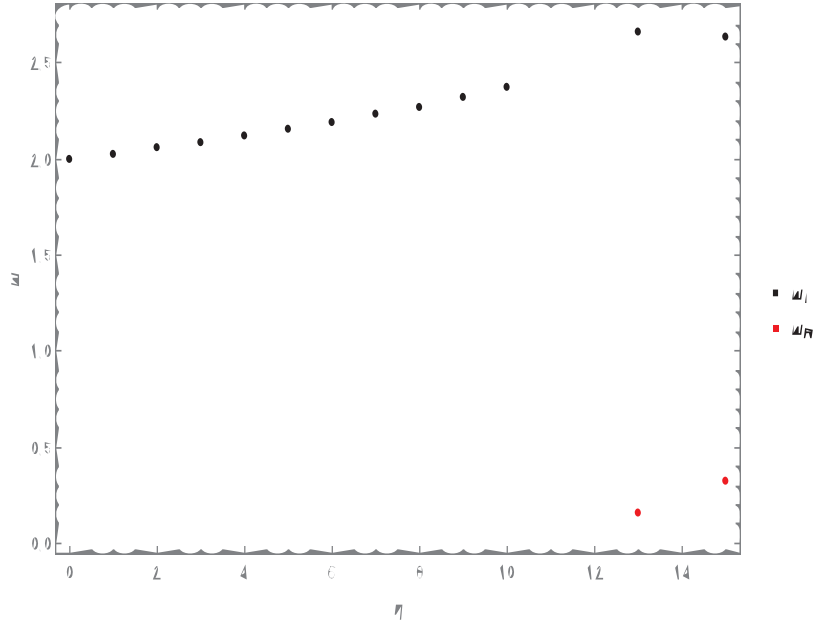, width=0.8\linewidth, height=0.6\linewidth}}
\caption{Gráfico de -$\omega_I$,-$\omega_R$ x $\eta$ para $r_+=100$, $l=10$ e $M=100$ }
\end{figure}
O que se verifica com os gráficos é que para $\eta$ pequeno as frequências são puramente imaginárias. À medida que cresce, há uma transição e começa a aparecer a parte real. Exatamente no ponto onde a parte real aparece é onde se tem o maior valor de $\omega_I$. A partir deste ponto a parte imaginária começa a decrescer e vai tendendo a zero, enquanto a parte real só vai aumentando.

Como dito, para o \emph{5Dz1} a convergência não é tão boa, e os gráficos podem não ser tão claros dada a explicação acima, mas, para o \emph{6Dz1}, onde a convergência é bem melhor, esse comportamente é manifesto bem mais claramente nos gráficos.

\subsection{Evolução Eletromagnética}
A evolução do campo eletromagnético foi analisada em separado para o caso onde $\kappa=0$ e $\kappa\neq0$.

\subsubsection{Sem número de onda $\kappa=0$}
Quando não se tem a influência das dimensões a frequência se comporta de maneira sempre crescente. A análise escolhida foi para $\kappa=0$, mas poderia ser escolhida também para $\kappa=1$, porque os dois gráficos se comportam exatamente da mesma maneira, ficando sobrepostos.
\begin{figure}[H]
\centering
{\epsfig{file=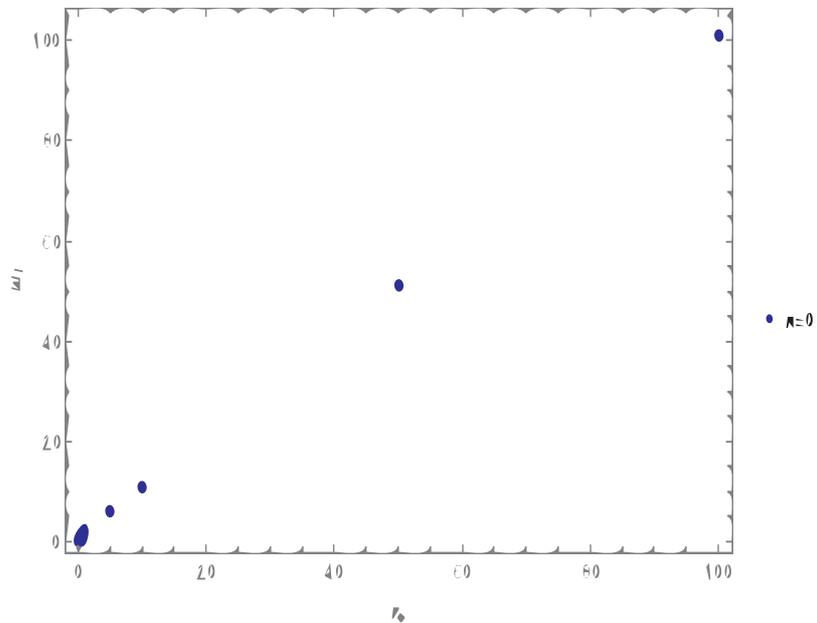, width=0.8\linewidth, height=0.6\linewidth}}
\caption{Gráfico de -$\omega_I$ x $r_+$ para $\kappa=0$ com $l=1$}
\end{figure}
Aqui ocorre também a transição de puramente imaginário, mas, não é apresentado o gráfico da parte real pois para $\kappa=0$ não foi encontrada a transição. Para $\kappa=1$ é possível observar a transição mas em uma faixa bem pequena, o que não ficaria evidente no gráfico.

\subsubsection{Com número de onda $\kappa\neq 0$}
A evolução do campo eletromagnético com número de onda diferente de zero apresenta a transição de puramente imaginário. Foi colocada aqui a tabela para se ver o quão grande deve ser $\kappa$ para ser manifesta essa transição.
\begin{table}[H]
\centering
\caption{MQN eletromagnético, \emph{5Dz1}}
\vspace{0.5cm}
\begin{tabular}{|c|c|c|c|c|c|c|c|c|}\hline
& \multicolumn{2}{|c|}{$r_+=5$} & \multicolumn{2}{|c|}{$r_+=10$} & \multicolumn{2}{|c|}{$r_+=50$} & \multicolumn{2}{|c|}{$r_+=100$}  \\ \hline
$\kappa$ & $\omega_R$ & $\omega_I$ & $\omega_R$ & $\omega_I$ & $\omega_R$ & $\omega_I$ & $\omega_R$ & $\omega_I$ \\ \hline
0        & 0          & -9.3400    & 0          & -19.7074   & 0          & -99.9400   & 0          & -199.9701 \\ \hline
1        & 0          & -9.3411    & 0          & -19.7084   & 0          & -99.9400   & 0          & -199.9701 \\ \hline
10       & 0          & -10.2536   & 0          & -19.8073   & 0          & -99.9408   & 0          & -199.9702 \\ \hline
100      & -50.0000   & -45.0000   & 0          & -30.0117   & 0          & -100.0200  & 0          & -199.9801 \\ \hline
1000     & nc         & nc         & nc         & nc         & 0          & -107.9412  & 0          & -200.9652 \\ \hline
10000    & nc         & nc         & nc         & nc         & nc         & nc         & 0          & -300.0000  \\ \hline
\end{tabular}
\end{table}
O símbolo ``nc" representa locais onde o método não foi capaz de convergir. Vê-se que deve haver um equilíbrio entre $r_+$ e $\kappa$, pois, a medida que o número de onda é muito maior que o raio do horizonte de eventos o método não se comporta bem. E, à medida que $r_+$ aumenta, sua transição acontece para um $\kappa$ ainda maior, o que dificulta ainda mais a convergência. Nota-se também que para $r_+$ grandes, quando $\kappa$ é pequeno sua influência é quase imperceptível.

Foram feitos gráficos da parte imaginária de $\omega$ por $\kappa$, para $r_+=5$, $10$, $50$ e $100$, e separados, colocando juntos, primeiramente, $r_+=5$ e $10$, e depois, para $r_+=50$ e $100$ para melhor visualização dos resultados.
\begin{figure}[H]
\centering
{\epsfig{file=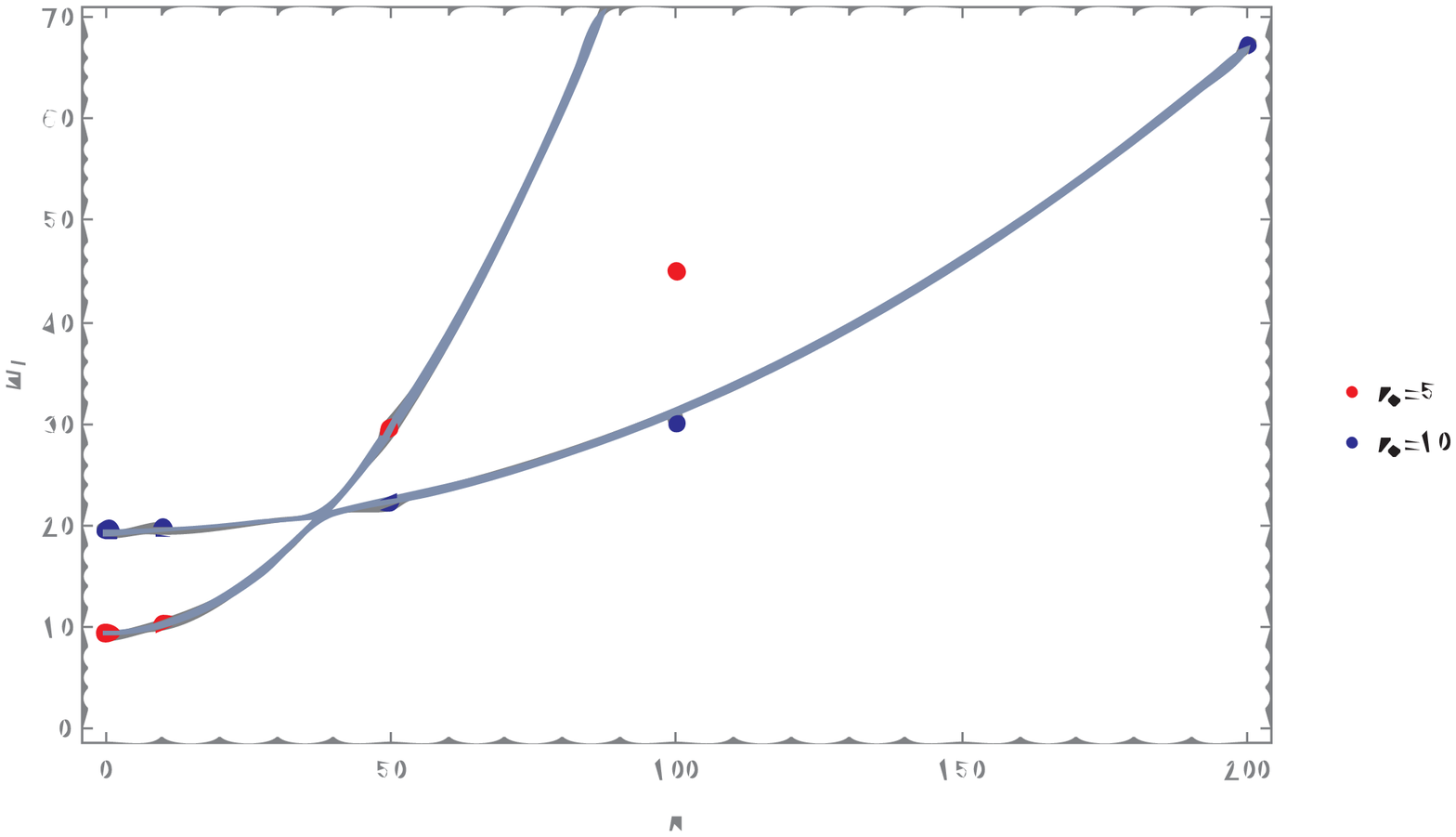, width=0.8\linewidth, height=0.6\linewidth}}
\caption{Gráfico de -$\omega_I$ x $\kappa$ para $r_+=5$ e $10$ com $l=1$}
\end{figure}
\begin{figure}[H]
\centering
{\epsfig{file=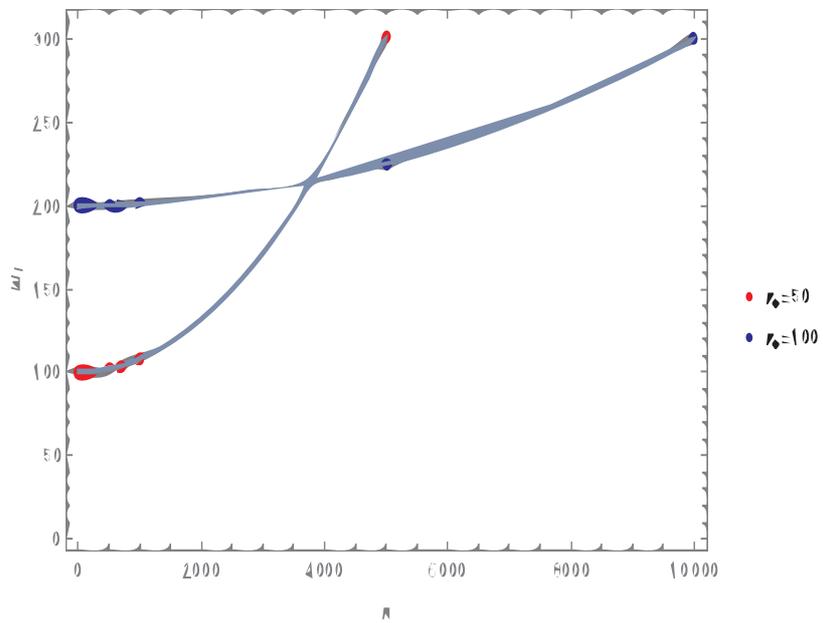, width=0.8\linewidth, height=0.6\linewidth}}
\caption{Gráfico de -$\omega_I$ x $\kappa$ para $r_+=50$ e $100$ com $l=1$}
\end{figure}
Para todos os casos foram feitos ajustes quadráticos, da forma $ax^2+b$, e colocados juntamente com os respectivos pontos.

\section{\emph{Buraco Negro 6Dz0}}
O buraco negro \emph{6Dz0} apresentou vários comportamentos interessantes e bom comportamento com o método HH, com boa convergência quando variados os parâmetros. Veremos primeiro a análise da evolução do campo escalar e depois do campo eletromagnético.

\subsection{Evolução Escalar}
A evolução do campo escalar para este buraco negro, utilizando o HH, apresentou resultados melhores que para \emph{5Dz1}. Os programas convergiam melhor e, assim, apresentou gráficos com resultados e comportamentos mais claros.

\subsubsection{Sem acoplamento $\eta=0$}
Abaixo são apresentados 2 gráficos de $\omega_I$ e $\omega_R$ x $\kappa$. O primeiro para $r_+=5$ e o segundo para $r_+=10$.

Os dados obtidos são apresentados na tabela abaixo e verifica-se que quando $r_+$ é da mesma ordem que $\kappa$ o método funciona bem. Quando $\kappa\gg r_+$ não há convergência e quando $r_+ \gg \kappa$, sua influência é quase despercebida.
\begin{table}[H]
\centering
\caption{MQN escalar $\eta0$ \emph{6Dz0}}
\vspace{0.5cm}
\begin{tabular}{|c|c|c|c|c|c|c|c|c|}\hline
& \multicolumn{2}{|c|}{$r_+=5$} & \multicolumn{2}{|c|}{$r_+=10$} & \multicolumn{2}{|c|}{$r_+=50$} & \multicolumn{2}{|c|}{$r_+=100$}\\ \hline
$\kappa$ & $\omega_R$ & $\omega_I$ & $\omega_R$ & $\omega_I$  & $\omega_R$ & $\omega_I$  & $\omega_R$ & $\omega_I$  \\ \hline
0      & 0        & -5.9999 &   0      & -10.4174 & 0         & -50.0801 & 0         & -100.0400   \\ \hline
10     & -1.7183  & -7.0318 & 0        & -10.5319 & 0         & -50.0809 & 0         & -100.0401      \\ \hline
$10^2$ & -12.2616 & -2.8199 & -8.2164  & -11.9222 & 0         & -50.1002 & 0         & -100.0500 \\ \hline
$10^3$ & nc       & nc      & -47.3699 & -0.5406  & 0         & -63.5448 & 0         & -101.0583  \\ \hline
$10^4$ & nc       & nc      & nc       & nc       & -121.4885 & -27.9316 & -80.3966  & -118.8903    \\ \hline
$10^5$ & nc       & nc      & nc       & nc       & nc        & nc       & -473.3769 & -5.3870 \\ \hline
\end{tabular}
\end{table}
\begin{figure}[H]
\centering
{\epsfig{file=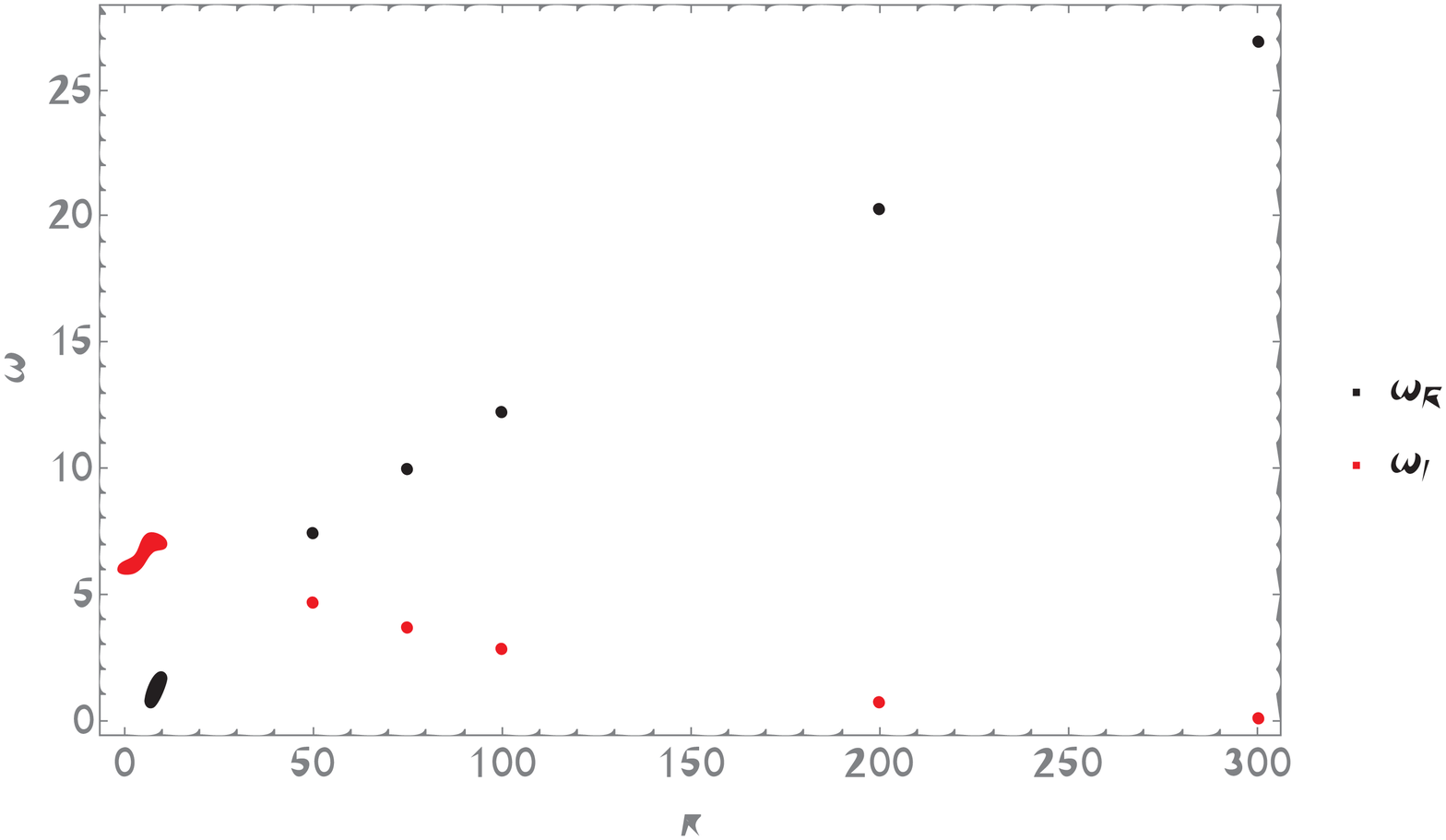, width=0.8\linewidth, height=0.6\linewidth}}
\caption{Gráfico de -$\omega_I$,-$\omega_R$ x $\kappa$ para $r_+=5$ com $l=1$ e $\eta=0$}
\end{figure}
\begin{figure}[H]
\centering
{\epsfig{file=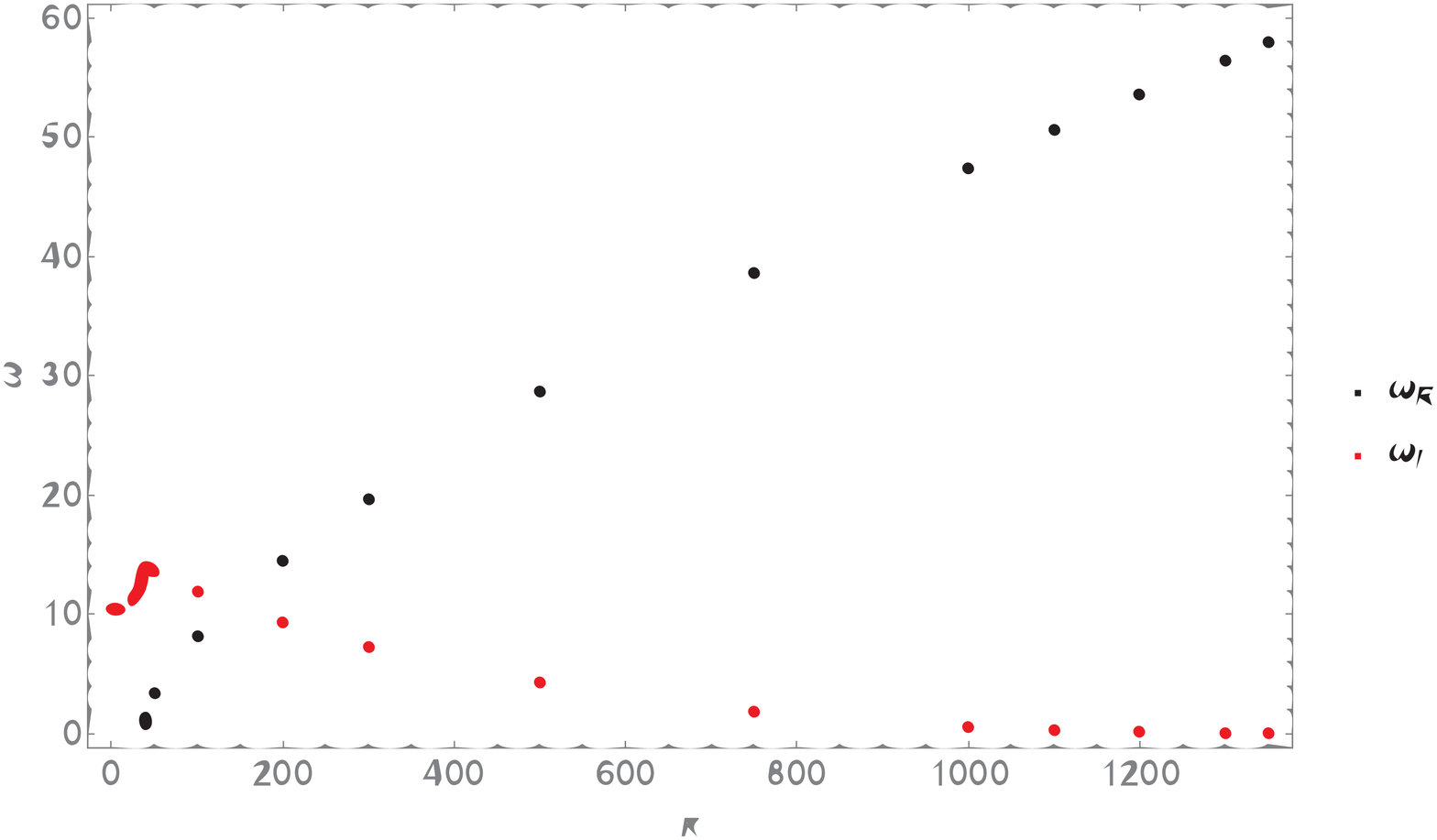, width=0.8\linewidth, height=0.6\linewidth}}
\caption{Gráfico de -$\omega_I$,-$\omega_R$ x $\kappa$ para $r_+=10$ com $l=1$ e $\eta=0$}
\end{figure}
O que se verifica com os gráficos é que para $\kappa$ pequeno as frequências são puramente imaginárias. À medida que cresce, há uma transição e começa a aparecer a parte real. Exatamente no ponto onde a parte real aparece é onde se tem o maior valor de $\omega_I$. A partir deste ponto a parte imaginária começa a decrescer e vai tendendo a zero, enquanto a parte real só vai aumentando.

A seguir há o gráfico de $\omega_I$ por $r_+$.
\begin{figure}[H]
\centering
{\epsfig{file=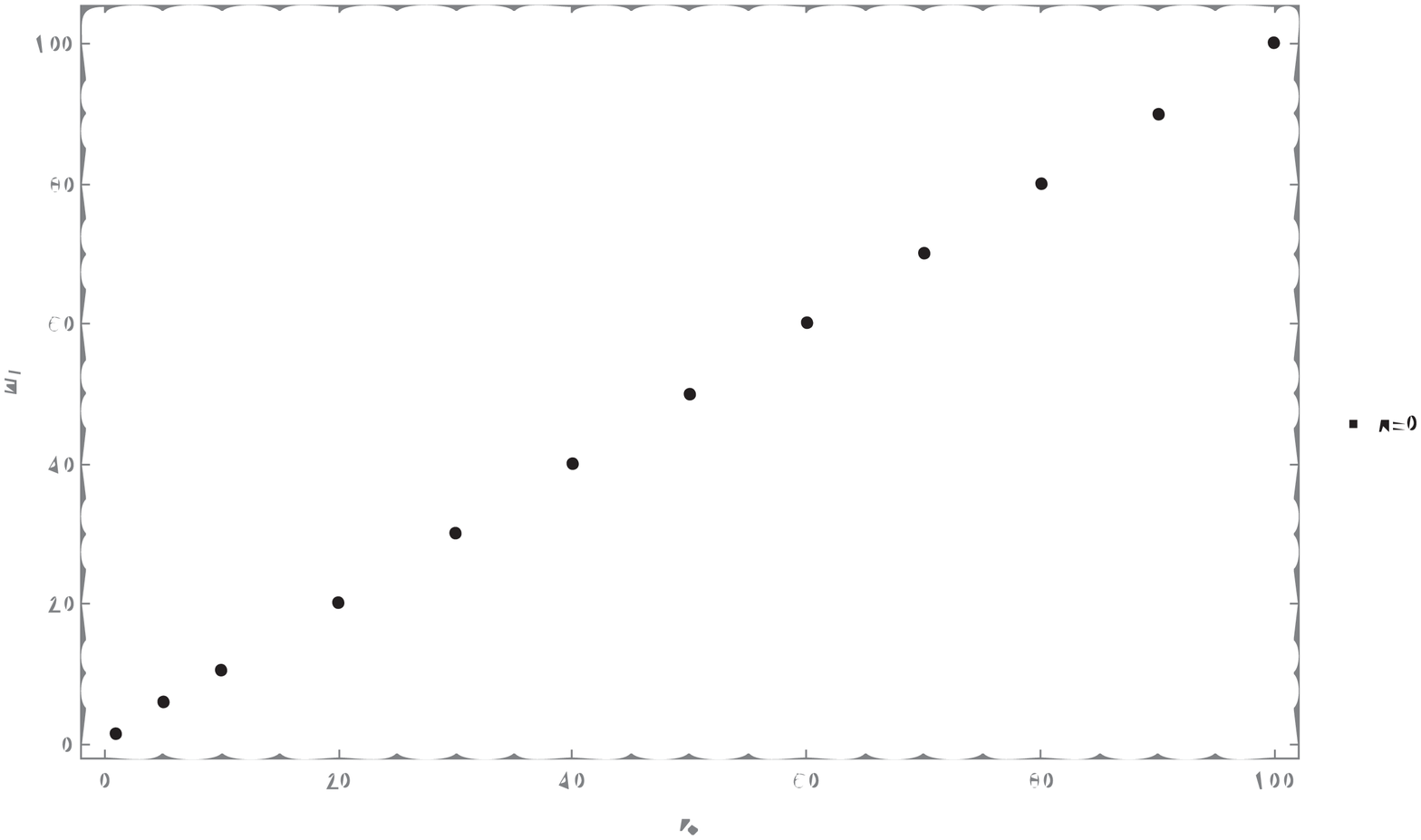, width=0.8\linewidth, height=0.6\linewidth}}
\caption{Gráfico de -$\omega_I$ x $r_+$ para $\kappa=0$ com $l=1$}
\end{figure}
Como discutido para o buraco negro anterior, este comportamento linear da frequência imaginária com o raio do horizonte de eventos poderia representar um tempo de escala para o equilíbrio térmico em uma correspondência gravidade/calibre. Porém, a temperatura deste buraco negro, como visto no capítulo 2, é independente de $r_+$. Isso é de fato muito curioso, já que para o \emph{5Dz1} isso não ocorre.

\subsubsection{Com acoplamento $\eta\neq 0$}
A análise do campo escalar com acoplamento para este buraco negro também se mostrou bastante desafiadora e promissora. Foram escolhidos valores específicos de $r_+$, $l$ e $M$ para que se obtivessem convergências. Foi escolhido $r_+=5$, $10$, $50$ e $100$, todos com $l=10$ e $M=0.25$, $1$, $25$ e $100$, respectivamente.
\begin{figure}[H]
\centering
{\epsfig{file=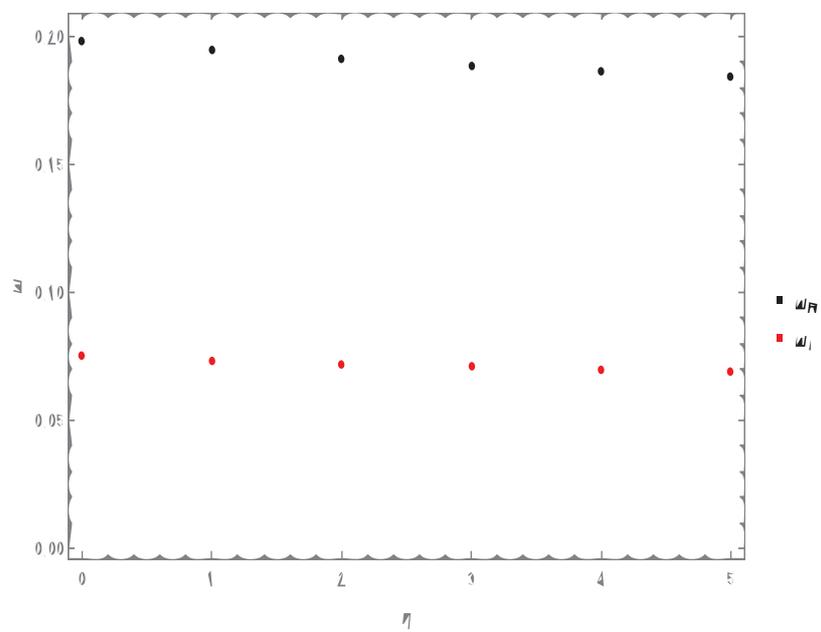, width=0.8\linewidth, height=0.6\linewidth}}
\caption{Gráfico de -$\omega_I$,-$\omega_R$ x $\eta$ para $r_+=5$ com $l=10$ e $M=0.25$}
\end{figure}
\begin{figure}[H]
\centering
{\epsfig{file=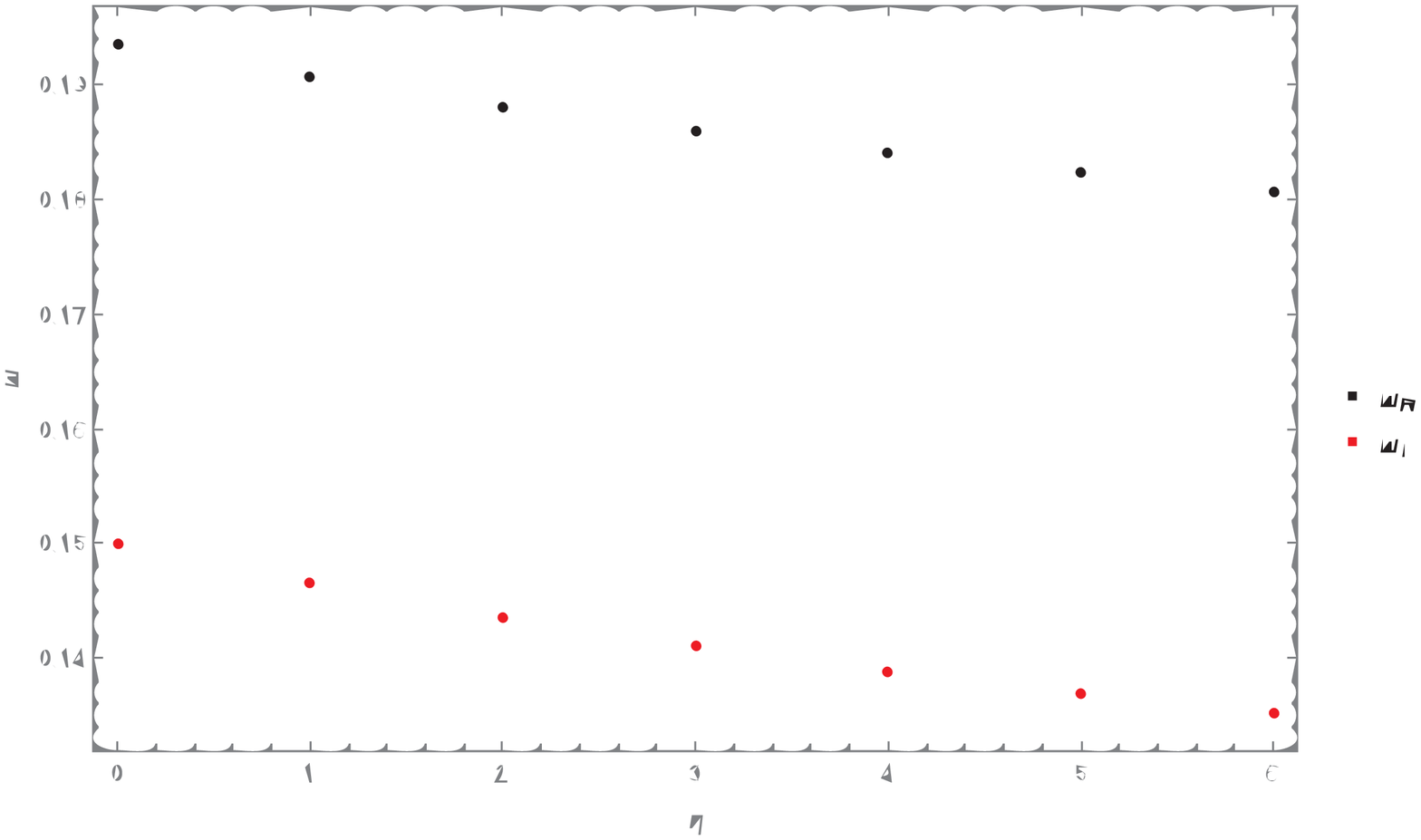, width=0.8\linewidth, height=0.6\linewidth}}
\caption{Gráfico de -$\omega_I$,-$\omega_R$ x $\eta$ para $r_+=10$ com $l=10$ e $M=1$}
\end{figure}
\begin{figure}[H]
\centering
{\epsfig{file=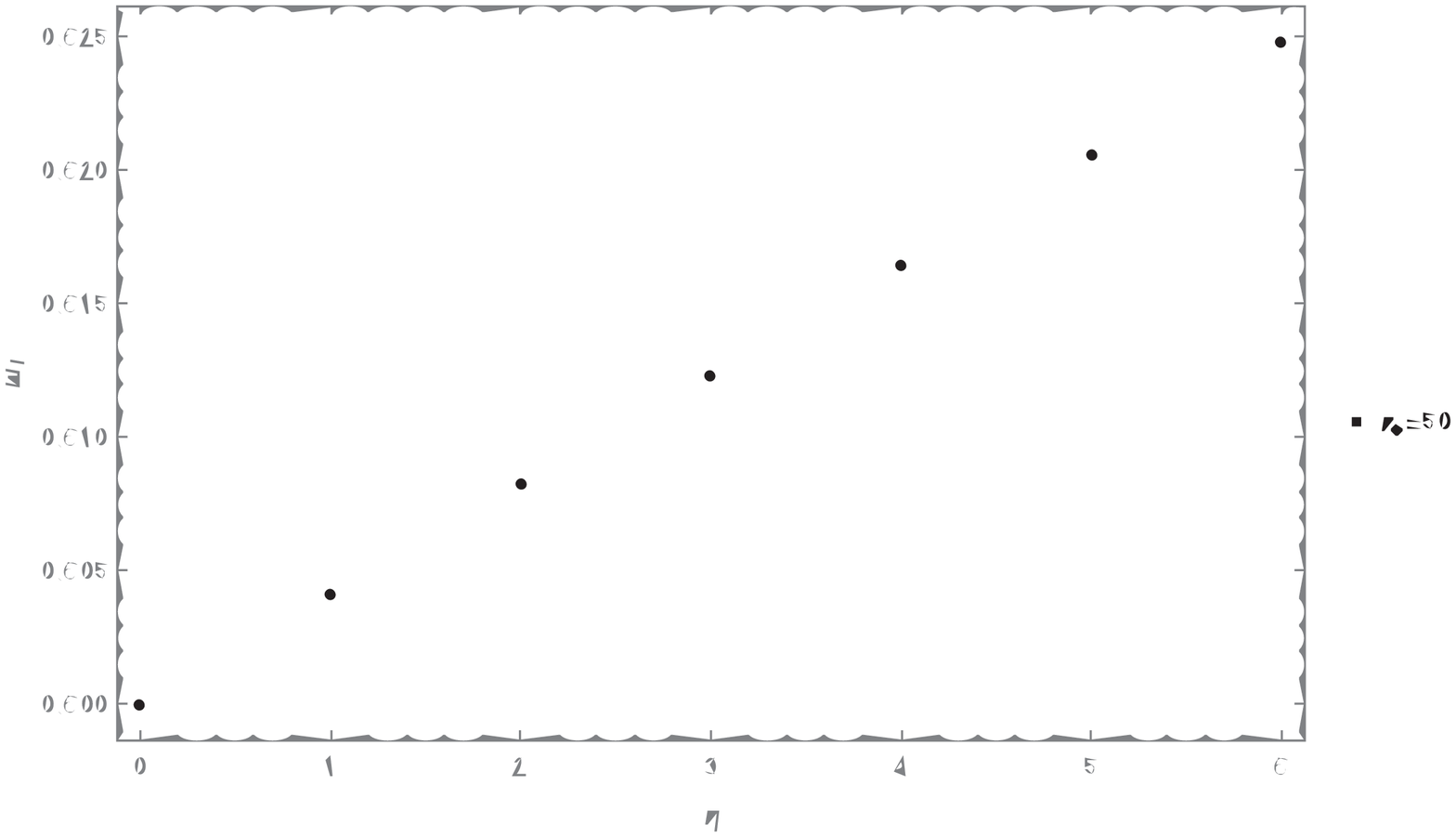, width=0.8\linewidth, height=0.6\linewidth}}
\caption{Gráfico de -$\omega_I$ x $\eta$ para $r_+=50$ com $l=10$ e $M=25$}
\end{figure}
\begin{figure}[H]
\centering
{\epsfig{file=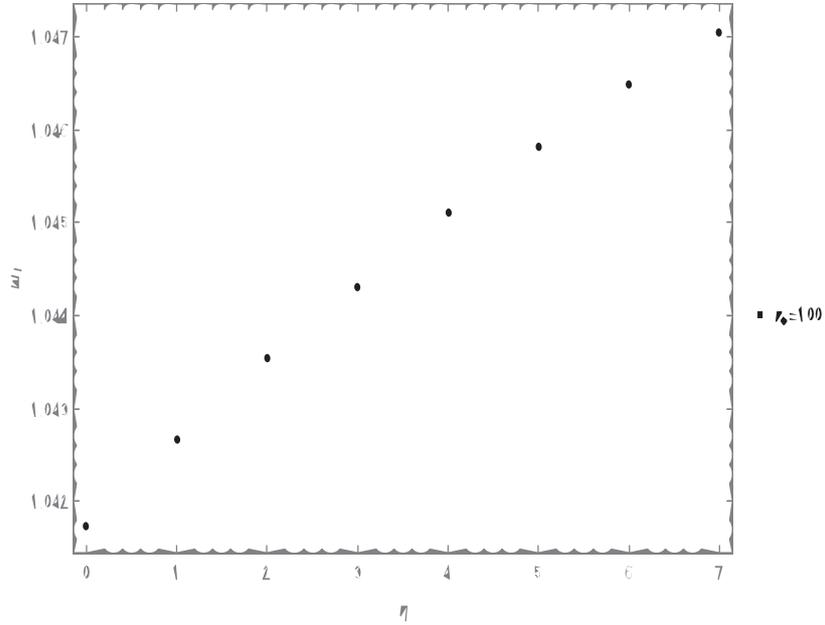, width=0.8\linewidth, height=0.6\linewidth}}
\caption{Gráfico de -$\omega_I$ x $\eta$ para $r_+=100$ com $l=10$ e $M=100$}
\end{figure}
O que se observa é que para $\eta$ grande o método não converge. Para $r_+=5$ e $10$, enquanto $\eta$ é da mesma ordem já aparece uma frequência com parte real e imaginária e à medida que $\eta$ cresce o método não converge e não há certeza sobre a transição para puramente imaginário.

Curiosamente para $r_+=50$ e $100$, para $\eta$ pequeno nós temos resultados puramente imaginários. À medida que $\eta$ cresce o método deixa de convergir, mas pode ser feita a análise para $\eta\to\infty$ e o resultado é também puramente imaginário.

\subsection{Evolução Eletromagnética}
A evolução do campo eletromagnético foi analisada em separado para o caso onde $\kappa=0$ e $\kappa\neq0$.

\subsubsection{Sem número de onda $\kappa=0$}
Quando não se tem a influência das dimensões a frequência se comporta de maneira sempre crescente. A análise escolhida foi para $\kappa=0$, mas poderia ser escolhida também para $\kappa=1$, porque os dois gráficos se comportam exatamente da mesma maneira, ficando sobrepostos.
\begin{figure}[H]
\centering
{\epsfig{file=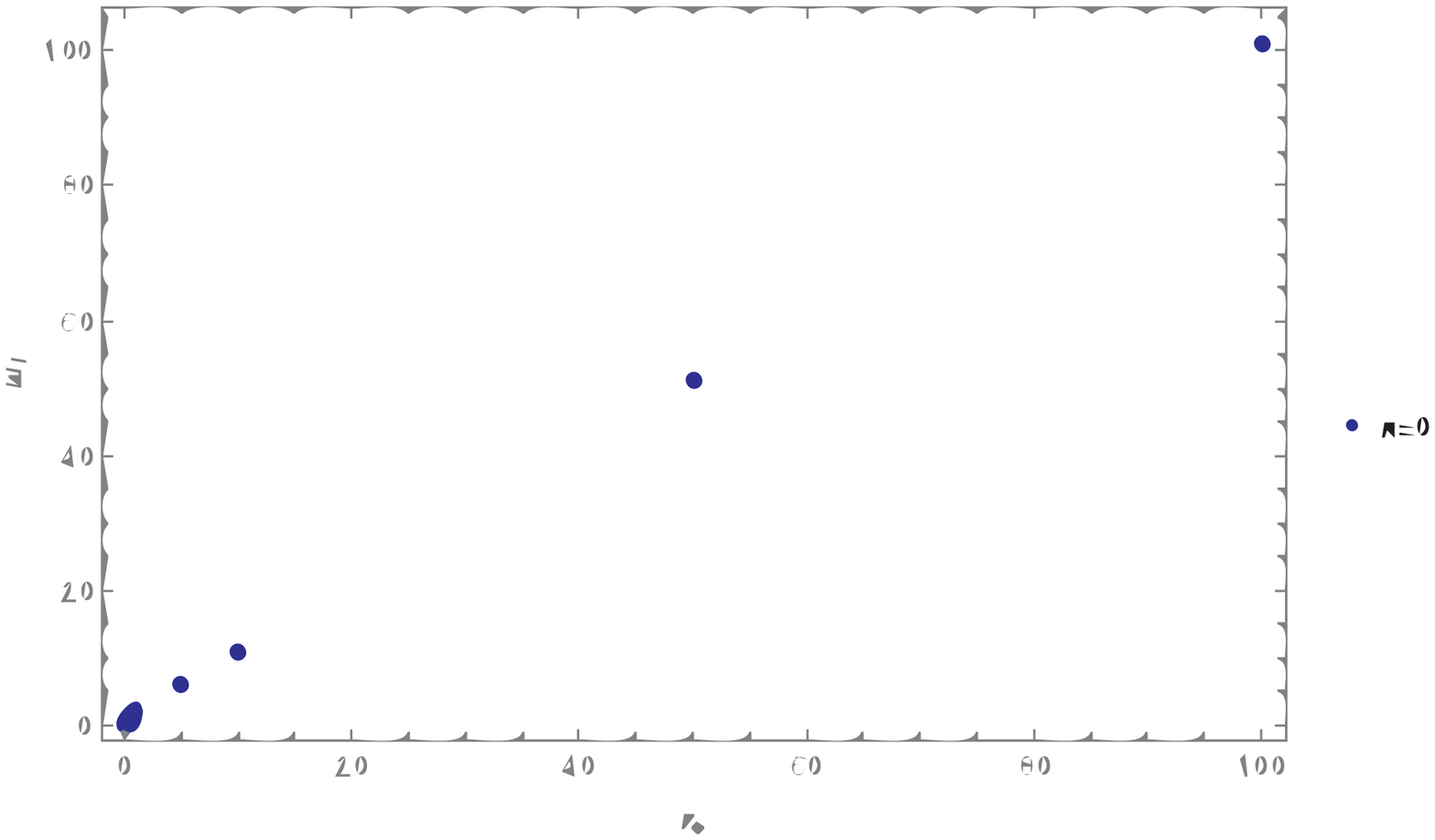, width=0.8\linewidth, height=0.6\linewidth}}
\caption{Gráfico de -$\omega_I$ x $r_+$ para $\kappa=0$ com $l=1$}
\end{figure}
Aqui ocorre também a transição de puramente imaginário, mas, não é apresentado o gráfico da parte real pois para $\kappa=0$ não foi encontrada a transição. Para $\kappa=1$ é possível observar a transição mas em uma faixa bem pequena, o que não ficaria evidente no gráfico.

\subsubsection{Com número de onda $\kappa\neq 0$}
A evolução do campo eletromagnético com número de onda diferente de zero apresenta a transição de puramente imaginário. Foi colocada aqui a tabela para se ver o quão grande deve ser $\kappa$ para ser manifesta essa transição.
\begin{table}[H]
\centering
\caption{MQN eletromagnético \emph{6Dz0}}
\vspace{0.5cm}
\begin{tabular}{|c|c|c|c|c|c|c|c|c|}\hline
 & \multicolumn{2}{|c|}{$r_+=5$} & \multicolumn{2}{|c|}{$r_+=10$} & \multicolumn{2}{|c|}{$r_+=50$} & \multicolumn{2}{|c|}{$r_+=100$}\\ \hline
$\kappa$ & $\omega_R$ & $\omega_I$ & $\omega_R$ & $\omega_I$ & $\omega_R$ & $\omega_I$ & $\omega_R$ & $\omega_I$  \\ \hline
0      & 0       & -5.9230  & 0        & -10.9564 & 0 & -50.9902904 & 0 & -100.99507370 \\ \hline
1      & 0       & -5.9238  & 0        & -10.9565 & 0 & -50.9902905 & 0 & -100.99507378 \\ \hline
10     & 0       & -6.0017  & 0        & -10.9613 & 0 & -50.990298  & 0 & -100.995074   \\ \hline
100    & -7.1994 & -13.6463 & 0        & -11.4470 & 0 & -50.9910    & 0 & -100.9951     \\ \hline
1000   & nc      & nc       & -42.1128 & -42.7942 & 0 & -51.0695    & 0 & -101.00005    \\ \hline
10000  & nc      & nc       & nc       & nc       & nc& nc          & 0 & -101.4927     \\ \hline
100000 & nc      & nc       & nc       & nc       & nc& nc          & 0 & -151.2573     \\ \hline
\end{tabular}
\end{table}
O símbolo nc representa locais onde o método não foi capaz de convergir. Vê-se que deve haver um equilíbrio entre $r_+$ e $\kappa$, pois, a medida que o número de onda é muito maior que o raio do horizonte de eventos o método não se comporta bem. E, à medida que $r_+$ aumenta, sua transição acontece para um $\kappa$ ainda maior, o que dificulta ainda mais a convergência. Nota-se também que para $r_+$ grandes, quando $\kappa$ é pequeno sua influência é quase imperceptível.

Foram feitos gráficos da parte imaginária de $\omega$ por $\kappa$, para $r_+=5$, $10$, $50$ e $100$, e separados, colocando juntos, primeiramente, $r_+=5$ e $10$, e depois, para $r_+=50$ e $100$ para melhor visualização dos resultados.
\begin{figure}[H]
\centering
{\epsfig{file=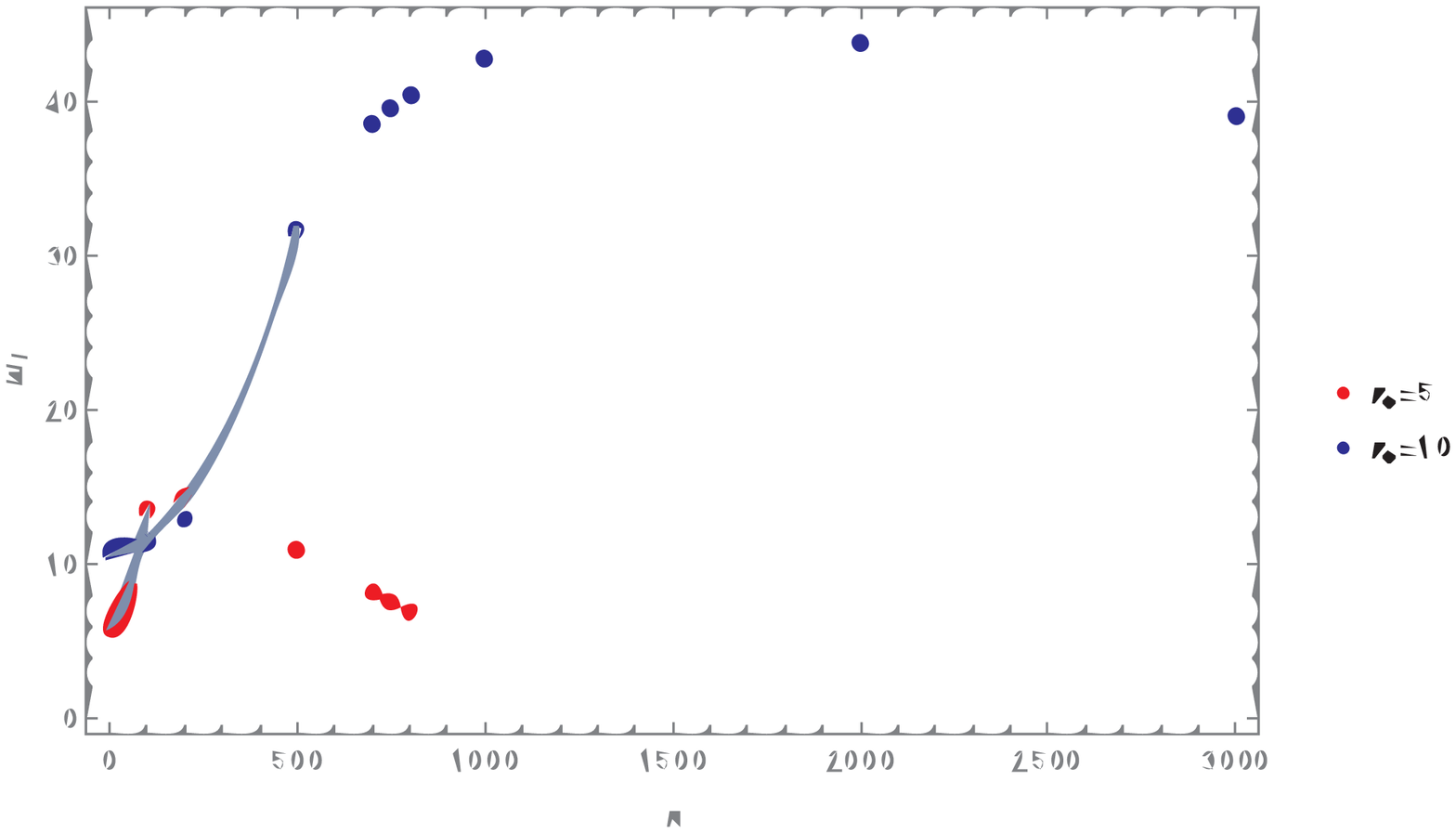, width=0.8\linewidth, height=0.6\linewidth}}
\caption{Gráfico de -$\omega_I$ x $\kappa$ para $r_+=5$ e $10$ com $l=1$}
\end{figure}
\begin{figure}[H]
\centering
{\epsfig{file=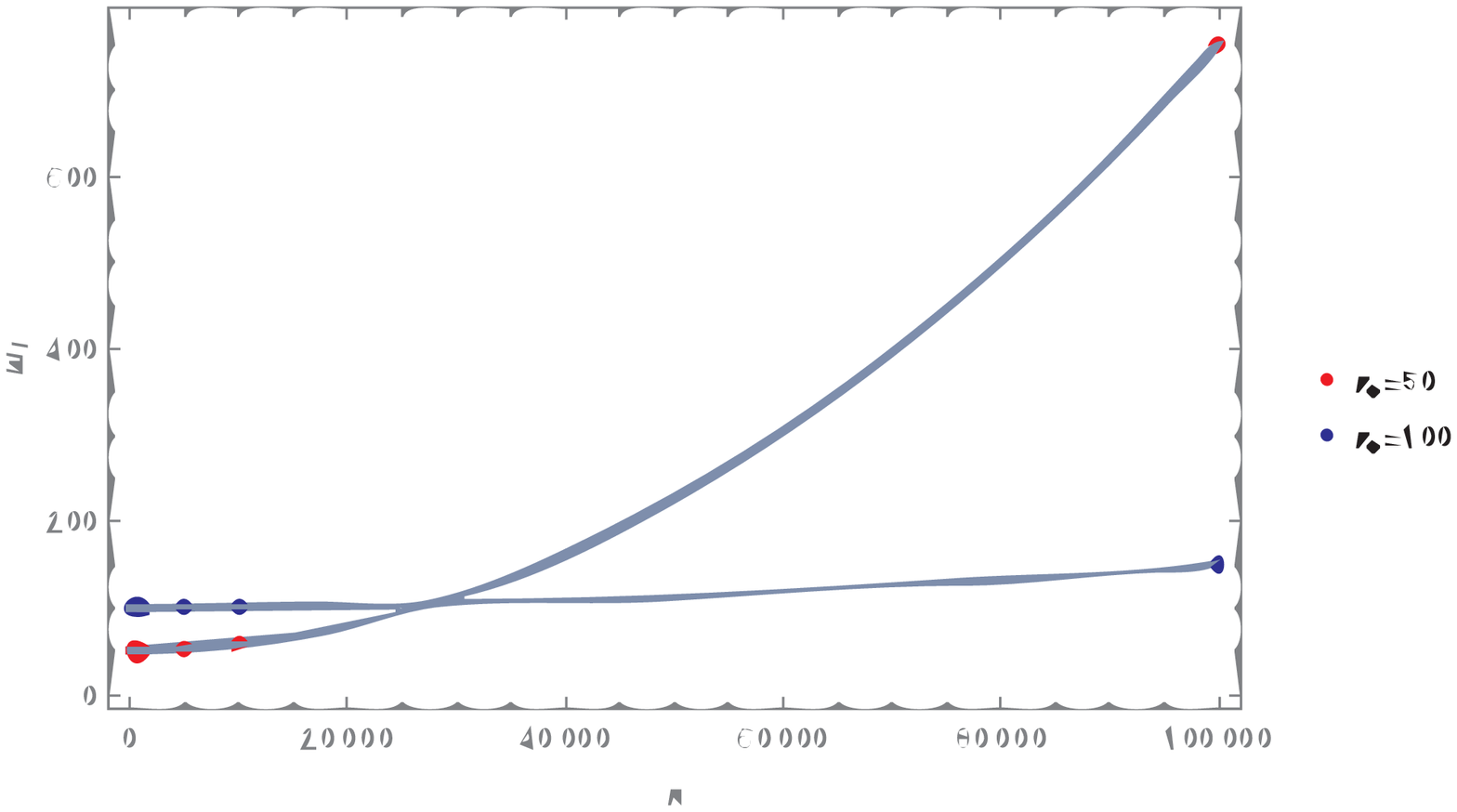, width=0.8\linewidth, height=0.6\linewidth}}
\caption{Gráfico de -$\omega_I$ x $\kappa$ para $r_+=50$ e $100$ com $l=1$}
\end{figure}
Para todos os casos foram feitos ajustes quadráticos, da forma $ax^2+b$, e colocados juntamente com os respectivos pontos.

\chapter{Conclusões}
Neste estudo foi proposta a análise da evolução de dois campos de matéria em buracos negros planarmente simétricos $D$-dimensionais com simetria de Lifshitz. Os campos inspecionados foram o campo escalar e o eletromagnético.

Estes campos, evoluídos nos buracos negros \emph{5Dz1} e \emph{6Dz0}, pelo menos no regime em que trabalhamos, não apresentaram instabilidades, nem referentes aos acoplamentos nem aos espaços-tempos.

Foram observados, para ambos buracos negros, dentro do limite de convergência, regiões com modos puramente imaginários e uma transição para modos que apresentem parte real e imaginária. Esta transição é semelhante aquela que aparece no oscilador harmônico amortecido, com algum valor crítico que não pôde-se especificar nessa investigação.

Nestes casos analisados foi encontrado um comportamento interessante para $\omega_I$ quando $k\neq0$ e $r_+<10$. Em geral, $\omega_I$ é crescente com $k$, mas aqui encontramos um crescimento quadrático para $k\sim r_+$ com um decréscimo quando $k>>r_+$. Esse comportamento pode ser relevante no contexto da dualidade gravidade/calibre.

A aplicação do método AIM se mostrou bastente promissora também para buracos negros AdS, para os quais ele não era geralmente usado. Conseguimos recuperar os valores dos modos quasinormais para Schwarzschild-AdS, BTZ e AGGH e obter um padrão de convergência. Este resultado é muito importante visto que o AIM se comportou bem quando $r_+<1$, região esta na qual o HH geralmente apresenta dificuldades.
\chapter{Desenvolvimentos futuros}
\begin{itemize}
\item Cálculo dos modos quasinormais de oscilação de BTZ e AGGH para $\eta \neq 0$, $\xi\neq 0$ e $m\neq 0$ utilizando HH e AIM.

\item Cálculo dos modos quasinormais de oscilação para \emph{5Dz1} e \emph{6Dz0} utilizando AIM.

\item Análise termodinâmica da estabilidade dos buracos negros BTZ, AGGH e Lifshitz através do Método de Poincarè com utilização das séries lineares representativas de cada ensemble e, em parceria com a Prof. Dr. Berta, comparação destes resultados com outro método termodinâmico para estabilidade de buracos negros AdS.
\end{itemize}


\appendix
\label{apendice}

\chapter{Invariantes escalares de curvatura e tensor de Einstein}

Neste apêndice apresentamos os invariantes escalares de curvatura e as componentes do tensor de Einstein dos espaços-tempos que foram analisados ao longo deste trabalho

\section{\emph{Espaço-tempo de Lifshitz puro $D\ge 3$}}
Escalar de Ricci $\mathcal{R}$
\begin{equation}
\mathcal{R}=-\frac{2}{l^2}\left[\sum_{i=3}^D(i-2)+(D-2)z+z^2\right]
\end{equation}
Contração do Tensor de Ricci
\begin{equation}
R_{\mu\nu}R^{\mu\nu}=\frac{2}{l^4}\left(D-2+z^2\right)\left[\sum_{i=3}^{D}(i-2)+z(D-2+z)\right]
\end{equation}
Escalar de Kretschmann
\begin{equation}
R_{\alpha\beta\mu\nu}R^{\alpha\beta\mu\nu}=\frac{4}{l^4}\left[\sum_{i=3}^{D}(i-2)+(D-2)z^2+z^4\right]
\end{equation}
Tensor de Einstein $G^{\mu\nu}$
\begin{eqnarray}
G^{tt}&=&-\frac{l^{2z-2}}{l^{2z}}\sum_{i=3}^{D}(i-2)\\
G^{rr}&=&\frac{r^2}{l^4}\left[(D-2)z+\sum_{i=3}^{D}(i-3)\right]\\
G^{xx}&=&\frac{1}{r^2}\left[z^2+(D-3)z+\sum_{i=3}^{D}(i-3)\right]
\end{eqnarray}
Contração do Tensor de Einstein
\begin{equation}
G_{\mu\nu}G^{\mu\nu}=\frac{D-2}{l^4}\left[X[D]+z(D-4+z)\right]\left[\sum_{i=3}^{D}(i-2)+(D-2)+z^2\right]
\end{equation}
O termo X acima é dependente da dimensão, porém, não conseguimos encontrar uma forma geral para ele. Sabemos que $X[3]=1$, $X[4]=2$, $X[5]=4$, $X[6]=7$ e $X[7]=11$.

Escalar de Weyl
\begin{equation}
C_{abcd}C^{abcd}=Y[D]\frac{(z-1)^2z^2}{l^4}
\end{equation}
Aqui o termo Y é também dependente da dimensão e não foi possível encontrar uma forma geral para ele. Sabemos que $Y[3]=0$, $Y[4]=4/3$, $Y[5]=2$, $Y[6]=12/5$, $Y[7]=8/3$ e $Y[8]=20/7$.

\section{\emph{Buraco Negro de Lifshitz 5Dz1}}
Escalar de Ricci $\mathcal{R}$
\begin{equation}
\mathcal{R}=-\frac{20}{l^2}+\frac{6M}{r^2}
\end{equation}
Contração do Tensor de Ricci
\begin{equation}
R_{\mu\nu}R^{\mu\nu}=\frac{80}{l^4}+\frac{12M^2}{r^4}-\frac{48M}{l^2r^2}
\end{equation}
Escalar de Kretschmann
\begin{equation}
R_{\alpha\beta\mu\nu}R^{\alpha\beta\mu\nu}=\frac{40}{l^4}+\frac{12M^2}{r^4}-\frac{24M}{l^2r^2}
\end{equation}
Tensor de Einstein $G^{\mu\nu}$
\begin{eqnarray}
G^{tt}&=&\frac{3Ml^2-6r^2}{r^4-Ml^2r^2}\\
G^{rr}&=&-\frac{9M}{l^2}+\frac{3M^2}{r^2}+\frac{6r^2}{l^4}\\
G^{xx}&=&\frac{6r^2-Ml^2}{r^4}
\end{eqnarray}
Contração do Tensor de Einstein
\begin{equation}
G_{\mu\nu}G^{\mu\nu}=3\left(\frac{60}{l^4}+\frac{7M^2}{r^4}-\frac{36M}{l^2r^2}\right)
\end{equation}
Escalar de Weyl
\begin{equation}
C_{abcd}C^{abcd}=\frac{2M^2}{r^4}
\end{equation}

\section{\emph{Buraco Negro de Lifshitz 6Dz0}}
Escalar de Ricci $\mathcal{R}$
\begin{equation}
\mathcal{R}=-\frac{20}{l^2}+\frac{8M}{r^2}
\end{equation}
Contração do Tensor de Ricci
\begin{equation}
R_{\mu\nu}R^{\mu\nu}=\frac{80}{l^4}+\frac{24M^2}{r^4}-\frac{80M}{l^2r^2}
\end{equation}
Escalar de Kretschmann
\begin{equation}
R_{\alpha\beta\mu\nu}R^{\alpha\beta\mu\nu}=\frac{40}{l^4}+\frac{56M^2}{r^4}-\frac{6M}{l^2r^2}
\end{equation}
Tensor de Einstein $G^{\mu\nu}$
\begin{eqnarray}
G^{tt}&=&\frac{10r^2-6Ml^2}{Ml^4-l^2r^2} \\
G^{rr}&=&\frac{6r^2}{l^4}+\frac{2M^2}{r^2}-\frac{8M}{l^2}  \\
G^{xx}&=&\frac{6r^2-2Ml^2}{r^4}
\end{eqnarray}
Contração do Tensor de Einstein
\begin{equation}
G_{\mu\nu}G^{\mu\nu}=\frac{280}{l^4}+\frac{56M^2}{r^4}-\frac{240M}{l^2r^2}
\end{equation}
Escalar de Weyl
\begin{equation}
C_{abcd}C^{abcd}=\frac{192M^2}{5r^4}
\end{equation}

\chapter{Transformação para Equação no Formato Schrodinger - AIM}
Para utilização do método AIM a equação de evolução geral deve estar no formato Schrodinger.

Utilizando a Lagrangeana dada pela Eq.(\ref{lagrangeana}), evolui-se o campo escalar com acoplamento com Tensor de Einstein e Escalar de Ricci, Eq.(\ref{gordon}), em uma métrica $D$-dimensional, esfericamente simétrica e planarmente simétrica.

A equação inicialmente estará no formato
\begin{eqnarray}
C_1\ddot{R}+C_2\dot{R}+\omega^2R-V(r)R=0
\end{eqnarray}
e será transformada para
\begin{eqnarray}
\label{eqtransformada}
\bar{R}^{''}+\omega^2\bar{R}-\bar{V}(r)\bar{R}=0
\end{eqnarray}
onde $\dot{R}$ significa derivada com respeito a coordenada $r$ e $\bar{R}^{'}$ signifca a derivada com respeito a coordenada tartaruga $r_*$. Para isso são feitas as seguintes transformações
\begin{eqnarray}
\label{transfR}
R&=&\bar{R}(r)b(r), \\
\label{transfH}
H(r)&=&\frac{h^{rr}}{\bar{h}^{tt}}, \\
\label{transfr}
r&=&r(r_*).
\end{eqnarray}
Aqui utilizou-se a definição que $\bar{h}^{tt}=-h^{tt}$. A equação com as transformações das Eqs.(\ref{transfR}, \ref{transfH}, \ref{transfr}) fica escrita como segue
\begin{eqnarray}
H\frac{\partial r_{*}}{\partial r}\frac{\partial}{\partial r_{*}}\left[\frac{\partial r_{*}}{\partial r}\frac{\partial(\bar{R}b)}{\partial r_{*}}\right]&+&H\frac{\partial r_{*}}{\partial r}\frac{\partial(\bar{R}b)}{\partial r_{*}}\left[\frac{\partial r_{*}}{\partial r}\frac{\partial}{\partial r_{*}}\ln(h^{rr}\sqrt{-g})\right]+ \nonumber\\
&+&\omega^2\bar{R}b-V(r)\bar{R}b=0.
\end{eqnarray}
Expandindo a equação e dividindo por $b(r)$ e se obtém
\begin{eqnarray}
&&H\dot{r}_*^2\bar{R}^{''}+\bar{R}^{'}\left\{H\dot{r}_*\dot{r}_*^{'}+2H\dot{r}_*^2\frac{b^{'}}{b}+H\dot{r}_*^2\left[\ln(h^{rr}\sqrt{-g})\right]^{'}\right\}+\nonumber\\
&+&\bar{R}\left\{H\dot{r}_*\dot{r}_*^{'}\frac{b^{'}}{b}+H\dot{r}_*^2\frac{b^{''}}{b}+H\dot{r}_*\frac{b^{'}}{b}\left[\ln(h^{rr}\sqrt{-g})\right]^{'}+\omega^2-V(r)\right\}=0,\nonumber\\
\end{eqnarray}
que é uma equação com a seguinte forma
\begin{eqnarray}
C_3\bar{R}^{''}+C_4\bar{R}^{'}+C_5\bar{R}+\omega^2\bar{R}-V(r)\bar{R}=0.
\end{eqnarray}
Agora compara-se tal equação com a Eq.(\ref{eqtransformada}). Disto vem as seguintes condições
\begin{eqnarray}
\label{condicao1}
C_3&=&1,\\
\label{condicao2}
C_4&=&0, \\
\bar{V}&=&V-C_5,
\end{eqnarray}
onde
\begin{eqnarray}
\label{c5}
C_5=H\dot{r}_*\dot{r}_*^{'}\frac{b^{'}}{b}+H\dot{r}_*^2\frac{b^{''}}{b}+H\dot{r}^2_*\frac{b^{'}}{b}\left[\ln(h^{rr}\sqrt{-g})\right]^{'}.
\end{eqnarray}
Logo, vê-se que este termo $C_5$ será subtraído ao potencial ao se fazer a transformação. Da condição da Eq.(\ref{condicao1}) obtém-se
\begin{eqnarray}
\label{condicao11}
\dot{r}_*=\sqrt{\frac{\bar{h}^{tt}}{h^{rr}}} \ \ \ \ \to \ \ \ \ \frac{d}{dr_*}=\sqrt{\frac{h^{rr}}{\bar{h}^{tt}}}\frac{d}{dr}.
\end{eqnarray}
Antes de prosseguir percebe-se que a métrica efetiva pode ser escrita como $h^{\mu\nu}$, de maneira mais conveniente. A função $\bar{h}^{tt}$ pode ser escrita como
\begin{eqnarray}
\bar{h}^{tt}=\bar{g}^{tt}-\eta G^{tt}=\bar{g}^{tt}[1-\alpha(r)],
\end{eqnarray}
onde $\bar{g}^{tt}=-g^{tt}$ e a função $\alpha(r)$ geral. A função $h^{rr}$ escrita como
\begin{eqnarray}
h^{rr}=g^{rr}+\eta G^{rr}=g^{rr}[1+\beta(r)],
\end{eqnarray}
onde a função $\beta(r)$ é geral. Com essa maneira mais conveniente de escrever agora se aplica a condição da Eq.(\ref{condicao11}) à condição da Eq.(\ref{condicao2}). Isto levará a seguinte equação
\begin{eqnarray}
-\frac{\dot{\alpha}}{2(1-\alpha)}+\frac{\dot{\beta}}{2(1+\beta)}+\frac{2\dot{b}}{b}+\frac{D-2}{r}=0,
\end{eqnarray}
que é uma equação para qualquer dimensão $D$. Esta equação leva a uma forma geral para a função $b(r)$, a qual é
\begin{eqnarray}
\label{bgeral}
b=\frac{r^{\frac{2-D}{2}}}{[(1-\alpha)(1+\beta)]^{1/4}}.
\end{eqnarray}
Agora utilizando deste resultado pode-se calcular o termo $C_5$ que será subtraído do potencial $V(r)$ para se tornar o potencial $\bar{V}(r)$. Então, substituindo Eq.(\ref{bgeral}) em Eq.(\ref{c5}) obtém-se
\begin{eqnarray}
C_5=\frac{H}{b}\left[\ddot{b}+\dot{b}\frac{d}{dr}\ln(h^{rr}\sqrt{-g})\right].
\end{eqnarray}
Nota-se que esta foi uma análise feita para uma métrica $D$-dimensional, tanto esfericamente simétrica, com sua parte espacial dada por
\begin{eqnarray}
r^2d\Omega^2=r^2\left[d\chi_2^2+\sin^2\chi_2 d\chi_3^2+\left(\prod_{l=2}^{n-1}\sin^2\chi_l\right) d\chi_n^2\right],
\end{eqnarray}
ou para planarmente simétrica, onde a parte espacial é representada por
\begin{equation}
\frac{r^2}{l^2}d\vec{x}^2=\frac{r^2}{l^2}\sum_{i=1}^{D-2} dx_i^2.
\end{equation}
Agora, para o cálculo direto das equações será avaliado cada caso separadamente.

\section{Esfericamente Simétrica $D$-dimensional}
Para o caso $D$ dimensional pode-se escrever de forma geral o tensor de Einstein. Aqui ele foi separado em parte temporal, radial e parte angular. O índice latino $a$ ($a=2,3,...,D-2$) foi utilizado denotar as componentes angulares.
\begin{eqnarray}
\label{Gtgeral}
G^{tt}&=&\frac{D-2}{2}\frac{\dot{B}}{AB^2r}+\frac{(D-3)(B-1)}{ABr^2} \\
\label{Grgeral}
G^{rr}&=&\frac{D-2}{2}\frac{\dot{A}}{AB^2r}+\frac{(D-3)(1-B)}{B^2r^2} \\
\label{Gfgeral}
G^{aa}&=&-\frac{g^{aa}}{4B^2}\left\{\frac{\dot{A}^2B}{A^2}+\frac{2(D-3)[(D-4)B(B-1)+\dot{B}r]}{r^2}+\right.\nonumber\\
&+&\left.\frac{\dot{A}\dot{B}r-2B[(D-3)\dot{A}+\ddot{A}r]}{Ar} \right\}
\end{eqnarray}
O Escalar de Ricci $\mathcal{R}$ de maneira geral, para $D\ge 2$, é escrito como
\begin{eqnarray}
\label{riccigeral}
\mathcal{R}=\frac{1}{2B^2}\left\{\frac{B\dot{A}^2}{A^2}+\frac{2(D-2)[(D-3)(B-1)B+r\dot{B}]}{r^2}+\frac{\dot{A}\dot{B}-2B[(D-2)\dot{A}/r+\ddot{A}]}{A}\right\}\nonumber\\
\end{eqnarray}
onde as funções $\alpha$ e $\beta$ são definidas, dependentes da dimensão, como
\begin{eqnarray}
\alpha(r)=\eta\left[\frac{D-2}{2}\frac{\dot{B}}{rB^2}+\frac{(D-3)(B-1)}{r^2B}\right]
\end{eqnarray}
\begin{eqnarray}
\beta(r)=\eta\left[\frac{D-2}{2}\frac{\dot{A}}{rAB}-\frac{(D-3)(B-1)}{r^2B}\right]
\end{eqnarray}

\chapter{Transformação da Equação do Formato Schrodinger para o formato do método HH}
A seguinte equação no formato Schrodinger será transformada
\begin{equation}
\label{hh1}
\bar{R}^{''}+\omega^2\bar{R}-\bar{V}\bar{R}=0
\end{equation}
onde $\bar{R}^{'}$ representa a derivação com respeito a coordenada tartaruga $r_*$. Primeiro supõe-se uma solução para $\bar{R}$ do tipo
\begin{equation}
\bar{R}(r_*,t)=\hat{R}(r_*)e^{-i\omega r_*}
\end{equation}
o que quando substituido na Eq.(\ref{hh1}) a ser transformada fica
\begin{eqnarray}
\label{hh3}
\hat{R}^{''}-2i\omega \hat{R}^{'}-\bar{V}\hat{R}=0
\end{eqnarray}
Antes lembrando das definições que
\begin{eqnarray}
H=\frac{h^{rr}}{\bar{h}^{tt}},
\end{eqnarray}
onde
\begin{eqnarray}
h^{rr}&=&g^{rr}+\eta G^{rr},\\
\bar{h}^{tt}&=&\bar{g}^{tt}-\eta G^{tt}.
\end{eqnarray}
Calculando a coordenada tartaruga para fazer a transformação de $r_*\to r$.
\begin{eqnarray}
r_*=\int \frac{dr}{\sqrt{H}} \ \ \ \to \ \ \ dr_*=\frac{dr}{\sqrt{H}} \ \ \ \to \ \ \ \frac{d}{dr_*}=\sqrt{H}\frac{d}{dr}
\end{eqnarray}
Realizando essa transformação na Eq.(\ref{hh3})
\begin{eqnarray}
\sqrt{H}\ddot{\hat{R}}+\left(\frac{d}{dr}\sqrt{H}-2i\omega\right)\dot{\hat{R}}-\frac{\bar{V}}{\sqrt{H}}\hat{R}=0
\end{eqnarray}
onde define-se o potencial $\hat{V}=\bar{V}/\sqrt{H}$, assim
\begin{eqnarray}
\sqrt{H}\ddot{\hat{R}}+\left(\frac{d}{dr}\sqrt{H}-2i\omega\right)\dot{\hat{R}}-\hat{V}\hat{R}=0
\end{eqnarray}
onde $\dot{\hat{R}}$ representa a derivação com respeito a coordenada $r$.

Para o caso particular tratado no artigo HH recupera-se o resultado original, pois, fazendo $\eta=0$, $h^{rr}=g^{rr}$ e $\bar{h}^{tt}=\bar{g}^{tt}$. Assim, $g^{rr}=1/B(r)=f(r)$ e $\bar{g}^{tt}=1/A(r)=1/f(r)$. Desta forma $\sqrt{H}=\sqrt{f^2(r)}=f(r)$ e a equação fica

\begin{eqnarray}
f(r)\ddot{\hat{R}}+(\dot{f}(r)-2i\omega)\dot{\hat{R}}-\hat{V}\hat{R}=0
\end{eqnarray}

\end{document}